\documentclass[
superscriptaddress,
nofootinbib,
 amsmath,amssymb,
 aps,
]{revtex4-2}

\usepackage{xcolor}
\usepackage{graphicx}
\usepackage{dcolumn}
\usepackage{bm}
\usepackage{hyperref}
\usepackage[mathlines]{lineno}
\usepackage{algorithm}
\usepackage{diagbox}
\usepackage{booktabs}
\usepackage{algpseudocode}
\usepackage{comment}

\definecolor{mpl_blue}{HTML}{1F77B4}
\definecolor{mpl_orange}{HTML}{FF7F0E}
\definecolor{mpl_green}{HTML}{2CA02C}
\definecolor{mpl_red}{HTML}{D62728}

\newcommand{\half}{\tfrac{N_t}{2}}

\newcommand{\diff}{\mathop{}\!\mathrm{d}}

\DeclareMathOperator{\supp}{supp}
\algrenewcommand\algorithmicrequire{\textbf{Input:}}
\algrenewcommand\algorithmicensure{\textbf{Output:}}

\begin{document}

\preprint{APS/123-QED}

\title{The WDM Time-Frequency Transform in Gravitational-Wave Data Analysis I: Formalism}

\author{Aaron D. Johnson}
 \email{aaronj@caltech.edu}
\affiliation{NASA Marshall Space Flight Center, Huntsville, AL 35812, USA}
\affiliation{Science and Technology Institute, Universities Space Research Association, Huntsville, AL 35805, USA}
\affiliation{TAPIR, California Institute of Technology, Pasadena, CA 91125, USA}%

\author{Katerina Chatziioannou}
 \email{kchatziioannou@caltech.edu}
\affiliation{TAPIR, California Institute of Technology, Pasadena, CA 91125, USA}%
\affiliation{LIGO Laboratory, California Institute of Technology, Pasadena, CA 91125, USA}

\author{Jake Summers}
 \email{jsummers@caltech.edu}
\affiliation{TAPIR, California Institute of Technology, Pasadena, CA 91125, USA}
\affiliation{LIGO Laboratory, California Institute of Technology, Pasadena, CA 91125, USA}

\date{\today}

\begin{abstract}
For slowly-varying noise, time-frequency methods offer a natural middle ground between the efficiency of the frequency domain and the generality of the more expensive time domain.
Despite growing adoption, such methods remain less well documented and less familiar in the gravitational-wave literature compared to the ubiquitous frequency domain.
Aimed at gravitational-wave analysts, in this paper we present a self-contained account of time-frequency methods, detail derivations for a specific basis, namely the Wilson-Daubechies-Meyer (WDM) basis, and share intuition and lessons learned.
We document key concepts: the orthogonality and good time-frequency localization of the basis, the edge effects at the DC and Nyquist frequencies, the forward and inverse transforms and their practical implementation, and the noise covariance matrix and likelihood in the time-frequency domain.
\end{abstract}

\maketitle


\section{Introduction}
\label{sec:intro}

Time-frequency analyses are rooted in Gabor's seminal paper that applied the mathematics of quantum mechanics to signal processing~\cite{Gabor1946TheoryCommunication}.
Gabor constructed a transform that uses a system of windowed complex exponentials to produce a time-frequency representation on a time-frequency grid (which he called the ``information plane'') consisting of a set of (not necessarily orthogonal) ``quanta of information,'' or \textit{logons}.
A Gabor system is a collection of time-frequency shifted ``atoms'' of the form
\begin{equation}
    g_{mn}(t) = e^{2\pi i m b t}g(t - na)\,,
    \label{eq:gabor}
\end{equation}
where $m, n \in \mathbb{Z}$ and $a, b$ are step sizes for translation in time and frequency respectively such that $g_{mn}$ is centered at $(an,bm)$.
The sampling density is related to the product $ab$. 
Tiling the atoms so that they exactly cover the time-frequency plane without overlapping yields a critically-sampled system with $ab = 1$.\footnote{In more familiar notation that we will adopt later, this corresponds to $\Delta T \Delta F = 1$ for the complex Gabor basis, where $\Delta T$ and $\Delta F$ are the bin sizes in time and frequency respectively. For WDM, $\Delta T \Delta F = 1/2$ as its coefficients are real.}
If $ab>1$, the system is incomplete and there are signals that cannot be represented by the atoms. 
If $ab<1$, the system is overcomplete: the atoms are not orthonormal and a 
given signal has no unique representation.
Critical sampling comes at high cost: the Balian-Low theorem~\cite{Balian1981UnPrincipe, Low1985CompleteSets} prohibits critically-sampled, orthonormal Gabor bases that are also well-localized, i.e., have a finite variance, in both time and frequency, 
\begin{align}
    \int_{-\infty}^{\infty} t^2 |g(t)|^2 \diff t = \infty \quad \mathrm{or} \quad
    \int_{-\infty}^{\infty} f^2 |\tilde{g}(f)|^2\,\diff f = \infty\,.
\end{align}
This means that either the time or frequency domain (or both) window must have poor localization.
This situation is familiar from the common Fourier transform where a box function (well-localized in time) is transformed to a sinc function (spans all frequencies).\footnote{
The Balian-Low theorem applies to the entire basis of Gabor atoms, not to its individual elements. Each atom may saturate the Fourier uncertainty principle and be well localized in both time and frequency. The theorem states that a basis constructed from those atoms will not be simultaneously orthogonal and critically sampled.}

Three escape routes around the conditions of the Balian-Low theorem achieve simultaneous good time and frequency localization:
\begin{itemize}
    \item \textbf{Gabor frames at overcritical density, $ab<1$,} give up orthonormality and accept a redundant representation. In a data analysis context, this complicates the likelihood calculation as generically the covariance matrix includes both the noise covariance and the overlap between atoms. An example of overcritical density is the short-time Fourier transform with overlapping windows.
    \item \textbf{Wavelet transforms} replace the uniform time-frequency grid with a non-uniform dyadic one, thus gaining localization at the cost of frequency-dependent resolution.
    \item \textbf{Wilson bases} keep both critical sampling and orthonormality by replacing the complex exponentials of the Gabor system with real trigonometric functions on a half-shifted lattice.
\end{itemize}
The Wilson-Daubechies-Meyer (WDM) transform takes the third route.

Originally proposed in the context of the coherent WaveBurst pipeline that searches for unmodeled excess power~\cite{Klimenko:2005xv}, the WDM transform combines three insights.
First, in 1987 \citet{Wilson1987} proposed real cosine and sine modulations on a half-shifted lattice, rather than the complex exponential of Eq.~\eqref{eq:gabor}, as a way to evade the Balian-Low theorem at critical density.
Second, in 1991 \citet{doi:10.1137/0522035} proved that if a window that satisfies certain conditions generates a Gabor system at twice the critical sampling density, then the equivalent Wilson system is an orthonormal basis with rapid time-domain decay (exponential for smooth windows, algebraic for compact windows).
The critically-sampled and orthonormal Wilson basis is constructed from a Gabor frame that is overcomplete and thus falls outside the conditions of the Balian-Low theorem.
Together, the above results constitute the ``Wilson-Daubechies'' component of the WDM transform.
Third, the Meyer scaling function~\cite{Meyer1992} is a smooth window function that is compact in frequency and satisfies the conditions of Ref.~\cite{doi:10.1137/0522035}.
The Wilson-Daubechies basis guarantees orthonormality, critical sampling, and good time and frequency localization.
The Meyer window guarantees compact frequency support such that a signal with limited frequencies appears in few time-frequency cells and reduces false-alarms in excess power searches.
Implementation-wise, the WDM transform requires $\mathcal{O}(N \log N)$ operations from the time domain or $\mathcal{O}(N \log N_t)$ operations from the frequency domain~\cite{Cornish:2020odn}.

More recently, \citet{Cornish:2020odn} proposed the WDM transform in the context of modeled searches and inference as a way to handle nonstationary noise.
Such analyses rely on noise-weighted inner products of the form
\begin{equation}
    (\mathbf{a}|\mathbf{b}) \equiv \mathbf{a}^\dagger \mathbf{C}^{-1} \mathbf{b}\,,
\end{equation}
where $\mathbf{a}$ and $\mathbf{b}$ are data representations in some (yet unspecified) basis and $\mathbf{C}$ is the noise covariance matrix in the same basis.
The inner product can be evaluated in any basis, including the frequency~\cite{Cutler:1994ys}, time~\cite{Isi:2021iql}, or WDM~\cite{Cornish:2020odn}, as long as the data and noise covariance are calculated self-consistently.
For stationary noise, the Fourier domain is the clear choice as the noise covariance is diagonal~\cite{Wiener1930,Khinchin1934} and the Gaussian likelihood reduces to the Whittle likelihood~\cite{Whittle1957}.
However, in theory nothing precludes a frequency-domain non-diagonal covariance and associated inner product. 
In practice, therefore, the choice of basis is driven by practical considerations such as orthonormality and computational efficiency~\cite{Necula_2012} or compactness of the signal and noise representation.
The WDM basis is convenient given its orthogonality and $\mathcal{O}(N \log N)$ cost, while also guaranteeing that signals whose power is concentrated along a track in time-frequency such as gravitational-wave chirps have sparse representations~\cite{Cornish:2020odn}.
In the context of LISA~\cite{Colpi:2024xhw}, \citet{Digman:2022jmp} showed that it can model the galactic white dwarf confusion background's
cyclostationarity caused by the satellite's motion around the sun, while \citet{Digman:2022igm} inferred the properties of stellar-origin binaries assuming a known, but nonstationary, noise.

The motivation for this work is to provide a self-contained description of the WDM transform that can serve as an aid toward implementation, discuss its advantages and disadvantages and relation to the more familiar time- and frequency-domain representations, and impart intuition in light of its rising popularity.
Prior treatments leave the derivation, discretization, edge bands, and covariance normalization implicit; we make them explicit and reproducible.
This paper is the first in a series of two: here we describe the formalism and stationary noise, while in the companion~\cite{WDMII} we tackle evolutionary and nonstationary noise along with physically motivated examples.

In Sec.~\ref{sec:WDMtheory} we describe the theory, filter, and basis for the WDM transform.
In Sec.~\ref{sec:transform} we derive the forward and inverse transform of the data.
In Sec.~\ref{sec:PSD} we derive the transform of the noise covariance.
In Sec.~\ref{sec:logL} we derive the likelihood and inner product in the WDM domain.
In Sec.~\ref{app:faq}, we conclude by addressing common questions.
Appendix~\ref{app:POU} proves partition-of-unity, a defining property of the filter function in the Wilson-Daubechies transform. 
Appendix~\ref{app:orthogonality} proves orthogonality of the WDM basis.
Appendix~\ref{app:nyquist} shows that finite-sampled data have a frequency-domain periodicity of twice the Nyquist frequency.
In Appendix~\ref{app:algorithms} we provide algorithm blocks for the forward and inverse WDM transform.
In Appendix~\ref{app:correspondence} we derive the correspondence between the WDM and frequency domains in the appropriate limit.
In Appendix~\ref{app:SFT} we discuss the relation between the WDM and short-time Fourier transforms as well as the implications of a non-orthogonal frame. 
In Appendix~\ref{app:toeplitz} we discuss the (parity-modulated) Toeplitz structure of the WDM covariance matrix.
In Appendix~\ref{app:WDM_Heisenberg} we construct the WDM transform from the Heisenberg group.

\subsection{Notation and Fourier Conventions}
\label{sec:notation}

Throughout, we denote a frequency-domain function with an overhead tilde and a continuous argument of $f$ or discrete index $k$.
The corresponding time-domain function has a continuous argument of $t$ or discrete index $j$.
We use $m$ as the frequency index and $n$ as the time index in the WDM domain.
The notation $(mn)$ denotes a flattened array over both index values so that all pairs fall in a single vector.
Bold fonts denote vectors or matrices in the sense of a collection of numbers such as a timeseries or the noise covariance.
In index notation, elements of the vectors or matrices are denoted with square brackets, e.g., $x[j]$.
However, we keep subscripts for WDM domain quantities to declutter the notation.

Unless stated otherwise, we adopt the Fourier convention using frequency (instead of angular frequency)
\begin{equation}
\begin{split}
    \tilde{x}(f) &= \int_{-\infty}^{\infty}x(t) e^{-2\pi if t} \diff t\,, \\
    x(t) &= \int_{-\infty}^\infty \tilde{x}(f)e^{2\pi if t}\diff f\,.
    \label{eq:FFT_physical_frequency}
\end{split}
\end{equation}
In discrete form, each integral turns into a Riemann sum as
\begin{equation}
\begin{split}
    \tilde{x}[k] &= \Delta t\sum_{j=0}^{N-1}x[j] e^{-2\pi i j k/N}\,, \\
    x[j] &= \Delta f\sum_{k=0}^{N-1} \tilde{x}[k]e^{2\pi i j k/N} = \frac{1}{N\Delta t}\sum_{k=0}^{N-1} \tilde{x}[k]e^{2\pi i j k/N}\,.
    \label{eq:FFT_asymmetric_discrete}
\end{split}
\end{equation}
The \textsc{numpy} fast Fourier transform (FFT) convention involves no normalization prefactor on the forward transform and a division by the number of points for the inverse:
\begin{equation}
\begin{split}
    \tilde{x}_{\mathrm{np}}[k]&=\sum_{j=0}^{N-1} x[j] e^{-2 \pi i j k / N}\,,\\
    x[j]&=\frac{1}{N} \sum_{k=0}^{N-1} \tilde{x}_{\mathrm{np}}[k] e^{2 \pi i j k / N}\,.
    \label{eq:FFT_numpy_discrete}
\end{split}
\end{equation}
In practical computations, the difference between the two conventions comes down to the decision to include the increment $\Delta t$ in the Fourier transform definition.
Whichever way is chosen, the final result should be unaltered assuming appropriate normalization quantities are included in different places.
Numerical algorithms such as those in Appendix~\ref{app:algorithms} are presented in the \textsc{numpy} convention in order to match what is practically implemented.
Correspondingly, we use $\mathrm{FFT}()$ and $\mathrm{IFFT}()$ as operators to indicate the FFT and its inverse respectively in the \textsc{numpy} convention.
Where possible, we will discuss how formulas need to be adjusted to switch conventions.
Because the convention of Eqs.~\eqref{eq:FFT_physical_frequency} and ~\eqref{eq:FFT_asymmetric_discrete} is used in \textsc{LISA Analysis Tools}~\cite{Katz:2024oqg} 
(and also \textsc{LALSuite}~\cite{lalsuite} and \citet{Creighton:2011zz}), we call it the ``LT convention'' from here on.

In Sec.~\ref{sec:logL}, we also invoke the discrete unitary convention
\begin{equation}
\begin{split}
    \tilde{x}_\mathrm{unit}[k] &= \frac{1}{\sqrt{N}}\sum_{j=0}^{N-1}x[j]e^{-2\pi ijk/N}\,,\\
     x[j]&= \frac{1}{\sqrt{N}}\sum_{k=0}^{N-1}\tilde{x}_\mathrm{unit}[k]e^{2\pi ijk/N}\,.
     \label{eq:FFT_unitary_discrete}
\end{split}
\end{equation}
The purpose of this is conciseness: many normalization factors that are absorbed into definitions or cancel anyways, explicitly disappear.

The continuous noise PSD is denoted by $S$ rather than $S_n$ to avoid confusion with the WDM time cell index $n$.

\section{The WDM transform}
\label{sec:WDMtheory}

\begin{figure}[]
    \centering
    \includegraphics{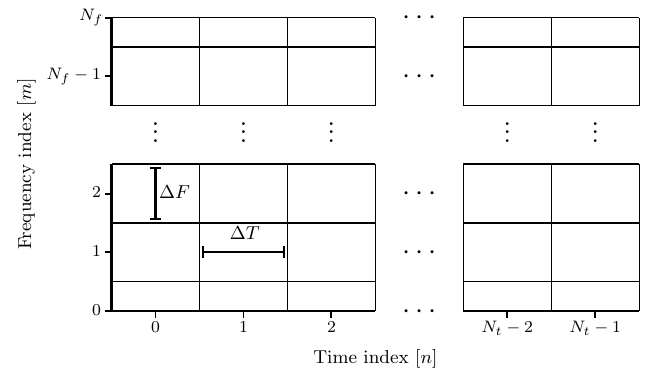}
    \caption{Uniform time-frequency grid of the WDM transform. Time is indexed by $n$ and cells have temporal width $\Delta T$. Frequency is indexed by $m$ and cells have height $\Delta F$. The cell area saturates the uncertainty principle, $\Delta F \Delta T = 1/2$. For $N$ time-domain data points, there are $N - N_t$ full-sized and $2N_t$ half-sized time-frequency cells that correspond to the DC and Nyquist frequencies. 
    Similar to the frequency domain, half of the edge bins are identically zero. The total number of degrees of freedom on the entire grid is thus $N -N_t + (2N_t)/2 = N$, matching the number of time-domain data points.}
    \label{fig:tf_grid}
\end{figure}

Throughout, we consider a set of discrete, real time-domain data $x[j]$ which are sampled from a function $x(t)$ at intervals of $\Delta t$.
With this sampling rate, the time-domain data consist of $N$ samples at times $(0,\Delta t, 2 \Delta t, \ldots, (N - 1)\Delta t)$, for a total observation time of $T = N\Delta t$.
In the frequency domain, there are $N/2+1$ complex data points $\tilde{x}[k]$ at frequencies $(0,\Delta f, 2 \Delta f, \ldots, (N/2)\Delta f)$ with $\Delta f=1/T$.
As in Gabor's construction, the WDM transform breaks the time-frequency plane up into cells with temporal extent $\Delta T$ and bandwidth $\Delta F$ as shown in Fig.~\ref{fig:tf_grid}.
We use indices $n$ and $m$ for the time and frequency coordinates of the cells respectively.

The total number of degrees of freedom remains the same in the time, frequency, and time-frequency domain representations.
This preservation of information can be traced to the fact that all transforms are complete and orthonormal. 
In the time domain, the data are a real one-dimensional vector of length $N$.
In the Fourier domain, we have a one-dimensional complex vector of length $N/2 + 1$.
This still adds up to a total of $N$ degrees of freedom because the $N/2 - 1$ interior bins each consist of two degrees of freedom (real and imaginary part), while the DC and Nyquist bins are purely real.
In the time-frequency domain, the time axis is tiled with $N_t$ time bins, while the frequency axis is tiled with $N_f-1$ interior frequency bins and two edge half-bins, corresponding to the DC and Nyquist frequencies. 
Each interior cell contributes one degree of freedom, but half the edge cells are empty. 
Overall, the total number of degrees of freedom adds up to $(N_f-1)N_t + (2N_t)/2=N_fN_t=N$.
There is no loss of information in performing the WDM transform except when $N$ needs to be truncated so that it is divisible by $N_f$.

The behavior of the edge frequency bins in the WDM domain is qualitatively similar to the frequency domain. 
In the frequency domain, the DC bin is purely real, a simplification that can be traced to the fact that a constant (zero frequency) has an amplitude but not a phase.
Mathematically, the sine basis of the Fourier transform evaluated at zero frequency vanishes identically. 
In the WDM domain, both the basis norm and the basis coefficient vanish identically for $m=0$ and odd-$n$ cells.
The behavior of the Nyquist bin is a consequence of finite sampling and aliasing. 
As there are only $2$ data points per cycle, a phase shift of the Fourier mode results only in a rescaling of the amplitude.
Mathematically, the sine Fourier basis vanishes on all time-domain discrete data where it reduces to $\sin(j\pi)$.
A similar result is reached for the WDM basis for cells with $m=N_f$ and odd $N_f+n$.

With the time-frequency grid in hand, we can define some useful quantities that enter the formalism.
The total time of observation $T$ covers the horizontal span of the time-frequency grid.
Since we have subdivided it into $N_t$ points,
\begin{equation}
    \Delta T = T / N_t = \Delta t N_f\,,
\end{equation}
where we have used $T = N \Delta t$ and $N = N_t N_f$.
Because we are saturating the uncertainty principle $\Delta T \Delta F = 1/2$,
\begin{equation}
    \Delta F = 1 / (2 N_f \Delta t)\,.
\end{equation}
The time and frequency extent of WDM cells is larger than the corresponding time- and frequency-domain bins, $\Delta T>\Delta t$ and $\Delta F>\Delta f$.
Each cell encloses $\Delta T/\Delta t = N_f$ of the time-domain samples and $\Delta F/\Delta f = N_t/2$ of the frequency-domain samples.

\subsection{The generalized Meyer scaling function}

\begin{figure*}[]
    \centering
        \includegraphics[width=0.48\textwidth]{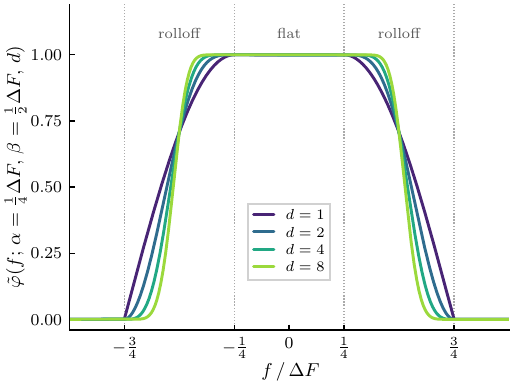}
        \includegraphics[width=0.48\textwidth]{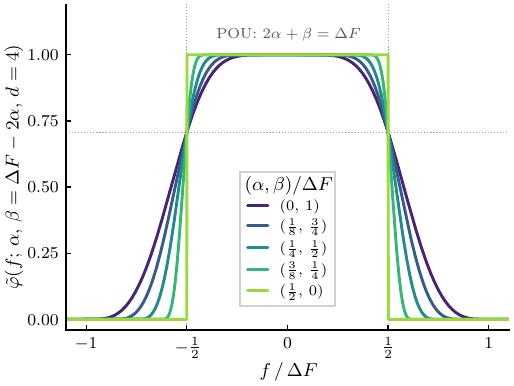}
    \caption{The generalized Meyer filter as used in the WDM basis, varying $d$ (left) and $\alpha$ (right). The filter consists of a central flat region and two rolloff regions that overlap with adjacent bands. The filter is completely determined by parameters $(d, \alpha,\beta)$ with the constraint that $2\alpha + \beta = \Delta F$ required for the partition-of-unity (POU) property. The filter vanishes identically beyond the rolloff regions. The parameter $d$ adjusts the polynomial order of the rolloff region resulting in sharper falloff with increasing $d$, while $\alpha$ and $\beta$ adjust the width of the flat and rolloff regions.
    }
    \label{fig:meyer_filter}
\end{figure*}

\begin{figure}[]
    \centering
    \includegraphics{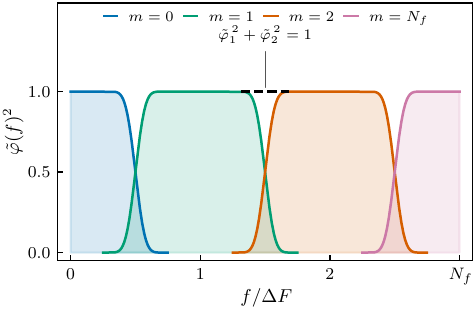}
    \caption{Scaled Meyer filters displayed for a WDM transform with $N_f = 3$ frequency bins. The DC ($m = 0$) and Nyquist ($m = N_f$) bins have half the size and use half the filter function. Overlapping segments must satisfy the partition-of-unity property to ensure uniform power distribution. This means that for any value of $f$, the squared filters must sum to one. We use $\alpha=\Delta F/4, \beta=\Delta F/2$, for which any interior filter will overlap 25\% with its neighbor.}
    \label{fig:wdm_partition}
\end{figure}

The Wilson basis uses a pair of window functions at symmetrically spaced positive and negative frequencies, centered at $m \Delta F$ and $-m\Delta F$.
The \citet{doi:10.1137/0522035} construction requires a real window $\tilde{\phi}$ that is sufficiently smooth and satisfies
\begin{equation}
\sum_{p \in \mathbb{Z}} \tilde{\phi}(f+p \Delta F) \tilde{\phi}(f+(p+2 j) \Delta F)=\frac{\delta_{j0}}{\Delta F}\,, \quad \forall j \in \mathbb{Z}\,.
\label{eq:DJJcondition}
\end{equation}
The case $j=0$ amounts to partition-of-unity depicted in Fig.~\ref{fig:wdm_partition}.
The case $j\neq0$ imposes cross-band cancellations.
Any filter function that satisfies these conditions can be used to construct an orthogonal Wilson basis.

For the WDM basis, the window in the frequency domain is a scaled version of the Meyer scaling function, $\tilde{\phi}(f) = \left(1 / \sqrt{\Delta F}\right)\, \tilde{\varphi}(f)$,
with
\begin{equation}
    \tilde{\varphi}(f) =
    \begin{cases}
        1 & \text{if } |f| < \alpha\\
        \cos \left[\frac{\pi}{2}\nu_d\left(\frac{|f|-\alpha}{\beta}\right) \right] & \text{if } \alpha \leq |f| < \alpha + \beta\\
        0 & \text{otherwise}
    \end{cases}\,,
    \label{eq:Meyer-def}
\end{equation}
and
\begin{align}
    \label{eqn:rolloff}
    \nu_d(x) &= I_x(d, d)= \frac{(2d - 1)!}{[(d - 1)!]^2}\sum_{k=0}^{d-1}(-1)^k \binom{d - 1}{k}\frac{x^{d+k}}{d+k}\,,
\end{align}
where $I_x(d, d)$
is the regularized beta function which is a polynomial of degree $2d - 1$ for positive integer $d$.

The filter has compact support, $\supp \tilde{\phi} = \supp \tilde{\varphi} = [-(\alpha + \beta), (\alpha + \beta)]$.
The free parameters $\alpha,\beta,d$ are subject to the constraint $2\alpha + \beta = \Delta F$, discussed further below.
The filter function for different values of its free parameters is shown in Fig.~\ref{fig:meyer_filter}.
It is characterized by a flat region that is exactly $1$ (whose width is determined by $\alpha$), a cosine-type rolloff (whose width is determined by $\beta$ and shape by $d$), and vanishes exactly outside its area of support. 
Equation~\eqref{eqn:rolloff} generically satisfies $\nu_d(0) = 0$ and $\nu_d(1) = 1$, hence the boundaries of the falloff region vary from $\cos(0)= 1$  to $\cos\left(\pi/2\right) = 0,$ which match the adjacent regions and make the function continuous and $C^{d-1}$, i.e., differentiable to order $d-1$, for $\beta\neq0$. 
Each Meyer scaling function covers one frequency cell and up to half of the two adjacent cells.
Hence the maximum support of a filter centered at $f=0$ is $[-\Delta F,\Delta F]$, achieved for $\alpha=0,\beta=\Delta F$.
Despite this overlapping support, the resulting WDM basis is orthonormal.
The compact support of the Meyer function means that the window automatically satisfies the $j\neq0$ condition of Eq.~\eqref{eq:DJJcondition}.

The Meyer filter further satisfies the $j=0$ condition of Eq.~\eqref{eq:DJJcondition}: partition-of-unity, depicted in Fig.~\ref{fig:wdm_partition}.
Partition-of-unity requires that the overlapping filters squared add up to unity even in the overlap regions:
\begin{equation}
    \sum_{p\in\mathbb{Z}}\tilde{\varphi}(f - p\Delta F)^2 = 1\,.
    \label{eqn:partition}
\end{equation}
The partition-of-unity condition ensures a smooth handoff between neighboring Meyer bands. 
As one filter falls off, the next ramps up so that the total power response remains constant, preventing gaps or double counting in frequency space.
Schematically, consider $f>0$ and the $m=0$ filter centered at $f=0$ and the $m=1$ filter centered at $f=\Delta F$. 
The $m=0$ filter is constant and equal to $1$ until $f=\alpha$. 
At that point, the $m=0$ filter starts turning off and the $m=1$ filter starts turning on.
At $f=\alpha+\beta$ the $m=0$ filter reaches a value of zero, coinciding with the place where the $m=1$ filter reaches a value of one, as $\alpha+\beta=\Delta F-\alpha \Leftrightarrow 2\alpha+\beta=\Delta F$.
At that point, the handoff between the two filters is complete, the $m=1$ filter remains at a value of one until $f=\Delta F+\alpha$, and so on. 
Mathematically, partition-of-unity relies on two conditions: the identity
\begin{equation}
    \label{eqn:part_unity}
    \nu_d(x) + \nu_d(1 - x) = 1\,,
\end{equation}
that follows from Eq.~\eqref{eqn:rolloff},
and the parameter constraint $2\alpha+\beta=\Delta F$ that ensures that adjacent filters have overlapping rolloff regions. 
See Appendix~\ref{app:POU} for the derivation.

Subsequent derivations will be presented for generic parameters $\alpha, \beta$ and $d$.
Numerical implementations and figures, unless otherwise stated, employ $\alpha=\Delta F/4$,  $\beta=\Delta F/2$, and $d=4$.
For this choice, the filter has support in $f/\Delta F \in [-3/4, 3/4]$ and $25\%$ overlap with both of its neighbors.
Moreover, for $d = 4$,
\begin{equation}
    \nu_4(x) = 35x^4 - 84x^5 + 70x^6 - 20x^7\,.
\end{equation}

Before moving on to the full WDM transform, we clarify that use of the Meyer scaling function (which also appears in canonical wavelet transforms) does not qualify WDM to be a ``wavelet transform.''
The latter formally uses translations and dilations, typically with a dyadic scaling,\footnote{A dyadic scaling via dilation with powers of $2$ would rescale the wavelet $j$ by $2^j$ with respect to the mother wavelet.} to produce an efficient and well-localized transform, see Sec.~\ref{sec:wavelet-vs-WDM} for further discussion.
For example, the standard square window is used in the short-time Fourier transform and can also be related to the Haar wavelet, but these transforms have very different properties.
Specifically, the Haar wavelet transform does not guarantee an exact frequency associated with each basis function, while the short-time Fourier transform is either overcomplete or has poor frequency localization as both simultaneously are prohibited by the Balian-Low theorem.
The short-time Fourier transform is further discussed in Appendix~\ref{app:SFT}.

\subsection{The WDM basis}

The WDM basis tiles the time-frequency plane with a scaled Meyer window function that covers positive and negative frequencies centered at $m\Delta F$ and $-m\Delta F$ respectively. 
In the frequency domain,
\begin{equation}
    \tilde{g}_{mn} (f) = A_m e^{-2\pi i n f \Delta T}\left[C_{mn}^* \tilde{\phi}(f + m\Delta F) + C_{mn}\tilde{\phi}(f - m\Delta F)\right]\,,
    \label{eq:gmunu}
\end{equation}
where we have defined
\begin{equation}
     A_{m} =
    \begin{cases}
        \frac{1}{2} & m=0,N_f\\
        \frac{1}{\sqrt{2}} & \mathrm{~otherwise}
    \end{cases}\,,\qquad \mathrm{and} \qquad
    C_{mn} =
    \begin{cases}
        1 & m + n \mathrm{~even}\\
        i & m + n \mathrm{~odd}
        \label{eq:Cmunu}
    \end{cases}\,.
\end{equation}
Orthogonality of this basis is shown in Appendix~\ref{app:orthogonality}:
\begin{equation}
    \int_{-\infty}^\infty \tilde{g}_{mn}(f)\tilde{g}^*_{pq}(f)\diff f = \delta_{mp}\delta_{nq}
    \begin{cases}
        0 & m=0,n\mathrm{~odd}\\
        0 & m=N_f, N_f+n \mathrm{~odd}\\
        1 & \mathrm{~otherwise}
    \end{cases}\,.
    \label{eqn:orthonormality}
\end{equation}
The vanishing cells at the DC and Nyquist frequencies correspond to the half-degrees of freedom of these bands.

The positive/negative frequency coverage is the crucial step in Wilson's construction that distinguishes Eq.~\eqref{eq:gmunu} from a Gabor system of Eq.~\eqref{eq:gabor}.
The inverse Fourier transform of the frequency-shifted window is
\begin{equation}
    \tilde{\phi}(f \pm m\Delta F) \xrightarrow{\text{IFFT}} e^{\mp 2\pi i m \Delta F t} \phi (t)\,,
\end{equation}
and adding the time shift leads to 
\begin{equation}
    e^{-2\pi i n f \Delta T} \tilde{\phi}(f \pm m\Delta F) \xrightarrow{\text{IFFT}} e^{\mp 2\pi i m \Delta F (t-n\Delta T)} \phi (t-n\Delta T)\,.
\end{equation}
The basis then becomes
\begin{align}
    \tilde{g}_{mn} (f) &= 
     \begin{cases}
        A_m e^{-2\pi i n f \Delta T}\left[\tilde{\phi}(f + m\Delta F) + \tilde{\phi}(f - m\Delta F)\right] & m + n \mathrm{~even}\\
        - A_m e^{-2\pi i n f \Delta T} i \left[\tilde{\phi}(f + m\Delta F) - \tilde{\phi}(f - m\Delta F)\right] & m + n \mathrm{~odd}
    \end{cases} \quad \xrightarrow{\text{IFFT}} \nonumber\\
    {g}_{mn} (t) &= 
     \begin{cases}
        A_m \left[e^{-2\pi i m \Delta F (t-n\Delta T)} + e^{+ 2\pi i m \Delta F (t-n\Delta T)} \right] \phi (t-n\Delta T) & m + n \mathrm{~even}\\
        - A_m i \left[e^{- 2\pi i m \Delta F (t-n\Delta T)} - e^{+2\pi i m \Delta F (t-n\Delta T)}\right] \phi (t-n\Delta T) & m + n \mathrm{~odd}
    \end{cases} \nonumber\\
     &= 
     \begin{cases}
        2 A_m \cos \left[2\pi m \Delta F (t-n\Delta T)\right] \phi (t-n\Delta T) & m + n \mathrm{~even}\\
        - 2 A_m \sin \left[2\pi m \Delta F (t-n\Delta T)\right] \phi (t-n\Delta T) & m + n \mathrm{~odd}
    \end{cases} \,.
    \label{eq:gmunu_int}
\end{align}
Since $\tilde{\phi}(f)$ is both real and even, $\phi(t)$ is real as well, which makes the time-domain basis real.

Contrasted with the Gabor system of Eq.~\eqref{eq:gabor}, this basis replaces the complex exponentials with real sine and cosine modulations.
The modulation alternates across the lattice: at a fixed frequency, i.e., fixed $m$, consecutive time cells counted by $n$ alternate between sine and cosine.
The alternating time cells are $\pi/2$ apart in phase and the lattice is half-shifted by $\Delta T/2$.
Still, for a given $m$ the two combined real half-shifted lattices contain the same information as the complex Gabor lattice. 

Intuitively, the Wilson construction is based on the following idea.
A Wilson basis has support at both $+m\Delta F$ and $-m\Delta F$, while the Gabor frame of Eq.~\eqref{eq:gabor} occupies a single frequency. 
Additionally, a Gabor frame at \emph{twice} the critical density evades the Balian-Low theorem.
The Wilson construction then folds the overcomplete Gabor grid via the real sine/cosine pairing on the Wilson half-shifted lattice, thus going from an overcomplete, complex Gabor frame to a critically-sampled, real Wilson basis. 
The overcompleteness by a factor of $2$ is exactly canceled out by the sine/cosine pairing that reduces the complex frame into a real basis.
Correspondingly, the critical-sampling condition for the real Wilson grid is $\Delta T\Delta F = 1/2$ contrasted to $\Delta T\Delta F = 1$ for the complex Gabor grid.

\begin{figure}
    \centering
    \includegraphics{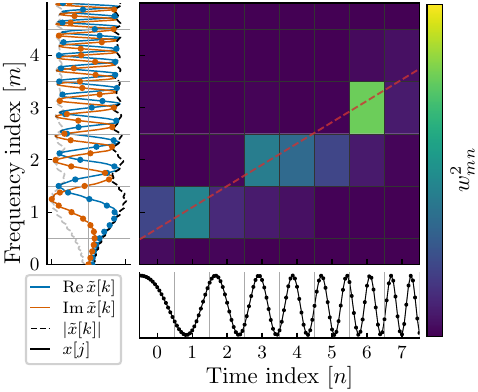}
    \caption{WDM transform for a linear chirp signal. The middle panel shows a scalogram with a heat map corresponding to the power (squared coefficients) in each time-frequency cell. The analytical frequency as a function of time is shown as a red, dashed line. The horizontal axis gives the time-series representation of the data along with partitions of size $\Delta T$.
    The vertical axis gives the real, imaginary, and absolute value of the frequency-series with partitions of size $\Delta F$. 
    }
    \label{fig:time_freq_wdm}
\end{figure}

As shown in Fig.~\ref{fig:wdm_partition}, the full WDM basis in the frequency domain is a set of $N_f - 1$ full-width filters for the frequency band interior and $2$ half-width filters for the edges.
Basis functions covering the two edges of the frequency band at the DC and Nyquist frequencies have a prefactor smaller by $\sqrt{2}$, necessary for normality, see Appendix~\ref{app:orthogonality}.
This is because the positive and negative frequency segments of the basis overlap for the DC and Nyquist bins.
As real time-domain data have Hermitian symmetry under the Fourier transform, we restrict the figure to positive frequencies.
An example of the time, frequency, and WDM-domain data for a simple linear chirp signal is given in Fig.~\ref{fig:time_freq_wdm}.

\section{Forward and inverse transform of the data}
\label{sec:transform}

The WDM transform is more efficiently calculated from frequency-domain data, because the Meyer filter has compact support in the frequency domain but it is only algebraically decaying (but formally stays nonzero) in the time domain.
Going directly from the time domain to the WDM domain would thus require truncation of the filter function.
If we start from time-domain data, we first Fourier transform them and then compute the WDM transform.
Below, we present this route of the transform as well as its inverse.
The frequency index runs $m\in[0,N_f]$, with $m = 0$ the DC band and $m = N_f$ the Nyquist band for a total of $N_f+1$ indices.
The time index runs $n \in[0,N_t-1]$ for a total of $N_t$ indices. 
The two band edges contribute $N_t/2$ degrees of freedom each, preserving the count $N_fN_t = N$.

\subsection{Frequency domain to WDM domain}
\label{sec:forward}

We carry out the derivation for a continuous function, $\tilde{x}(f)$, and discretize the final answer to connect to numerical implementations.
Once in the Fourier domain, the data can be expanded in the WDM basis of Eq.~\eqref{eq:gmunu}
\begin{equation}
    \tilde{x}(f) = \sum_{m=0}^{N_{f}}\sum_{n=0}^{N_{t} - 1} w_{mn} \tilde{g}_{mn}(f)\,,
    \label{wdm_decomp}
\end{equation}
where $w_{mn}$ are the real WDM coefficients, with $m$ the frequency and $n$ the time index of a given time-frequency cell.
The coefficients are calculated in the standard way of multiplying by the conjugate of a basis element and invoking orthogonality:
\begin{equation}
\int_{-\infty}^{\infty}\tilde{x}(f)\tilde{g}_{pq}^*(f)\,\diff f = \int_{-\infty}^{\infty}\sum_{m,n} w_{mn}\,
   \tilde{g}_{mn}(f)\tilde{g}_{pq}^*(f)\diff f\Rightarrow\\
    w_{mn} = \int_{-\infty}^{\infty} \tilde{x}(f) \tilde{g}^*_{mn}(f) \diff f\,.
    \label{eq:wmn_integral}
\end{equation}

Substituting Eq.~\eqref{eq:gmunu} in Eq.~\eqref{eq:wmn_integral}, and using the compact support of the Meyer filter, we find
\begin{equation}
w_{mn} =
A_m
\left[
\int_{(-m - 1)\Delta F}^{(-m + 1)\Delta F}
C_{mn}\tilde{x}(f)\tilde{\phi}(f + m\Delta F)
e^{2\pi inf\Delta T}\,\diff f
+ \int_{(m - 1)\Delta F}^{(m+1)\Delta F}
C_{mn}^*\tilde{x}(f)\tilde{\phi}(f - m\Delta F)
e^{2\pi inf\Delta T}\,\diff f
\right] \,.
\label{eq:wmnsub1}
\end{equation}
The limits of integration have been restricted to the maximum possible support of the Meyer filter.
Substituting $f \rightarrow -f$ in the first integral,
\begin{equation}
    w_{mn} = A_m \left[\int_{(m-1)\Delta F}^{(m+1)\Delta F} C_{mn}\tilde{x}(-f)\tilde{\phi}\left(-f + m \Delta F\right) e^{-2\pi inf\Delta T} \diff f + \int_{(m-1)\Delta F}^{(m+1)\Delta F} C_{mn}^*\tilde{x}(f)\tilde{\phi}\left(f - m \Delta F\right) e^{2\pi inf\Delta T} \diff f\right]\,.
\end{equation}
Since the time series is real, the data have Hermitian symmetry, $\tilde{x}(-f) = \tilde{x}^*(f)$. 
We use the even symmetry of the filter function, $\tilde{\phi}(f) = \tilde{\phi}(-f)$, to rewrite the first integral as
\begin{equation}
    w_{mn} = A_m \left[\int_{(m-1)\Delta F}^{(m+1)\Delta F} C_{mn}\tilde{x}^*(f)\tilde{\phi}\left(f - m \Delta F\right) e^{-2\pi inf\Delta T} \diff f + \int_{(m-1)\Delta F}^{(m+1)\Delta F} C_{mn}^*\tilde{x}(f)\tilde{\phi}\left(f - m \Delta F\right) e^{2\pi inf\Delta T} \diff f\right]\,.
\end{equation}
Since the first integral is the complex conjugate of the second, the coefficients are manifestly real: 
\begin{equation}
    \label{eqn:wmn_last}
    w_{mn} = 2 A_m\int_{(m-1)\Delta F}^{(m+1)\Delta F}\Re \left[C^*_{mn}\tilde{x}(f)\tilde{\phi}\left(f - m \Delta F\right) e^{2\pi inf\Delta T}\right]\diff f\,.
\end{equation}
There are two ways of looking at this equation.
As $m$ increments, either ``slide'' the filter $\tilde{\phi}$ across the data, or keep the filter fixed and move the data $\tilde{x}$ through it.
We choose the latter as it lines up with numerical implementations and fix the filter through the transformation $f' = f - m\Delta F$.
Since $\Delta F \Delta T = 1/2$, this can be written as 
\begin{align}
    w_{mn} &= 2 A_m\int_{-\Delta F}^{\Delta F}\Re \left[C^*_{mn}\tilde{x}(f' + m \Delta F)\tilde{\phi}\left(f'\right) e^{2\pi in(f' + m\Delta F)\Delta T}\right]\diff f' \nonumber\\
    &= 2 A_m (-1)^{mn}\int_{-\Delta F}^{\Delta F}\Re \left[C_{mn}^*\tilde{x}(f' + m \Delta F)\tilde{\phi}\left(f'\right) e^{2\pi inf'\Delta T}\right]\diff f'\nonumber\\
    &= 2 A_m(-1)^{mn}\int_{-\Delta F}^{\Delta F}\Re \left[C_{mn}^*\tilde{x}(f + m \Delta F)\tilde{\phi}\left(f\right) e^{2\pi inf\Delta T}\right]\diff f\,.
    \label{eq:wnm_final}
\end{align}
In the last line, we have relabeled $f' \rightarrow f$.
This derivation applies to the interior and edge $m=0,N_f$ bands alike. 
Numerically, the basis prefactor makes the edge coefficients a factor of $\sqrt{2}$ lower than the interior ones.

A crucial difference, however, is that Eq.~\eqref{eq:wnm_final} vanishes identically for edge cells with $m+n$ odd, thus leaving the edge bands with a total of $N_t/2$ degrees of freedom, compared to $N_t$ for the interior ones.
For the DC band, this is immediately seen from Eq.~\eqref{eq:wmnsub1} which for $m=0$ reduces to
\begin{align}
    w_{0n} &= \frac{1}{2}
    \left[
    \int_{- \Delta F}^{\Delta F}
    C_{0n}\tilde{x}(f)\tilde{\phi}(f)
    e^{2\pi inf\Delta T}\,\diff f
    + \int_{-\Delta F}^{\Delta F}
    C_{0n}^*\tilde{x}(f)\tilde{\phi}(f)
    e^{2\pi inf\Delta T}\,\diff f
    \right]\nonumber \\
    &= \int_{-\Delta F}^{\Delta F} \Re\left[ C_{0n}^*\right]\tilde{x}(f)\tilde{\phi}(f) e^{2\pi inf\Delta T}\diff f\,.
\end{align}
For $n$ odd, $\Re\left[ C_{0n}^*\right]=0$ and the WDM coefficients vanish for any data.

For the $m=N_f$ Nyquist band, the derivation cannot be done purely in the continuous domain and we need to consider the finite sampling condition. 
In Appendix~\ref{app:nyquist} we show that the finite sampling of the time-domain data results in a periodicity of twice the Nyquist frequency for the frequency-domain data.
Equivalently, $\tilde{x}(f - 2N_f\Delta F) =\tilde{x}(f)$ and $\tilde{x}(f - N_f\Delta F) =\tilde{x}(f + N_f\Delta F)$.
Returning to Eq.~\eqref{eq:wmnsub1} for $m=N_f$, we find
\begin{equation}
    w_{N_fn} = \frac{1}{2} \left[\int_{(-N_f - 1)\Delta F}^{(-N_f+1)\Delta F}  C_{N_fn}\tilde{x}(f)\tilde{\phi}(f+N_f\Delta F) e^{2\pi inf\Delta T}\diff f
    + \int_{(N_f - 1)\Delta F}^{(N_f+1)\Delta F} C_{N_fn}^*\tilde{x}(f)\tilde{\phi}(f-N_f\Delta F) e^{2\pi inf\Delta T} \diff f\right]\,.
\end{equation}
Focusing on the integral of the first term, $I_1$, we substitute $f' = f + 2N_f\Delta F$ and get
\begin{align}
   I_1&= \frac{1}{2} \int_{(N_f - 1)\Delta F}^{(N_f + 1)\Delta F}  C_{N_fn}\tilde{x}(f' - 2N_f\Delta F)\tilde{\phi}(f'-N_f\Delta F) e^{2\pi in(f' - 2N_f\Delta F)\Delta T}\diff f' \nonumber\\
    &= \frac{1}{2} \int_{(N_f - 1)\Delta F}^{(N_f + 1)\Delta F}  C_{N_fn}\tilde{x}(f')\tilde{\phi}(f'-N_f\Delta F) e^{2\pi inf'\Delta T}(e^{-2\pi i})^{n N_f} \diff f'\nonumber\\
    &= \frac{1}{2} \int_{(N_f - 1)\Delta F}^{(N_f + 1)\Delta F}  C_{N_fn}\tilde{x}(f)\tilde{\phi}(f-N_f\Delta F) e^{2\pi inf\Delta T} \diff f\,,
\end{align}
where in the second line we have used the $2f_\mathrm{Nyq}$ periodicity of the data and in the last line we have redefined $f' \rightarrow f$.
With this, the full WDM coefficient is
\begin{align}
    w_{N_fn} &= \frac{1}{2} \left[\int_{(N_f - 1)\Delta F}^{(N_f + 1)\Delta F}  C_{N_fn}\tilde{x}(f)\tilde{\phi}(f-N_f\Delta F) e^{2\pi inf\Delta T} \diff f+\int_{(N_f - 1)\Delta F}^{(N_f+1)\Delta F} C_{N_fn}^*\tilde{x}(f)\tilde{\phi}(f-N_f\Delta F) e^{2\pi inf\Delta T} \diff f\right]\nonumber\\
    &= \int_{(N_f - 1)\Delta F}^{(N_f + 1)\Delta F} \Re\left[C_{N_f n}^* \right]\tilde{x}(f)\tilde{\phi}(f-N_f\Delta F) e^{2\pi inf\Delta T}\diff f\,.
\end{align}
Similarly to the DC case, WDM coefficients for Nyquist cells with $N_f+n$ odd vanish.

\subsubsection{Discretization}
\label{sec:forwarddiscrete}

Equation~\eqref{eq:wnm_final} gives the WDM coefficients for the continuous transform. 
We now switch to practical implementation details and discretize it, also switching from the FFT convention of Eq.~\eqref{eq:FFT_physical_frequency} to the \textsc{numpy} discrete Fourier transform of Eq.~\eqref{eq:FFT_numpy_discrete}.
The discretization $f_k = k \Delta f$ has two effects: (i) $\tilde{x}(f_k) \rightarrow \Delta t \,\tilde{x}[k]$, and (ii) the WDM integral becomes a sum with increment $\diff f \rightarrow \Delta f$. 
Noting that $\Delta F/\Delta f= N_t/2$ and $\Delta f\Delta T=1/N_t$ yields
\begin{equation}
    w_{mn} = 2 A_m (-1)^{mn}\Delta f \Delta t \sum_{k=-N_t/2}^{N_t/2-1}\Re \left(C_{mn}^*\tilde{x}\left[k + m \frac{N_t}{2}\right]\tilde{\phi}\left[k\right] e^{2\pi i n k /N_t}\right)\,.
\end{equation}
The limits of $(-\Delta F, \Delta F)\rightarrow \{-N_t/2,N_t/2-1\}$ correspond to the maximum possible bandwidth of the filter function.\footnote{For maximal efficiency we could use a \textit{pruned} FFT~\cite{Markel1971} to avoid the zeros in the filter given the choice of filter parameters.
However, with our choice of $\alpha, \beta$, the filter is nonzero in $[-(3/4), (3/4)]\Delta F$, so this saves at most 25\% of computation time and depends heavily on the efficiency of the implementation.}

Next, we adjust the sum so that it centers both the filter and the $m$th band.
This keeps the frequencies in order from negative to positive, $k \in \left[0, N_t-1\right]$, which is more intuitive.
Defining $k' = k + N_t/2$ we get
\begin{align}
    w_{mn} &= 2 A_m (-1)^{mn}\Delta f \Delta t \sum_{k'=0}^{N_t - 1}\Re \left(C_{mn}^*\tilde{x}\left[k' + (m - 1) \frac{N_t}{2}\right]\tilde{\phi}\left[k' - \frac{N_t}{2}\right] e^{2\pi in (k' - N_t/2)/N_t}\right)\,.
    \label{eq:wnm_inter}
\end{align}
The exponential yields another factor of $(-1)^n$.
Exchanging the sum with the real operator and relabeling $k'\rightarrow k$, we define the windowed slice of the spectrum $\tilde{u}_m[k]$ of length $N_t$ as the data scaled by the $m$th filter 
\begin{equation}
\tilde{u}_m[k]=\tilde{x}\left[k + (m - 1) \frac{N_t}{2}\right]\tilde{\phi}\left[k - \frac{N_t}{2}\right]\,,
\end{equation}
with $k\in[0,N_t-1]$ and $y_m[n]$ as a vector of length $N_t$ 
\begin{equation}
    N_t y_m[n] = N_t\frac{1}{N_t}\sum_{k=0}^{N_t-1}\tilde{u}_m[k] e^{2\pi in k/N_t} = N_t\,\mathrm{IFFT}(\tilde{u}_m[k])\,,
    \label{eq:ym_defn}
\end{equation}
where the factor of $1/N_t$ accounts for the inverse FFT prefactor in the \textsc{numpy} convention as shown in Eq.~\eqref{eq:FFT_numpy_discrete}.
The remaining normalization terms of Eqs.~\eqref{eq:wnm_inter} and~\eqref{eq:ym_defn} simplify to 
\begin{equation}
    2 A_m (-1)^{mn}(-1)^n\Delta f \Delta t N_t = 2 A_m (-1)^{(m+1)n} \frac{N_t}{N} = 2 A_m (-1)^{(m+1)n} \frac{1}{N_f}\,,
\end{equation}
where in the first equality we have used $\Delta f \Delta t = 1/N$. Finally, from Eq.~\eqref{eq:wnm_inter} we arrive at the final form of the WDM coefficients for the interior bands in the \textsc{numpy} convention:
\begin{equation}
    w_{mn} = \frac{2 A_m}{N_f} (-1)^{(m+1)n}\Re \left(C_{mn}^*y_m[n]\right)\,.
    \label{eq:forward_WDM_fast}
\end{equation}
In the LT convention, the $1/N_f$ factor is hidden so that the prefactor in Eq.~\eqref{eq:forward_WDM_fast} is simply $2 A_m$.

\begin{figure}
    \centering
    \includegraphics[width=\textwidth]{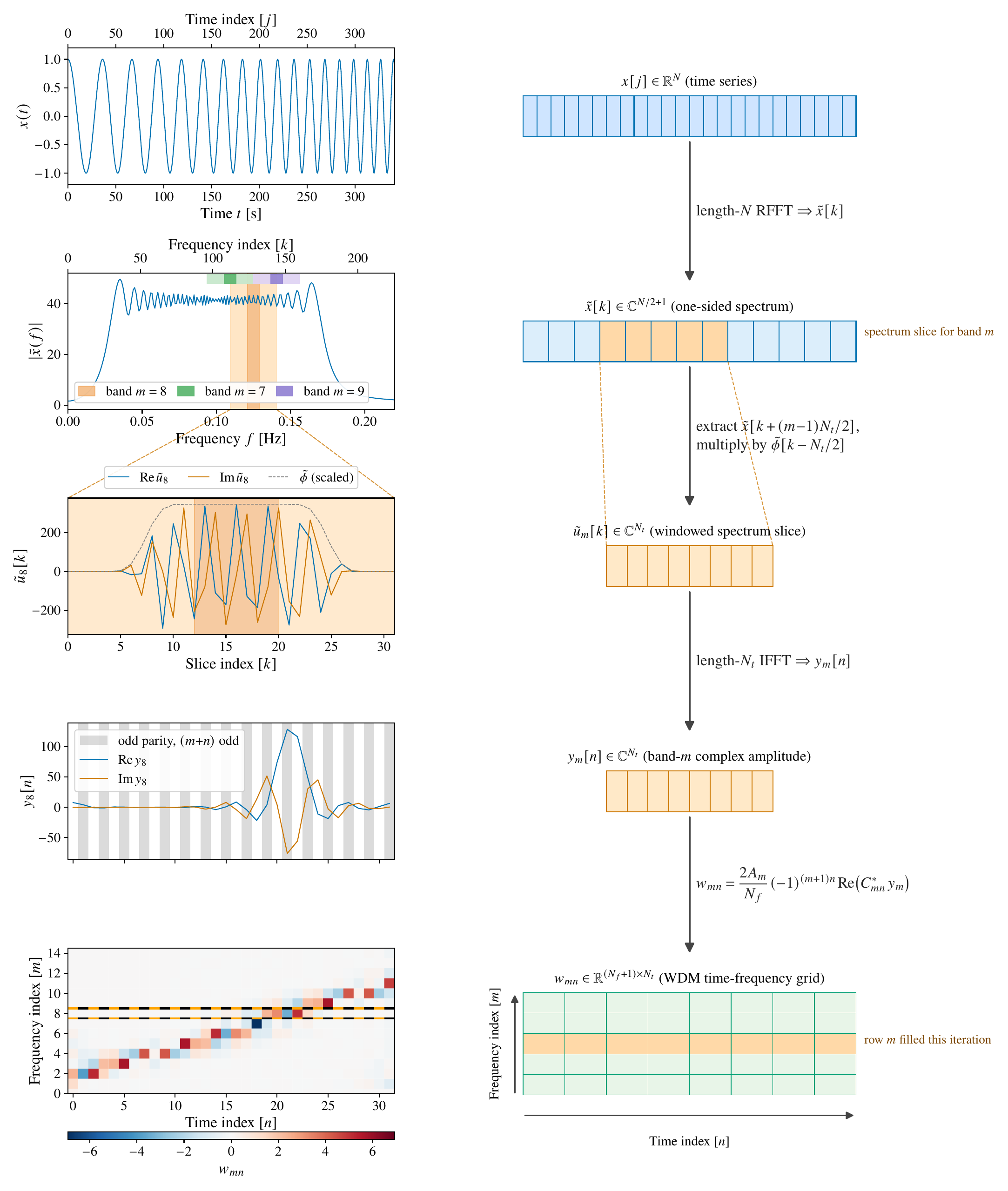}
    \caption{Schematic demonstration of the forward transform from the time to the WDM domain described in Sec.~\ref{sec:forward}, using the \textsc{numpy} convention.
    The left column plots the series in the relevant domain and the right column shows the corresponding bins or cells.
    First row: A real time series of length $N$.
    Second row: The complex FFT of length $N/2+1$ and the $m$th WDM band of length $N_t$.
    Third row: The real and imaginary parts of the spectrum slice $\tilde{u}_m$ (the data $\tilde{x}$ scaled by the $m$-centered filter $\tilde{\phi}$) of length $N_t$.
    Fourth row: The inverse Fourier transform of the sliced data of length $N_t$ with gray vertical striping indicating the $m + n$ odd parity segments.
    Fifth row: The WDM coefficients of the $m$th band.
    }
    \label{fig:wdm_forward_schematic}
\end{figure}

In practice, the prescription is to loop over rows of frequency index $m$ and compute all columns of time index $n$.
For each segment of data, the filter function $\tilde{\phi}[k]$ covers $N_t$ points of data $\tilde{x}[k]$ and the same number are output upon inverse Fourier transform.
We then precompute $C_{mn}^*$ and take the real or imaginary part of $y_m$
depending on $m+n$ being even or odd.
Appendix~\ref{app:algorithms} provides an algorithm block and Fig.~\ref{fig:wdm_forward_schematic} a schematic of the forward transform, all in the \textsc{numpy} convention: Fourier transform, the $m$th band slice, the $m$th windowed spectrum $\tilde{u}_m[k]$, its IFFT $y_m[n]$, the parity-modulated WDM coefficients.

\subsection{WDM domain to frequency domain}
\label{sec:inverse}

To recompose the data from the WDM back to the frequency domain we start by substituting Eq.~\eqref{eq:gmunu} for the basis in Eq.~\eqref{wdm_decomp} and restrict to the one-sided positive frequency data
\begin{equation}
    \tilde{x}(f) = \sum_{m=0}^{N_{f}}\sum_{n=0}^{N_{t} - 1} A_m w_{mn}e^{-2\pi i n f \Delta T}\left[C_{mn}^* \tilde{\phi}(f + m\Delta F) + C_{mn}\tilde{\phi}(f - m\Delta F)\right]\,.
    \label{eq:inversestart}
\end{equation} 
Negative frequencies can be trivially obtained via Hermitian symmetry. 
From here the calculation for the interior and edge bands proceeds differently.

\subsubsection{Interior frequency bands}
\label{sec:inverseinterior}

For the interior bands, $m\neq0,N_f$, the first term of Eq.~\eqref{eq:inversestart} drops out as $\tilde{\phi}(f + m\Delta F)$ has no support for $f>0$. 
The contribution of a single interior band $m$ on the frequency-domain data is
\begin{equation}
    \tilde{x}_m(f) = \frac{1}{\sqrt{2}}\sum_{n=0}^{N_{t} - 1} w_{mn}C_{mn} e^{-2\pi i n f \Delta T}\tilde{\phi}\left(f - m\Delta F\right)\,.
\end{equation}
Transforming $f' = f - m\Delta F$ (and relabeling $f'\rightarrow f$) yields
\begin{align}
    \tilde{x}_m\left(f + m\Delta F\right) &= \frac{1}{\sqrt{2}}\tilde{\phi}\left(f\right) \sum_{n=0}^{N_t-1}w_{mn} C_{mn} e^{-2\pi in(f+m\Delta F)\Delta T}\nonumber\\
     &= \frac{1}{\sqrt{2}}\tilde{\phi}\left(f\right) \sum_{n=0}^{N_t-1}(-1)^{mn}w_{mn} C_{mn} e^{-2\pi inf\Delta T}\,.
     \label{eq:inverseinterior_final}
\end{align}

\subsubsection{Edge frequency bands}
\label{sec:edgeinverse}

The contribution of the edge bands $m=0,N_f$ requires two modifications: the basis prefactor is $1/2$ instead of $1/\sqrt{2}$ and the positive/negative components of the basis coincide. 
The two terms in the square bracket of Eq.~\eqref{eq:inversestart} collapse to the same value. 
This follows immediately for $m=0$ as the filters actually coincide. 
For $m=N_f$ the filters do not coincide; instead the $2f_{\rm Nyq}$ periodicity of the discrete data shown in Appendix~\ref{app:nyquist}, together with a $f\to f-2N_f\Delta F$ change-of-variable in the first term, maps $\tilde\phi(f+N_f\Delta F)\to\tilde\phi(f-N_f\Delta F)$, and hence the two terms again coincide.

For both edge bands, their contribution to the frequency-domain data is
\begin{align}
    \tilde{x}_m(f) &= \frac{1}{2}\sum_{n=0}^{N_{t} - 1} w_{mn}\left[C^*_{mn}+C_{mn} \right]\tilde{\phi}\left(f - m\Delta F\right)e^{-2\pi i n f \Delta T}\nonumber\\
    &= \sum_{n=0}^{N_{t} - 1} w_{mn}\Re\left[C_{mn} \right]\tilde{\phi}\left(f - m\Delta F\right)e^{-2\pi i n f \Delta T}\,.
\end{align}
For $n+m$ odd the summand vanishes as $\Re\left[C_{mn} \right]=0$ but also $w_{mn}=0$.
Transforming $f' = f - m\Delta F$ (and relabeling $f'\rightarrow f$) yields
\begin{align}
    \tilde{x}_m(f+m\Delta F) 
    &= \tilde{\phi}\left(f\right) \sum_{n=0}^{N_{t} - 1} (-1)^{mn}w_{mn}\Re\left[C_{mn} \right]e^{-2\pi i n f \Delta T}\,.
\end{align}
Compared to Eq.~\eqref{eq:inverseinterior_final} for the interior bands, the contribution of the edge bands can be obtained by the same equation setting $m=0,N_f$ and multiplying by $\sqrt{2}$, noting that $w_{mn}$ vanishes if $m+n$ is odd and thus $\Re\left[C_{mn} \right]$ reduces to $C_{mn}$ for the nonvanishing terms and matches Eq.~\eqref{eq:inverseinterior_final}.
The factor of $\sqrt{2}$ is a combination of two effects: (i) division by $\sqrt{2}$ to account for the different basis prefactor, but also (ii) multiplication by $2$ for the coinciding positive/negative frequency filters.

\subsubsection{Discretization}
\label{sec:inversediscrete}

To discretize, we set $f_k = k \Delta f$ and $\tilde{x}_m(f)\rightarrow \Delta t\,\tilde{x}_m[k]$.
To cover all bands with a single equation, we define $\tilde{A}_m = 1/(2A_m)$ instead of the basis prefactor for the interior or edge bands.
Starting with Eq.~\eqref{eq:inverseinterior_final}
\begin{align}
    \Delta t\,\tilde{x}_m\left[k + m \frac{N_t}{2}\right] = \tilde{A}_m\tilde{\phi}[k]\sum_{n=0}^{N_t-1}(-1)^{mn}w_{mn} C_{mn} e^{-2\pi ink/N_t}\,.
\end{align}
Following similar steps to the forward transform, we center the filter and reindex from $k \in \left[-N_t/2, N_t/2-1\right]$ to $k \in \left[0, N_t-1\right]$, picking up a factor of $(-1)^n$.
We arrive at
\begin{equation}
    \tilde{x}_m\left[k + (m-1)\frac{N_t}{2}\right] = \frac{\tilde{A}_m}{\Delta t}\tilde{\phi}\left[k - \frac{N_t}{2}\right]\tilde{Z}_m[k]\,,
    \label{eq:xm-bandinverse}
\end{equation}
where we have defined 
\begin{equation}
    z_m[n]=w_{mn}C_{mn}(-1)^{(m+1)n}\,, \qquad \tilde{Z}_m = \mathrm{FFT}\left(z_m\right)[k]\,,
\end{equation}
again based on the conventions of Eq.~\eqref{eq:FFT_numpy_discrete}.
This is the contribution of a given WDM band $m$ to the frequency-domain data. 
For a given frequency, the data are obtained by summing over the $m$ bands that contribute to frequency $\ell$, i.e., $m$ such that $k=\ell-(m-1)N_t/2\in[0,N_t-1]$
\begin{equation}
\tilde{x}[\ell] =\sum_m \tilde{x}_m[\ell] = \sum_m \tilde{x}_m\left[k + (m-1)\left(\frac{N_t}{2}\right)\right]\,,
\label{eq:inverseinterior}
\end{equation}
where the index $k$ of each WDM band $m$ maps to the corresponding $\ell$ index of the frequency domain.

\begin{figure}
    \centering
    \includegraphics[width=\textwidth]{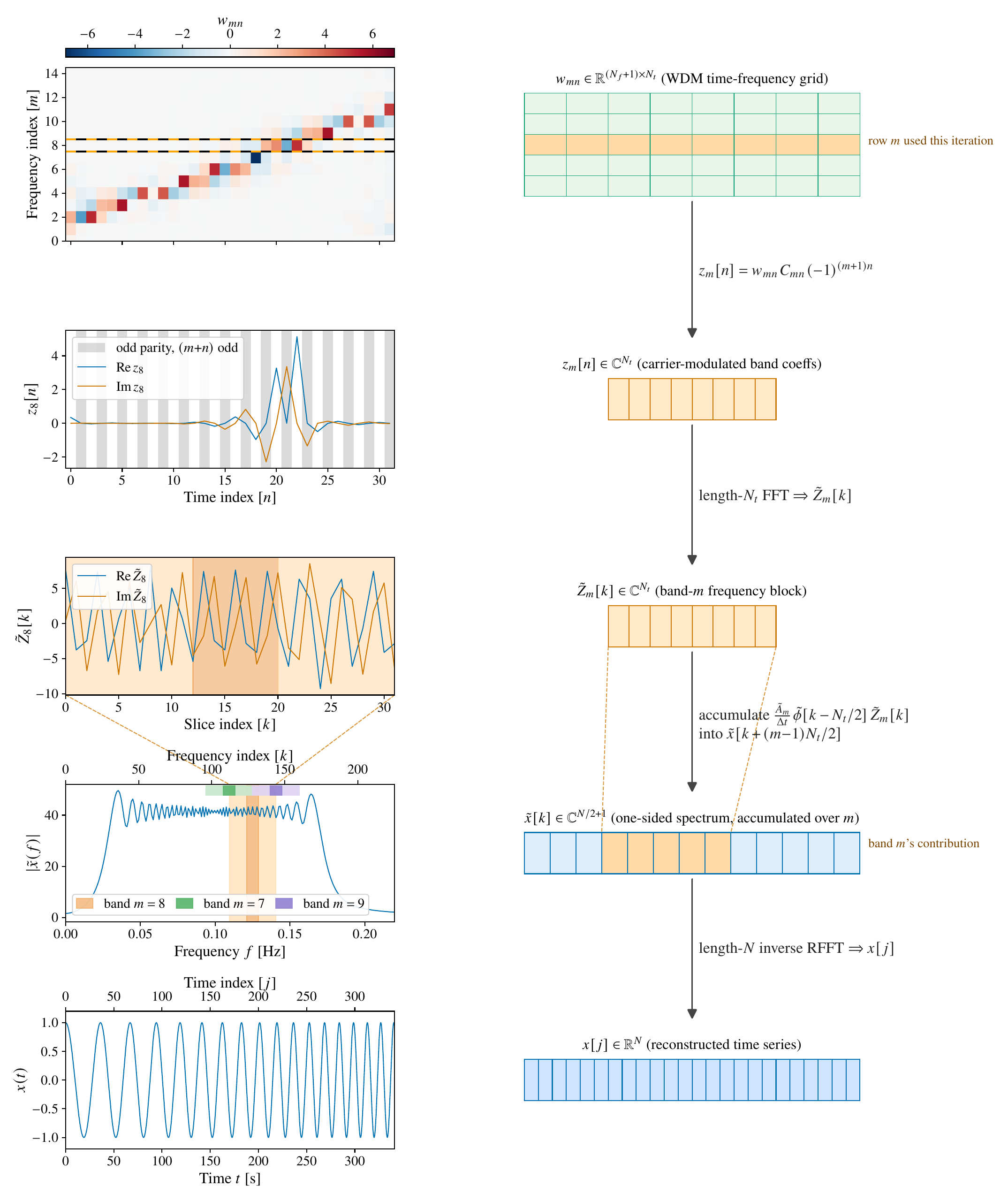}
    \caption{Schematic demonstration of the inverse transform from the WDM to the time domain described in Sec.~\ref{sec:inverse}.
    The left column plots the series in the relevant domain and the right column shows the corresponding bins or cells.
    First row: The WDM coefficients and the $m$th band highlighted.
    Second row: Real and imaginary part of the modulated coefficients in the $m$th band of length $N_t$ with gray vertical striping indicating odd $m + n$.
    Third row: Real and imaginary part of the Fourier transform of the modulated coefficients of length $N_t$.
    Fourth row: Spectrum accumulated band-by-band in $m$; the highlighted slice shows the $m$th band contribution.
    Fifth row: Inverse Fourier transform to a real time series of length $N$.
    }
    \label{fig:wdm_inverse_schematic}
\end{figure}

Appendix~\ref{app:algorithms} provides an algorithm block and
Fig.~\ref{fig:wdm_inverse_schematic} a schematic of the inverse transform, all in the \textsc{numpy} convention: WDM transform and the $m$th band, parity-modulation coefficients, FFT, place into the $m$th slice of the spectrum and accumulate over bands, inverse FFT to the time domain.
Converting Eq.~\eqref{eq:xm-bandinverse} to the LT convention requires multiplying: (i) the frequency-domain data by $\Delta t$, and (ii) $\tilde{Z}_m[k]$ by $\Delta T$.

\section{Transform of the noise Covariance}
\label{sec:PSD}

Besides the data, computing the likelihood in the WDM domain also requires the noise covariance.
In the WDM basis, the noise covariance is a matrix  $\Lambda_{(mn)(m'n')} = \langle n_{mn}n_{m'n'}^* \rangle$ where $n_{mn}$ is the WDM representation of the noise and angle brackets denote an expectation value.
The matrix flattens the $m$ and $n$ (and $m', n'$) indices into a single $(mn)$ index of size $(N_f+1)N_t=N+N_t$.
Thus nominally the covariance is a $(N+N_t)\times (N+N_t)$ matrix.
However, not all matrix elements are independent and half-occupied edge bands ($m=0,N_f$ with $m+n$ odd) render some of the matrix entries identically zero. 
The expected $N\times N$ effective matrix size matches the time- and frequency-domain covariances.

\subsection{Frequency domain to WDM domain}
\label{sec:PSDforward}

From Eq.~\eqref{eqn:wmn_last}, the noise in the WDM domain is
\begin{equation}
    n_{mn} = 2A_m\int_{(m-1)\Delta F}^{(m+1)\Delta F}\Re~\left[C_{mn}^*\tilde{\mathbf{n}}(f)\tilde{\phi}(f - m\Delta F)e^{2\pi inf \Delta T}\right]\diff f\,,
\end{equation}
and the conjugate is
\begin{equation}
    n^*_{mn} = 2A_m\int_{(m-1)\Delta F}^{(m+1)\Delta F}\Re~\left[C_{mn}\tilde{\mathbf{n}}^*(f)\tilde{\phi}(f - m\Delta F)e^{-2\pi inf \Delta T}\right]\diff f\,.
\end{equation}
Though $n_{mn}$ is real, expressing its complex conjugate in this form allows us to eventually form the frequency-domain covariance matrix, $\langle \tilde{\mathbf{n}}(f)\tilde{\mathbf{n}}^*(f')\rangle = \boldsymbol{\Sigma}(f, f')$.
The argument of the expectation value is then
\begin{equation}
    n_{mn}n^*_{m'n'} = 4 A_{m}A_{m'}\int_{(m'-1)\Delta F}^{(m'+1)\Delta F}\int_{(m-1)\Delta F}^{(m+1)\Delta F} \Re\left[X(f)\right]\Re\left[Y(f')\right]\diff f \diff f'\,,
\end{equation}
where
\begin{align}
    X(f) &= C_{mn}^*\tilde{\mathbf{n}}(f)\tilde{\phi}(f - m\Delta F)e^{2\pi inf \Delta T}\,,\\
    Y(f') &= C_{m'n'}\tilde{\mathbf{n}}^*(f')\tilde{\phi}(f' - m'\Delta F)e^{-2\pi in'f' \Delta T}\,.
\end{align}
Using $\Re(X)\Re(Y) = \frac{1}{2}\left[\Re(XY) + \Re(XY^*)\right]$ and the Hermitian symmetry of the Fourier-transformed noise, the expectation value of the integrand decomposes into two terms:
\begin{align}
    \langle X(f)Y(f')\rangle &= C_{mn}^*C_{m'n'} \langle \tilde{\mathbf{n}}(f)\tilde{\mathbf{n}}^*(f')\rangle\tilde{\phi}(f - m\Delta F)\tilde{\phi}(f'-m'\Delta F)e^{2\pi i(nf-n'f')\Delta T}\,,\\
    \langle X(f)Y^*(f') \rangle &= C_{mn}^* C_{m'n'}^*\langle\tilde{\mathbf{n}}(f) \tilde{\mathbf{n}}^*(-f')\rangle\tilde{\phi}(f - m\Delta F)\tilde{\phi}(f'-m'\Delta F)e^{2\pi i(nf + n'f')\Delta T}\,.
\end{align}
The noise contribution has reduced to the covariance matrix in the frequency domain, $\langle XY\rangle\propto \langle \tilde{\mathbf{n}}(f)\tilde{\mathbf{n}}^*(f')\rangle = \boldsymbol{\Sigma}(f, f')$ and $\langle XY^*\rangle \propto \langle \tilde{\mathbf{n}}(f)\tilde{\mathbf{n}}^*(-f')\rangle = \boldsymbol{\Sigma}(f, -f')$.
Overall,
\begin{equation}
    \Lambda_{(mn)(m'n')} = 2A_{m}A_{m'}\int_{(m'-1)\Delta F}^{(m'+1)\Delta F}\int_{(m-1)\Delta F}^{(m+1)\Delta F} \left(\Re~\langle X(f)Y(f')\rangle + \Re~\langle X(f)Y^*(f')\rangle\right)\diff f \diff f'\,.
    \label{eq:genericcovariance}
\end{equation}
For generic, non-stationary noise both terms are nonzero.
This fully generic form of the covariance matrix is not illuminating, and cannot be further simplified without making assumptions about the noise properties.
We will return to the covariance structure and provide examples in the companion paper~\cite{WDMII}.

The inverse covariance matrix, that directly enters the likelihood, transforms from the frequency to the WDM domain via the same formula, by replacing $\boldsymbol{\Sigma}\rightarrow\boldsymbol{\Sigma}^{-1}$.

\subsection{WDM domain to frequency domain}

The frequency-domain noise in terms of its WDM decomposition is
\begin{equation}
    \tilde{\mathbf{n}}(f) = \sum_{m=0}^{N_f}\tilde{n}_m(f)\,,
\end{equation}
where $\tilde{n}_m(f)$ is the contribution of each band $m$, given by Eq.~\eqref{eq:inverseinterior_final} for interior bands and requiring another factor of $\sqrt{2}$ for the DC and Nyquist bands.
The frequency-domain covariance is then
\begin{align}
        \boldsymbol{\Sigma}(f,f')& = \langle \tilde{\mathbf{n}}(f)\tilde{\mathbf{n}}^*(f')\rangle= \sum_{m=0}^{N_f}\sum_{m'=0}^{N_f} \langle\tilde{n}_m(f)\tilde{n}^*_{m'}(f')\rangle\,.
        \label{eq:cov-inverse}
\end{align}
Focusing on the interior bands, $m,m'\neq0,N_f$, the expectation value of the per-band noise is
\begin{align}
        \langle\tilde{n}_m(f)\tilde{n}^*_{m'}(f')\rangle &= \frac{1}{2} \tilde{\phi}(f - m\Delta F)\tilde{\phi}(f' - m'\Delta F)\sum_{n=0}^{N_t-1}\sum_{n'=0}^{N_t-1}C_{mn}C_{m'n'}^*\langle n_{mn}n_{m'n'}\rangle e^{-2\pi inf\Delta T +2\pi in'f'\Delta T}\nonumber\\
        &= \frac{1}{2}\tilde{\phi}(f - m\Delta F)\tilde{\phi}(f' - m'\Delta F)\sum_{nn'}C_{mn}C_{m'n'}^*  \Lambda_{(mn)(m'n')}e^{2\pi i (n'f'-nf)\Delta T}\,.
        \label{eq:nnstar-interior}
\end{align}
For edge bands, this expression is multiplied by $\sqrt 2$ for each of $m$ and $m'$ in $\{0,N_f\}$, before summing over bands to obtain the full covariance in Eq.~\eqref{eq:cov-inverse}.

\subsection{Stationary noise}
\label{sec:stationary}
For stationary noise, the covariance matrix in the time domain depends on time differences only
\begin{equation}
    \langle \mathbf{n}(t)\mathbf{n}^*(t')\rangle = R(t - t')\,,
\end{equation}
where $R(t)$ is the autocorrelation function.
The covariance matrix of the Fourier-transformed noise is
\begin{equation}
    \langle \tilde{\mathbf{n}}(f)\tilde{\mathbf{n}}^*(f')\rangle = \int\int\langle \mathbf{n}(t)\mathbf{n}^*(t')\rangle e^{-2\pi if t + 2\pi i f' t'}\diff t\diff t'= \int\int R(t - t') e^{-2\pi if t + 2\pi i f' t'}\diff t\diff t'\,,
\end{equation}
and using the transformation $\tau = t - t'$ we get
\begin{equation}
    \langle \tilde{\mathbf{n}}(f)\tilde{\mathbf{n}}^*(f')\rangle=\left[\int R(\tau) e^{-2\pi if \tau}\diff \tau\right]\int e^{2\pi i(f'- f) t'}\diff t' \equiv S_{2}(f)\delta(f - f')\,.
    \label{eq:increment_explain}
\end{equation}
The second integral is equal to the Dirac delta function, $ \delta(f - f ')$, and from the Wiener-Khinchin theorem, we identify the term in brackets as the two-sided noise power-spectral density (PSD), $S_{2}(f)$, defined as the Fourier transform of the autocorrelation function.
It is related to the one-sided PSD $S(f)$ through
\begin{equation}
\label{eq:two_sided_psd}
    S_2(f) = \frac{S(f)}{2}\,.
\end{equation}

Overall,
\begin{equation}
    \boldsymbol{\Sigma}(f, f') = \langle \tilde{\mathbf{n}}(f)\tilde{\mathbf{n}}^*(f')\rangle = \frac{1}{2}S(f)\delta(f-f')\,,
    \label{eq:stationary_noise}
\end{equation}
which is just the standard stationary noise condition of diagonal frequency-domain covariance.
Similarly, we obtain $\boldsymbol{\Sigma}(f,-f') = (1/2)S(f)\delta(f+f')$.

\subsubsection{Forward transform}

Returning to Eq.~\eqref{eq:genericcovariance}, we start with the first term,
\begin{align}
    I^{XY}_{(mn)(m'n')} &= 2A_{m}A_{m'}\int_{(m'-1)\Delta F}^{(m'+1)\Delta F}\int_{(m-1)\Delta F}^{(m+1)\Delta F} \Re~\langle X(f)Y(f')\rangle\diff f \diff f'\nonumber\\
    &=  A_{m}A_{m'}\Re\left[\int_{(m'-1)\Delta F}^{(m'+1)\Delta F}\int_{(m-1)\Delta F}^{(m+1)\Delta F} C_{m'n'}C_{mn}^* S(f)\delta(f - f') \tilde{\phi}(f - m\Delta F)\tilde{\phi}(f'-m'\Delta F)e^{2\pi i(nf-n'f')\Delta T}\diff f \diff f'\right]\nonumber\\
    &=  A_{m}A_{m'}\Re\left[ C_{m'n'}C_{mn}^* \int_{(m-1)\Delta F}^{(m+1)\Delta F}   S(f) \tilde{\phi}(f - m\Delta F)\tilde{\phi}(f-m'\Delta F)e^{2\pi i(n-n')f\Delta T}\diff f\right] \,.
    \label{eq:covariance-term1}
\end{align}
The ensuing calculation follows the orthogonality proof of Appendix~\ref{app:orthogonality}.
Transforming $f' = f - m\Delta F$ and then relabeling $f' \rightarrow f$, we get
\begin{align}
    I^{XY}_{(mn)(m'n')} &= A_{m}A_{m'} \Re\left[C_{m'n'}C_{mn}^* \int_{-\Delta F}^{\Delta F}   S(f + m\Delta F) \tilde{\phi}(f)\tilde{\phi}(f + (m-m')\Delta F)e^{2\pi i(n-n')(f + m\Delta F)\Delta T}\diff f\right]\nonumber \\
    &=  A_{m}A_{m'} \, (-1)^{(n-n')m}\Re\left[C_{m'n'}C_{mn}^* \int_{-\Delta F}^{\Delta F}   S(f + m\Delta F) \tilde{\phi}(f)\tilde{\phi}(f + (m-m')\Delta F)e^{2\pi i(n-n')f\Delta T}\diff f\right]\nonumber \\
    &\equiv A_{m}A_{m'} \, (-1)^{(n-n')m}\Re\left[C_{m'n'}C_{mn}^* {\cal{I}}_{(mn)(m'n')}\right]\,,
    \label{eq:covariance-term1-second}
\end{align}
where in the last line we have defined ${\cal{I}}_{(mn)(m'n')}$ as the integral over frequency for later convenience.

The second term of Eq.~\eqref{eq:genericcovariance} is
\begin{align}
    I^{XY^*}_{(mn)(m'n')} &= 2A_{m}A_{m'}\int_{(m'-1)\Delta F}^{(m'+1)\Delta F}\int_{(m-1)\Delta F}^{(m+1)\Delta F} \Re~\langle X(f)Y^*(f')\rangle\diff f \diff f'\nonumber\\
    &=  A_{m}A_{m'}\Re\left[\int_{(m'-1)\Delta F}^{(m'+1)\Delta F}\int_{(m-1)\Delta F}^{(m+1)\Delta F} C^*_{m'n'}C_{mn}^* S(f)\delta(f + f') \tilde{\phi}(f - m\Delta F)\tilde{\phi}(f'-m'\Delta F)e^{2\pi i(nf+n'f')\Delta T}\diff f \diff f'\right]\,.
    \label{eq:covariance-term2}
\end{align}
For interior bands, this term vanishes because $\delta(f + f')$ has support for $f = -f'$, but the integration domain is on $f \in \left[(m-1)\Delta F, (m+1)\Delta F\right]$ which has $f,f' > 0$ unless $m=0,N_f$.

As always, therefore, edge bands need to be handled separately.
For $m = 0$ or $m' = 0$ the integration domain includes negative frequencies, so $\delta(f + f')$ has nonzero support.
Carrying out the integral over $f'$ for the DC-$m'$ (where $m'\in\{0,1\}$) case we find
\begin{align}
    I^{XY^*}_{(0n)(m'n')} &=  A_{0}A_{m'}\Re \left[ C^*_{m'n'}C_{0n}^* {\cal{I}}^*_{(0n)(m'n')}\right]\,,
\end{align}
which is equal to $I^{XY}_{(0n)(m'n')}$ in Eq.~\eqref{eq:covariance-term1-second} if and only if $C_{0n}^*=C_{0n}$. 
This condition is satisfied on all non-zero cells that have $n$ even and thus $C_{0n}=1$. 
An analogous statement holds for the Nyquist-$m'$ (where $m'\in\{N_f,N_f-1\}$) terms: the negative-$f$ filter aliases onto the positive-$f$ one via the $2f_\mathrm{Nyq}$ periodicity from Appendix~\ref{app:nyquist}.
For the DC-DC and Nyquist-Nyquist terms, the overall doubling of the covariance from the $I^{XY^*}_{(mn)(m'n')}$ term is exactly canceled by the reduction in $A_m^2$ (from 1/2 to 1/4) at the edges.
Therefore both the interior and the edge diagonal terms have an overall numerical prefactor of $1/2$.
For edge off-diagonal terms, however, the prefactor is instead $2A_m A_{m'}\to 1/\sqrt{2}$.
To unify these expressions, we define $A_{mm'}$ to be $1/2$ for interior and the edge diagonal bands and $1/\sqrt{2}$ for edge off-diagonal bands, and reach the final expression for the covariance matrix of stationary noise: 
\begin{align}
    \Lambda_{(mn)(m'n')} 
    &= A_{mm'} \, (-1)^{(n-n')m}\,\Re\left[C_{m'n'}C_{mn}^* \int_{-\Delta F}^{\Delta F}   S(f + m\Delta F) \tilde{\phi}(f)\tilde{\phi}(f + (m-m')\Delta F)e^{2\pi i(n-n')f\Delta T}\diff f\right]\,.
    \label{eq:lambda_stationary_discretize}
\end{align}
The filter functions in the integrand overlap only for $|m - m'| \le 1$ and thus all other matrix elements vanish.
However, matrix elements with $m'=m\pm1$ are nonzero; the WDM covariance matrix is not diagonal even under perfectly stationary noise.

For a diagonal WDM covariance matrix, the noise needs to additionally be locally flat, at least to within the three WDM bins where the filters overlap.
Centering at bin $m$, the slow-varying noise assumption amounts to expanding the PSD to zeroth order $S(f + m\Delta F)\approx S(m\Delta F)$.
The covariance is then
\begin{equation}
    \Lambda_{(mn)(m'n')}\approx A_{mm'} \, (-1)^{(n-n')m}\,\Re\left[C_{m'n'}C_{mn}^* S(m\Delta F) \int_{-\Delta F}^{\Delta F}    \tilde{\phi}(f)\tilde{\phi}(f + (m-m')\Delta F)e^{2\pi i(n-n')f\Delta T}\diff f\right]\,.
\end{equation}
The integral is computed in Appendix~\ref{app:orthogonality} as part of the orthogonality proof, yielding a diagonal covariance 
\begin{equation}
    \Lambda_{(mn)(m'n')} \approx \frac{1}{2} S(m\Delta F)\delta_{mm'}\delta_{nn'}\,.
    \label{eq:stationary_slow_moving}
\end{equation}
The WDM diagonal covariance is proportional to the frequency-domain covariance via the PSD evaluated at the bin midpoint, $m\Delta F$, under the assumption that the noise does not change appreciably across adjacent bins.
\citet{Cornish:2020odn} instead evaluates an integral of the PSD over each filter band, yielding the diagonal elements of the full WDM noise covariance matrix while ignoring off-diagonal terms.
First-order corrections to the slowly-varying noise approximation and the off-diagonal terms were calculated in \citet{Cornish:2025awt}.
The effect of each of these approximations on inference is examined in the sequel~\cite{WDMII}.

\subsubsection{Inverse transform}

We now present the inverse transform to return from the WDM domain noise covariance matrix $\Lambda_{(mn)(m'n')}$ to the noise PSD in the frequency domain $S(f)$.
We start with Eq.~\eqref{eq:cov-inverse} and specializing to the stationary noise case, $f = f'$,
collapses the sums $m,m'$ to only the band-tridiagonal $m = m'$ and $m = m' \pm 1$.
Since the covariance matrix is symmetric, we can reduce this further to only the diagonal and one of the off-diagonals counted twice 
\begin{equation}
    \boldsymbol{\Sigma}(f,f) = \sum_{m=0}^{N_f}\langle \tilde{n}_m(f)\tilde{n}^*_{m}(f)\rangle + 2\sum_{m=0}^{N_f-1}\Re\langle \tilde{n}_m(f)\tilde{n}^*_{m + 1}(f)\rangle\,.
    \label{eq:diag_offdiag}
\end{equation}
The real operator in the second term is required because $\langle \tilde{n}_m(f)\tilde{n}^*_{m + 1}(f)\rangle$ is complex valued and conjugate symmetric, which can be seen by swapping the $\tilde{n}_m$ and $\tilde{n}^*_{m+1}$ indices.
Next we specialize Eq.~\eqref{eq:nnstar-interior} to stationary noise, moving from a coefficient of $1/2$ for the inner bands to $\tilde{A}_{m}\tilde{A}_{m'}$ (recall their definition around Eq.~\eqref{eq:xm-bandinverse}) to account for the edge bands
\begin{equation}
    \langle \tilde{n}_m(f)\tilde{n}^*_{m'}(f)\rangle = \tilde{A}_m\tilde{A}_{m'}\tilde{\phi}(f - m\Delta F)\tilde{\phi}(f - m'\Delta F)\sum_{nn'}C_{mn}C_{m'n'}^*  \Lambda_{(mn)(m'n')}e^{-2\pi i f r\Delta T}\,,
    \label{eq:nnexpected_stationary}
\end{equation}
where $r = n-n'$.
Though not immediately obvious due to the presence of the $C_{mn}C_{m'n'}^*$ term, the summand depends only on time differences and hence is Toeplitz in structure.
We prove this in Appendix~\ref{app:toeplitz}.
Therefore, the sum over $n,n'$ collapses to a sum over $r$ times a multiplicity $\mu$ for each lag, schematically
\begin{equation}
    \sum_{nn'}f[r] = \sum_{r}\mu f[r]\,,
\end{equation}
with
\begin{equation}
    \mu=\begin{cases}
        N_t& \text{interior}\\
        N_t/2 & \text{edge bands}
    \end{cases}\,.
\end{equation}
We define
\begin{equation}
    t_{mm'}[r] = \mu \,C_{mn}C_{m'n'}^*\Lambda_{(mn)(m'n')}\,.
    \label{eq:tmmprime}
\end{equation}
for any $n,n'$ with $r=n-n'$ and Eq.~\eqref{eq:nnexpected_stationary} is now 
\begin{equation}
    \langle \tilde{n}_m(f)\tilde{n}^*_{m'}(f)\rangle = \tilde{A}_m\tilde{A}_{m'}\,\tilde{\phi}(f - m\Delta F)\tilde{\phi}(f - m'\Delta F)\sum_{r}t_{mm'}[r]e^{-2\pi i f r\Delta T}\,.
\end{equation}
When inserted into Eq.~\eqref{eq:diag_offdiag}, we find
\begin{align}
    \boldsymbol{\Sigma}(f,f) = \sum_{m=0}^{N_f}&\tilde{A}_m^2\tilde{\phi}^2(f-m\Delta F)\sum_r t_{mm}[r]e^{-2\pi i f r \Delta T}\nonumber\\
    &+ 2\sum_{m=0}^{N_f-1}\Re \left[\tilde{A}_m\tilde{A}_{m+1}\tilde{\phi}(f-m\Delta F)\tilde{\phi}(f-(m+1)\Delta F)\sum_r t_{m(m+1)}[r]e^{-2\pi i f r \Delta T}\right]\,,
\end{align}
and after one final translation to $f \rightarrow f + m\Delta F$ we obtain,
\begin{align}
    \boldsymbol{\Sigma} &= \sum_{m=0}^{N_f}\tilde{A}_m^2\tilde{\phi}^2(f)\sum_r t_{mm}[r]e^{-2\pi i (f+m\Delta F)r \Delta T} + 2\sum_{m=0}^{N_f-1}\Re\left[\tilde{A}_m\tilde{A}_{m+1}\tilde{\phi}(f)\tilde{\phi}(f-\Delta F)\sum_r t_{m(m+1)}[r]e^{-2\pi i (f+m\Delta F) r \Delta T}\right]\nonumber\\
    &= \sum_{m=0}^{N_f}\tilde{A}_m^2\tilde{\phi}^2(f)\sum_r (-1)^{mr} t_{mm}[r]e^{-2\pi i fr \Delta T} + 2\sum_{m=0}^{N_f-1}\Re\left[\tilde{A}_m\tilde{A}_{m+1}\tilde{\phi}(f)\tilde{\phi}(f-\Delta F)\sum_r (-1)^{mr} t_{m(m+1)}[r]e^{-2\pi i f r \Delta T}\right]\,.
    \label{eq:sigma_to_discretize}
\end{align}
This equation gives the noise PSD in terms of the WDM noise covariance in continuous form.

\subsubsection{Discretization}

To discretize the covariance expression for Eq.~\eqref{eq:lambda_stationary_discretize}, we follow similar steps to Sec.~\ref{sec:forwarddiscrete}: $f \rightarrow k \Delta f$, $S(f) \rightarrow S[k]$, and $df \rightarrow \Delta f$.
The stationary PSD is evaluated at each of the bins $m\Delta F$, yielding
\begin{equation}
    \Lambda_{(mn)(m'n')} =  A_{mm'}\Delta f (-1)^{(n-n')m} \,\Re \left\{C_{m'n'}C_{mn}^* \sum_{k=-N_t/2}^{N_t/2-1}   S\left[k + m\frac{N_t}{2}\right] \tilde{\phi}[k]\,\tilde{\phi}\left[k+(m-m')\frac{N_t}{2}\right]e^{2\pi i(n-n')k/N_t}\right\}\,.
\end{equation}
Centering the bins with $k' = k + \frac{N_t}{2}$ and relabeling $k'$ back to $k$ gives
\begin{equation}
    \Lambda_{(mn)(m'n')} = A_{mm'}\Delta f (-1)^{(n-n')(m+1)} \Re \left\{C_{m'n'}C_{mn}^* \sum_{k=0}^{N_t-1}   \tilde{K}_{mm'}[k]e^{2\pi i(n-n')k/N_t}\right\}\,,
\end{equation}
where we discretize $\tilde{K}_{mm'}$,
\begin{equation}
    \tilde{K}_{mm'}[k]=S\left[k + (m - 1)\frac{N_t}{2}\right] \tilde{\phi}\left[k - \frac{N_t}{2}\right]\,\tilde{\phi}\left[k+(m-m' - 1)\frac{N_t}{2}\right]\,.
\end{equation}
The covariance depends only on $n - n'$ so it has a Toeplitz structure up to the parity factor induced by the term $C_{m'n'}C_{mn}^*$.
Since $n \in [0, N_t-1]$, then $r= n-n' \in [-(N_t - 1), (N_t - 1)]$.
With this definition,
\begin{align}
    \Lambda_{(mn)(m'n')} &= A_{mm'}\Delta f (-1)^{r(m+1)}\Re\left\{C_{m'n'}C_{mn}^* \sum_{k=0}^{N_t-1}   \tilde{K}_{mm'}[k]e^{2\pi irk/N_t}\right\}\nonumber\\
    &= A_{mm'} \Delta fN_t(-1)^{r(m+1)}\Re \left\{C_{m'n'}C_{mn}^* K_{mm'}[r]\right\}\,,
    \label{eq:covariance_stationary_discrete}
\end{align}
where 
\begin{equation}
    N_t K_{mm'}[r] = \frac{N_t}{N_t}\sum_{k=0}^{N_t-1} \tilde{K}_{mm'}[k]e^{2\pi irk/N_t} = N_t\, \mathrm{IFFT}(\tilde{K}_{mm'}[k])\,,
\end{equation}
where the $N_t$ factor is required for the IFFT as defined by \textsc{numpy}.
An algorithm block for this calculation is given in Appendix~\ref{app:algorithms}.

The likelihood calculation, explored in Sec.~\ref{sec:logL}, requires the inverse rather than the covariance itself.
In general, the inverse of a matrix with some degree of symmetry, see following discussion, will be dense.
However, in this case, the inverse can be trivially obtained by noting that for stationary noise, the frequency domain inverse is $\boldsymbol{\Sigma}^{-1}(f, f')\sim 1/S(f)$.
The WDM inverse covariance can therefore be obtained from Eq.~\eqref{eq:covariance_stationary_discrete} by setting $S(f)\rightarrow 1/S(f)$.
The WDM inverse covariance has the same symmetry structure as the covariance.

To discretize the backward transformation from the stationary WDM noise covariance to the noise PSD in the frequency domain, we start with Eq.~\eqref{eq:stationary_noise}
\begin{equation}
\begin{split}
{\boldsymbol{\Sigma}}[k,k'] &= \frac{1}{2}S[k]\delta\left[(k - k')\Delta f\right] = \frac{1}{2\Delta f}S[k]\delta_{kk'} = \frac{T}{2}S[k]\delta_{kk'}\,.
\end{split}
\end{equation}
This equation can be solved directly for $S[k]$
\begin{equation}
    S[k] = \frac{2}{T}\boldsymbol{\Sigma}[k,k]\,.
\end{equation}
Employing the discretized form of Eq.~\eqref{eq:sigma_to_discretize} we find
\begin{equation}
\begin{split}
    S[k] = \frac{2}{T}\Biggl\{ &\sum_{m=0}^{N_f}\tilde{A}_m^2\tilde{\phi}^2[k]\sum_r(-1)^{mr} t_{mm}[r]e^{-2\pi i r k /N_t} \\
    &+ 2\sum_{m=0}^{N_f-1}\Re\left[\tilde{A}_m\tilde{A}_{m+1}\,\tilde{\phi}[k]\tilde{\phi}\left[k-\frac{N_t}{2}\right]\sum_r (-1)^{mr} t_{mm+1}[r]e^{-2\pi i r k/N_t}\right] \Biggr\}\,.
\end{split}
\end{equation}
Finally, shifting $k\rightarrow k-N_t/2$ and using that $\tilde{\phi}[k - N_t] = \tilde{\phi}[k]$,
\begin{equation}
\begin{split}
    S[k] = \frac{2}{T}\Biggl\{ &\sum_{m=0}^{N_f}\tilde{A}_m^2\tilde{\phi}^2\left[k-\frac{N_t}{2}\right]\,\mathrm{FFT}( (-1)^{(m+1)r} t_{mm}[r]) \\
    &+ 2\sum_{m=0}^{N_f-1}\Re\left[\tilde{A}_m\tilde{A}_{m+1}\,\tilde{\phi}\left[k-\frac{N_t}{2}\right]\tilde{\phi}\left[k\right]\,\mathrm{FFT}((-1)^{(m+1)r} t_{mm+1}[r])\right] \Biggr\}\,.
\end{split}
\end{equation}
An implementation of this equation is given in Appendix~\ref{app:algorithms}.

Transforming from the noise PSD in the frequency domain to the block tri-diagonal WDM covariance matrix costs $\mathcal{O}(N \log N_t)$ to form $K_{mm'}[r]$ with a fast Fourier transform.
This means that the forward noise covariance to PSD transformation has cost $\mathcal{O}(N \log N_t)$.
The inverse transformation has two fast Fourier transformations per block, one for the diagonal and one for the off-diagonal.
Once again, the inverse transformation from the WDM noise covariance matrix to the frequency domain noise PSD has cost $\mathcal{O}(N \log N_t)$.
These costs only cover the special case of stationary noise, and generic cases will require more computation.

\subsubsection{Structure}
\label{sec:structure-stationary}

The noise covariance matrix of Eq.~\eqref{eq:covariance_stationary_discrete} has $(N_f+1)N_t\times (N_f+1)N_t$ entries that encode the $N/2+1$ degrees of freedom of stationary noise.
Its structure is simplified by a number of symmetries, but even then the number of nonzero, unique entries is greater than $N/2+1$.
This means that entries are in general not independent even when symmetries are taken into account, which complicates noise modeling. 
While in the frequency domain we could directly model each $S[i]$ as an independent number, we cannot do so with each WDM cell.  

Below we discuss some of the covariance matrix general structure and elucidate it with a schematic for the case $N_f=3, m\in[0,3]$ and $N_t=4, n\in [0,4)$. 
This is only illustrative and we focus on properties that apply generically to all values of $N_f$ and $N_t$.\footnote{For $N_t=4$ in particular, all cross-band blocks vanish, but that is not a generic property of other $N_t$ values and hence we do not show it here.}

\begin{enumerate} 
\item The covariance matrix is symmetric in $(mn)\leftrightarrow (m'n')$.
 \item It is block tri-diagonal, where the three diagonals correspond to $|m-m'|\le 1$ and the blocks correspond to $n$.
It therefore suffices to compute the $m = m'$ and $m = m' + 1$ elements and obtain the $m = m' - 1$ entries by symmetry.

In our example, the flattened matrix is block tri-diagonal with shape $(N_f+1)\times (N_f+1)=4\times 4$ blocks
\begin{equation}
\Lambda \;=\;
\begin{pmatrix}
D_0    & B_{01} & 0      & 0     \\
B^T_{01} & D_1    & B_{12} & 0     \\
0      & B^T_{12} & D_2    & B_{23}\\
0      & 0      & B^T_{23} & D_3
\end{pmatrix}\,.
\end{equation}
Each block is a $N_t\times N_t=4\times 4$ matrix.
 \item Up to parity each $(m,m')$ block depends on $n,n'$ only through $r=n-n'$ and therefore has a parity-corrected Toeplitz structure.
 
 \item The interior ($m,m'\neq 0, N_f$) diagonal ($m=m'$) blocks are symmetric, have a constant diagonal, and are Toeplitz for $r$ even and alternating-sign Toeplitz for $r$ odd.
 
 In our example, this corresponds to blocks $D_1$ and $D_2$, each of which is schematically 
\begin{equation}
D_{m} \;=\;
\begin{pmatrix}
d_0 & d_1 & d_2 & d_3 \\
d_1 & d_0 & -d_1 & d_2 \\
d_2 & -d_1 & d_0 & d_1 \\
d_3 & d_2 & d_1 & d_0\\
\end{pmatrix}\,.
\end{equation}

 \item The edge ($m,m'= 0, N_f$) diagonal ($m=m'$) blocks have similar properties as the interior ones and additionally rows/columns with $m+n$ odd vanish identically.
 
 In our example, this corresponds to blocks $D_0$ and $D_3$, the former of which is schematically 
\begin{equation}
D_{m} \;=\;
\begin{pmatrix}
d_0 & 0 & d_2 & 0 \\
0 & 0 & 0 & 0 \\
d_2 & 0 & d_0 & 0 \\
0 & 0 & 0 & 0\\
\end{pmatrix}\,.
\end{equation}

 \item The diagonal elements of the cross-band blocks vanish by parity arguments: for $m'=m+1$ and $r=n-n'=0$, $K_{m(m+1)}$ is purely real, while $C_{(m+1)n}C_{mn}^*$ is purely imaginary. 
Overall then, $\Re \left\{C_{(m+1)n}C_{mn}^* K_{m(m+1)}\right\}=0$.

 \item The interior ($m,m'\neq 0, N_f$) cross-band ($|m-m'|=1$) blocks are Toeplitz for $r$ odd and alternating-sign Toeplitz for $r$ even. The blocks are not fully symmetric, as odd-$r$ stripes are symmetric and even-$r$ stripes are antisymmetric.
 
  In our example, this corresponds to blocks $B_{12}$ and $B_{21}$, where schematically 
  \begin{equation}
B_{m(m+1)} \;=\;
\begin{pmatrix}
0      & b_{1} & -b_{2} & b_{3} \\
b_{1}    & 0      & b_{1} & b_{2} \\
b_{2}    & b_{1}    & 0      & b_{1} \\
b_{3}    & -b_{2}    & b_{1}    & 0
\end{pmatrix}\,.
\end{equation}

\item The edge ($m,m' = 0, N_f$) cross-band ($m=m'+1$) blocks have the same properties as the interior ones and additionally rows or columns with edge $m$ and $m+n$ odd vanish.
 
  In our example,  
  \begin{equation}
B_{01} \;=\;
\begin{pmatrix}
0      & b_{1} & -b_{2} & b_{3} \\
0    & 0      & 0 & 0 \\
b_{2}    & b_{1}    & 0      & b_{1} \\
0    & 0    & 0    & 0
\end{pmatrix}\,, \qquad
B_{23} \;=\;
\begin{pmatrix}
0      & b_{1} & 0 & b_{3} \\
0    & 0      & 0 & b_{2} \\
0    & b_{1}    & 0      & b_{1} \\
0    & -b_{2}    & 0    & 0
\end{pmatrix}\,.
\end{equation}

\end{enumerate}

\section{Likelihood and inner products}
\label{sec:logL}

With the data and covariance transforms in hand, in this section we derive inner products and the likelihood function in the WDM domain.
Derivations are simpler if we recast the transforms in terms of vectors and matrices, rather than the coefficients and integrals of the preceding sections.
We therefore start by establishing this notation before moving to the Gaussian likelihood, following Ref.~\cite{Pearson:2025wfd}.
Calculations are simplified under the unitary convention of Eq.~\eqref{eq:FFT_unitary_discrete}, which we adopt in Secs~\ref{subsec:notation},~\ref{subsec:covariance}, and~\ref{subsec:likelihood}.
In Sec.~\ref{subsec:stationary} we specialize to stationary noise and derive the standard frequency-domain Whittle likelihood and re-define the noise PSD, $S(f)$.
In the WDM domain, we obtain the analogous likelihood function but with an \textit{evolutionary} noise PSD, $S(t, f)$, that depends on frequency as well as time. 
For these results we revert to the more common convention of Eq.~\eqref{eq:FFT_asymmetric_discrete} to match familiar results.

\subsection{WDM transform under linear algebra notation}
\label{subsec:notation}

In matrix notation, the transform equations are formal and hold for any generic data or noise, making no assumption about noise stationarity or other simplifications.
As such, they are not meant for use in production analyses where inverting matrices would be prohibitively expensive, but instead as a tool for derivations.
In the $\mathbf{W}$ and $\tilde{\mathbf{W}}$ transform matrices introduced below, we pack the DC and Nyquist bins into $N$ elements by putting the nonzero DC elements in the $N_t/2$ even bins and the Nyquist elements into the $N_t/2$ odd bins.
By doing so, we avoid non-square matrices that are not unitary.
The same trick needs to also be performed on the Fourier domain transform matrix $\mathbf F$.
The matrices can be unpacked in the end to include the DC and Nyquist half-bins in their unpacked (singular) representations for user manipulation.

We define a vector $\mathbf{w}$ holding the WDM coefficients $w_{(mn)}$ in flattened form.
The corresponding time-domain and frequency-domain vectors are $\mathbf{x}$ and $\tilde{\mathbf{x}}$ respectively.
The time-to-WDM matrix operator $\mathbf{W}$ and the frequency-to-WDM matrix operator $\tilde{\mathbf{W}}$ are representations of the WDM transform from each of these domains to the WDM domain:
\begin{equation}
    \mathbf{w} = \mathbf{W}\mathbf{x} = \tilde{\mathbf{W}}\tilde{\mathbf{x}}\,.
\end{equation}
The $N$-point discrete Fourier transform is represented by operator $\mathbf{F}$ with entries $F_{jk}= \left(1/\sqrt{N}\right)e^{-2\pi i j k /N}$ and conjugate transpose $\mathbf{F}^\dagger$:
\begin{equation}
    \tilde{\mathbf{x}} = \mathbf{F x}\,, \qquad \mathbf{x} = \mathbf{F}^\dagger\tilde{\mathbf{x}}\,.
\end{equation}
In their packed form, both the WDM and the Fourier transform are unitary, so in the unitary conventions of Eq.~\eqref{eq:FFT_unitary_discrete} the operators satisfy
\begin{equation}
    \mathbf{W}\mathbf{W}^T = \mathbf{FF}^\dagger = \mathbf{I}\,,
\end{equation}
where $\mathbf{I}$ is the identity. 
From this, the frequency domain operator $\tilde{\mathbf{W}}$ can be written as
\begin{equation}
    \tilde{\mathbf{W}}\tilde{\mathbf{x}} = \mathbf{w} = \mathbf{Wx} = \mathbf{WF}^\dagger\tilde{\mathbf{x}} \Rightarrow \tilde{\mathbf{W}} = \mathbf{W}\mathbf{F}^\dagger\,.
\end{equation}
Thus $\tilde{\mathbf{W}}\tilde{\mathbf{W}}^\dagger = \mathbf{I}$ and $\tilde{\mathbf{W}}$ is also unitary.

Both the time-to-WDM $\mathbf{W}$ and the frequency-to-WDM $\tilde{\mathbf{W}}$ transformation matrix can be written down analytically.
While this exercise is pedagogical, multiplying a square matrix $\mathbf{M}\in  \mathbb{R}^{M\times M}$ and a vector $\mathbf{v}\in \mathbb{R}^M$ requires $\mathcal{O}(M^2)$ operations.
This is slower than performing the same calculation via (I)FFT as derived in Sec.~\ref{sec:transform} and is not the preferred method of transformation; however, this formulation provides a general-use prescription for transforming not only vectors, but also matrices such as the noise covariance matrix and products of vectors and matrices such as the likelihood.
This formulation of the WDM transform has found use in investigations of in-painting data gaps~\cite{Pearson:2025wfd} and will be used in the companion paper~\cite{WDMII}.

\subsection{Generic covariance}
\label{subsec:covariance}

In this notation, the covariance matrix for generic zero-mean noise is defined via the expectation value of the noise,
\begin{equation}
    \mathbf{C} = \langle\mathbf{n}\mathbf{n}^T\rangle\,.
\end{equation}
Using $\tilde{\mathbf{n}} = \mathbf{F}\,\mathbf{n}$, we can derive a transformation from the time to the frequency domain,
\begin{equation}
    \boldsymbol{\Sigma} = \langle \tilde{\mathbf{n}}\tilde{\mathbf{n}}^\dagger \rangle = \langle\mathbf{F}\,\mathbf{n}(\mathbf{F}\,\mathbf{n})^\dagger\rangle = \mathbf{F}\langle\mathbf{n}\mathbf{n}^T\rangle\mathbf{F}^\dagger = \mathbf{F}\,\mathbf{C}\,\mathbf{F}^\dagger\,.
\end{equation}
A similar expression can be derived between the time, frequency, and WDM noise covariance matrices. 
Denoting the noise vector in the WDM domain as $\mathbf{n_w}$ with elements $n_{(mn)}$, we find
\begin{equation}
    \boldsymbol{\Lambda} = \langle \mathbf{n_w}\mathbf{n_w}^\dagger\rangle = \langle\tilde{\mathbf{W}}\tilde{\mathbf{n}}(\tilde{\mathbf{W}}\tilde{\mathbf{n}})^\dagger\rangle=\tilde{\mathbf{W}}\langle\tilde{\mathbf{n}}\tilde{\mathbf{n}}^\dagger\rangle\tilde{\mathbf{W}}^\dagger = \tilde{\mathbf{W}} \boldsymbol{\Sigma} \tilde{\mathbf{W}}^\dagger = \mathbf{W}\mathbf{C}\mathbf{W}^T\,.
    \label{eq:WDM_covmat}
\end{equation}
Since $\tilde{\mathbf{W}}\tilde{\mathbf{W}}^\dagger = \mathbf{I}$, Eq.~\eqref{eq:WDM_covmat} can also be applied to the inverse covariance matrix, shown by taking the inverse of both sides to find
\begin{equation}
    \boldsymbol{\Lambda}^{-1} = \tilde{\mathbf{W}}\boldsymbol{\Sigma}^{-1}\tilde{\mathbf{W}}^\dagger\,.
    \label{eq:WDM_covmat_inv}
\end{equation}
Depending on the noise properties, it might be faster to obtain the covariance inverse in the frequency domain and compute the inverse WDM covariance with Eq.~\eqref{eq:WDM_covmat_inv} directly.
For example, for stationary noise the inverse covariance in the frequency domain is trivial to calculate, while inverting the WDM covariance involves inverting a block tri-diagonal matrix.

\subsection{Likelihood}
\label{subsec:likelihood}

As a scalar, the likelihood is invariant under these transforms, but we can compute it using data and the covariance matrix expressed in any domain.
As long as this is done self-consistently and without making domain-specific assumptions (such as diagonal frequency-domain covariance if the noise is not stationary), the final value will be the same.

We now combine the transformed data and noise covariance to compute the Gaussian likelihood in the WDM domain.
We have real data $\mathbf{d}$ in the time domain consisting of noise $\mathbf{n}$ and a signal $\mathbf{h}$\footnote{The subsequent equations can trivially be extended to multiple signals $\mathbf{H} = \sum_i \mathbf{h}_i$.} with parameters $\boldsymbol{\theta}$,
\begin{equation}
    \mathbf{d} = \mathbf{n} + \mathbf{h}(\boldsymbol{\theta})\,.
\end{equation}
Further, we assume that the noise follows a zero-mean multivariate Gaussian,
\begin{equation}
    \mathbf{n} \sim \mathcal{N}(\mathbf{0}, \mathbf{C}) = \frac{1}{\sqrt{(2\pi)^N\det \mathbf{C}}}\exp\left(-\frac{1}{2}\mathbf{n}^T\mathbf{C}^{-1}\mathbf{n}\right)\,,
\end{equation}
with covariance matrix $\mathbf{C}$.
Given an estimate of the noise covariance and signal parameters $\boldsymbol{\theta}$, the residual $\mathbf{r} = \mathbf{d} - \mathbf{h}(\boldsymbol{\theta})$ is equal to $\mathbf{n}$ if $\boldsymbol{\theta}$ are the true parameters.
The likelihood is then
\begin{equation}
    p(\mathbf{d}\mid\boldsymbol{\theta}) = \frac{1}{\sqrt{(2\pi)^N\det \mathbf{C}}}\exp\left(-\frac{1}{2}\mathbf{r}^T\mathbf{C}^{-1}\mathbf{r}\right)\,,
\end{equation}
or in the more common logarithmic form
\begin{equation}
\label{eq:time_domain_likelihood}
    \ln p(\mathbf{d}\mid \boldsymbol{\theta}) = -\frac{N}{2}\ln(2\pi) - \frac{1}{2}\ln \det \mathbf{C} - \frac{1}{2}\mathbf{r}^T\mathbf{C}^{-1}\mathbf{r} = -\frac{N}{2}\ln(2\pi) - \frac{1}{2}\ln \det \mathbf{C} - \frac{1}{2}(\mathbf{r}\mid \mathbf{r})_t\,,
\end{equation}
where $(\mathbf{a}\mid \mathbf{b})_t \equiv \mathbf{a}^T \mathbf{C}^{-1} \mathbf{b}$ is the time-domain noise-weighted inner product.

To obtain the frequency-domain inner product and likelihood function, we
transform the argument of the exponential as
\begin{equation}
    \mathbf{r}^T\mathbf{C}^{-1}\mathbf{r} = \mathbf{r}^\dagger \mathbf{C}^{-1} \mathbf{r} = \left(\mathbf{F}^\dagger \tilde{\mathbf{r}}\right)^\dagger \mathbf{C}^{-1}\left(\mathbf{F}^\dagger \tilde{\mathbf{r}}\right) = \tilde{\mathbf{r}}^\dagger \left(\mathbf{F}\mathbf{C}^{-1}\mathbf{F}^\dagger\right)\tilde{\mathbf{r}} = \tilde{\mathbf{r}}^\dagger\boldsymbol{\Sigma}^{-1}\tilde{\mathbf{r}} = (\mathbf{r}\mid\mathbf{r})_f\,,
\end{equation}
where $(\mathbf{a}\mid \mathbf{b})_f \equiv \mathbf{a}^\dagger \mathbf{\Sigma}^{-1} \mathbf{b}$ is the frequency-domain noise-weighted inner product.
The determinant of the covariance is
\begin{equation}
\det \mathbf{C} = \det \left(\mathbf{F}^\dagger \mathbf{\Sigma} \mathbf{F}\right) = \det \mathbf{F}^\dagger \det \mathbf{\Sigma} \det \mathbf{F}= \det  \mathbf{\Sigma}\,,
\label{eq:cov-det}
\end{equation}
since $\mathbf{F}$ is unitary and thus $\det\mathbf{F} \det\mathbf{F}^\dagger=\det\mathbf{F} \det(\mathbf{F})^{-1}=\det\mathbf{F} (\det\mathbf{F})^{-1}=1$.
Equation~\eqref{eq:cov-det} only applies under the unitary convention for which $\mathbf{FF}^\dagger = \mathbf{I}$, while under different conventions, it picks up constant prefactors.

Collecting all terms, the log-likelihood is then
\begin{equation}
    \ln p(\mathbf{d} \mid \boldsymbol{\theta}) = -\frac{N}{2} \ln\left(2\pi\right) - \frac{1}{2}\ln \det \boldsymbol{\Sigma} - \frac{1}{2}\tilde{\mathbf{r}}^\dagger\boldsymbol{\Sigma}^{-1}\tilde{\mathbf{r}} = -\frac{N}{2} \ln\left(2\pi\right) - \frac{1}{2}\ln \det \boldsymbol{\Sigma} - \frac{1}{2} (\mathbf{r}\mid\mathbf{r})_f\,.
\end{equation}
The frequency-domain expression, like the time-domain one above, is exact and applies even under non-stationary noise in which the frequencies are correlated as long as the inner product is calculated appropriately.
In that case, the nonstationary covariance matrix will be dense in general and require a more expensive inversion and multiplication than its stationary counterpart.
But as long as the full nonstationary covariance matrix is employed, the frequency-domain likelihood will be exact and equal to the time-domain one.

We obtain the likelihood in the WDM domain with similar steps, starting with the time-domain representation of Eq.~\eqref{eq:time_domain_likelihood} and the $\mathbf{W}$ transform matrix.
Starting with the exponential, we denote the WDM residual as $\mathbf{r_w}$, such that $\mathbf{r_w} = \mathbf{W} \mathbf{r}$ and $\mathbf{r} = \mathbf{W}^T \mathbf{r_w}$, and obtain
\begin{equation}
    \mathbf{r}^T\mathbf{C}^{-1}\mathbf{r} = (\mathbf{W}^T\mathbf{r_w} )^T\mathbf{C}^{-1} (\mathbf{W}^T\mathbf{r_w}) = \mathbf{r_w}^T\left(\mathbf{W} \mathbf{C}^{-1}\mathbf{W}^T\right) \mathbf{r_w} = \mathbf{r_w}^T\boldsymbol{\Lambda}^{-1}\mathbf{r_w} = (\mathbf{r_w}\mid \mathbf{r_w})_{tf}\,,
\end{equation}
where $(\mathbf{a}\mid \mathbf{b})_{tf} \equiv \mathbf{a}^T\boldsymbol{\Lambda}^{-1}\mathbf{b}$ is the WDM domain noise-weighted inner product.
Similarly, the determinant of the covariance is
\begin{equation}
\det \mathbf{C} = \det \left(\mathbf{W}^T \mathbf{\Lambda} \mathbf{W}\right) = \det \mathbf{W}^T \det \mathbf{\Lambda} \det \mathbf{W}= \det  \mathbf{\Lambda}\,,
\label{eq:cov-det-2}
\end{equation}
since $\mathbf{W}$ is orthogonal and thus $\det\mathbf{W} = \det\mathbf{W}^T=\pm 1$, a result that again applies only under the unitary convention. 

Collecting again all the terms, the log-likelihood is
\begin{equation}
    \ln p(\mathbf{d}\mid \boldsymbol{\theta}) = -\frac{N}{2}\ln(2\pi) - \frac{1}{2}\ln\det \boldsymbol{\Lambda} - \frac{1}{2}  \mathbf{r_w}^T\boldsymbol{\Lambda}^{-1}\mathbf{r_w} = -\frac{N}{2}\ln(2\pi) - \frac{1}{2}\ln\det \boldsymbol{\Lambda} - \frac{1}{2}  (\mathbf{r_w}\mid\mathbf{r_w})_{tf}\,.
    \label{eq:wdm_gauss_analytic}
\end{equation}
Thus the likelihood is the same, regardless of which domain it is computed in.

\subsection{Stationary noise and the Whittle likelihood}
\label{subsec:stationary}

Under stationary noise, where the noise covariance in the frequency domain is diagonal, the frequency-domain likelihood reduces to the standard Whittle likelihood.
To match familiar expressions in the literature in this subsection we return to the LT FFT convention of Eq.~\eqref{eq:FFT_asymmetric_discrete}. 

The stationary noise condition, Eq.~\eqref{eq:stationary_noise}, can be discretized as
\begin{equation}
\begin{split}
{\Sigma}[k,k'] &= \frac{1}{2}S[k]\delta\left[(k - k')\Delta f\right] = \frac{1}{2\Delta f}S[k]\delta_{kk'} = \frac{T}{2}S[k]\delta_{kk'}\,,\\
{\Sigma}[k] & = \frac{T}{2}S[k] = \sigma^2[k]\,,
\end{split}
\label{eq:frequency_T}
\end{equation}
where $\sigma^2[k]$ is the noise variance per frequency bin, and from here on we return to our standard convention.
The inner product is then
\begin{equation}
    \tilde{\mathbf{r}}^{\dagger}\boldsymbol{\Sigma}^{-1}\tilde{\mathbf{r}} = \sum_{k=0}^{N-1}\frac{|\tilde{r}[k]|^2}{\Sigma[k]}=\left[\frac{|\tilde{r}[0]|^2}{\sigma^2[0]} + 2\, \sum_{k=1}^{N/2 - 1}\frac{|\tilde{r}[k]|^2}{\sigma^2[k]} + \frac{|\tilde{r}[N/2]|^2}{\sigma^2[N/2]}\right] = 2\,\sum_{k=0}^{N/2}\epsilon[k]\frac{|\tilde{r}[k]|^2}{\sigma^2[k]}\,,
    \label{eq:discrete_inner_prod}
\end{equation}
where in the last expression we have used the Hermitian symmetry to sum over only the DC and positive frequencies and
\begin{equation}
    \epsilon[k] = \begin{cases}
        1 \qquad &k\in[1,\frac{N}{2}-1]\,\\
        \frac{1}{2} \qquad &k \in \{0, \frac{N}{2}\}
    \end{cases}\,.
\end{equation}
The DC and Nyquist components are real and hence do not get doubled with only one degree of freedom instead of the two associated with the inner bands.
In the above sum, these terms are multiplied by $1/2$ relative to the inner bands to account for this.
This is almost always neglected when writing this expression in paper and often overlooked even in software implementations of the inner product.
Taking the continuum limit of Eq.~\eqref{eq:discrete_inner_prod} from a sum to an integral as $\sum_k\rightarrow T\int \diff f = (1/\Delta f) \int \diff f$, leads to the standard frequency domain noise-weighted inner product
\begin{equation}
    \langle\mathbf{a}\mid\mathbf{b}\rangle= 4\,\Re \int_{0}^{\infty} \frac{\tilde{a}^*(f)\tilde{b}(f)}{S(f)}\diff f\,.
\end{equation}
The overall factor of $4$ comes from two effects: combining positive/negative frequencies and the one-sided PSD definition.

Under this convention, the determinant of the covariance matrix is no longer the same in the time and frequency domains, instead 
\begin{equation}
\det \boldsymbol{\Sigma} = (N \Delta t^2)^N\det \boldsymbol{C} = (T \Delta t)^N\det \boldsymbol{C}\,.
\end{equation}

Finally, we arrive to the Whittle likelihood in the frequency domain as
\begin{equation}
\begin{split}
    \ln p(\mathbf{d} \mid \boldsymbol{\theta}) &=-\frac{N}{2} \ln\left(2\pi\right) + \frac{N}{2} \ln (T \Delta t) - \frac{1}{2}\ln \det \boldsymbol{\Sigma} - \frac{1}{2}\tilde{\mathbf{r}}^\dagger \boldsymbol{\Sigma}^{-1}\tilde{\mathbf{r}} \\
    &= -\frac{N}{2} \ln\left(2\pi\right) + \frac{N}{2} \ln (T \Delta t) - \frac{1}{2}\cdot2 \sum_{k=0}^{N/2} \epsilon[k]\ln \sigma^2[k] - \frac{1}{2}\left[2\,\sum_{k=0}^{N/2}\epsilon[k]\frac{|\tilde{r}[k]|^2}{\sigma^2[k]}\right] \\
    &= -\frac{N}{2} \ln\left(2\pi\right) + \frac{N}{2} \ln (T \Delta t) - \sum_{k=0}^{N/2} \epsilon[k]\ln \sigma^2[k] - \frac{1}{2}\langle\mathbf{r}\mid\mathbf{r}\rangle \\
    & \sim - \sum_{k=0}^{N/2} \epsilon[k]\ln \sigma^2[k] - \frac{1}{2}\langle\mathbf{r}\mid\mathbf{r}\rangle\,.
\end{split}
\label{eq:freq_whittle_likelihood}
\end{equation}
Up to normalization, this matches the common Whittle likelihood used in the gravitational-wave data analysis literature (e.g., Eq.~1 of Ref.~\cite{Katz:2024oqg}), though the per-bin variance $\sigma^2[k]$ is sometimes displayed as the noise PSD (typically $S_n(f)$ elsewhere, but $S(f)$ here).
The noise PSD should be used if the Whittle likelihood is written in the continuous form; otherwise, one should use the per-bin variance as we have defined it here.

In the WDM domain, the stationary noise covariance is not diagonal, see Sec.~\ref{sec:stationary}.
For a diagonal covariance, we need to additionally assume that the noise is white, though corrections can be small for a slowly varying spectrum~\cite{Cornish:2025awt}.

More generically, the evolutionary noise PSD augments the standard stationary PSD $S(f)$ with a time dependence.
We defer a formal definition and exploration of the continuous evolutionary noise PSD $S(t,f)$ and its discretized form to the companion paper~\cite{WDMII}.
For now a working definition will suffice: an extension of the noise covariance at the center of time intervals of width $\Delta T$ that is both stationary and slowly varying with $f$
\begin{equation}
    \langle \mathbf{\tilde{n}}(f)\mathbf{\tilde{n}}^*(f')\rangle \approx \frac{1}{2}S(n\Delta T, f)\delta(f-f')\,.
    \label{eq:slow_stationary_noise}
\end{equation}
Assuming that the time dependence is slow and negligible on the scale of $\Delta T$, the covariance reduces to
\begin{equation}
    \Lambda_{(mn)(m'n')} \approx\frac{1}{2}S(n\Delta T, m \Delta F)\delta_{mm'}\delta_{nn'} = \frac{1}{2}S_{mn}\delta_{mm'}\delta_{nn'}\,,
    \label{eq:mystery_T}
\end{equation}
where the evolutionary PSD $S_{mn}$ is evaluated at the middle of the time-frequency bins.
Equation~\eqref{eq:mystery_T} lacks a factor of $T$ that appears in the frequency-domain covariance of Eq.~\eqref{eq:frequency_T} when the delta function $\delta(f-f')$ is discretized.
This is because the WDM covariance includes a frequency integral of the frequency-domain covariance, e.g., Eq.~\eqref{eq:covariance-term1}, which absorbs the delta function before discretization.
The inner product is then
\begin{equation}
    \mathbf{r_w}^T\boldsymbol{\Lambda}^{-1}\mathbf{r_w}=\sum_{mn}\frac{r^2_{mn}}{\Lambda_{(mn)(mn)}} = 2\sum_{mn}\frac{r_{mn}^2}{S_{mn}}\,,
\end{equation}
and the covariance determinant becomes
\begin{equation}
\det \boldsymbol{\Lambda} = (\Delta t)^N \det \boldsymbol{C}\,.
\end{equation}

Finally, we assemble the full likelihood
\begin{equation}
\begin{split}
    \ln p(\mathbf{d} \mid \boldsymbol{\theta}) &= -\frac{N}{2} \ln\left(2\pi\right) + \frac{N}{2} \ln\left(\Delta t\right) - \frac{1}{2}\sum_{mn} \ln \left(\frac{1}{2}S_{mn}\right) - \sum_{mn}\frac{r_{mn}^2}{S_{mn}}\\
    &= -\frac{N}{2} \ln\left(2\pi\right) + \frac{N}{2} \ln\left(\Delta t\right) +\frac{N}{2}\ln2- \frac{1}{2}\sum_{mn} \ln S_{mn} - \sum_{mn}\frac{r_{mn}^2}{S_{mn}}\\
    &= -\frac{N}{2} \ln\left(\pi\right) + \frac{N}{2} \ln\left(\Delta t\right) - \frac{1}{2}\sum_{mn} \ln S_{mn} - \sum_{mn}\frac{r_{mn}^2}{S_{mn}}\\
    &\sim - \frac{1}{2}\sum_{mn} \ln S_{mn} - \sum_{mn}\frac{r_{mn}^2}{S_{mn}}\,.
    \label{eq:WDM_whittle_likelihood}
\end{split}
\end{equation}
Compared to the frequency-domain Whittle likelihood of Eq.~\eqref{eq:freq_whittle_likelihood}, the WDM-domain Whittle likelihood lacks the factor of 2 that comes from negative/positive frequencies in the Fourier domain and does not require special treatment of the edge bins.

Equations~\eqref{eq:freq_whittle_likelihood} and~\eqref{eq:WDM_whittle_likelihood} are the Whittle likelihood in the frequency and WDM domains in the LT FFT convention.
When self-consistently evaluated (including all the constants), they should yield the same result.
We have verified so numerically to machine precision (and also against the time-domain likelihood).
The same is true when evaluating the likelihood $\ln p(\mathbf{d} \mid \boldsymbol{\theta})$ across different FFT conventions, thanks to the extra prefactors of the determinants.
Thus in the form written, these equations are the likelihood of the data, and not the likelihood of the frequency vector $\tilde{\mathbf r}$ or $\mathbf{r_w}$ that would instead differ by the Jacobian of the convention transformation.
In practice, these Jacobians are always a constant and can be neglected during inference even when the noise covariance is not fixed.

A more formal definition and exploration of the properties of the evolutionary noise PSD will be presented in the companion paper~\cite{WDMII}.

\section{Common Questions}
\label{app:faq}

To conclude, we address commonly asked questions and potential points of confusion about time-frequency analyses and the WDM transform.

\subsection{Is there any loss of information when using the WDM transform?}

No. The WDM basis is complete and orthonormal, Appendix~\ref{app:orthogonality}.
The transform maps $N$ real time-domain samples to $N=N_fN_t$ real WDM coefficients.
The WDM degrees of freedom thus exactly match the time-domain and frequency-domain ones.
In the WDM domain, the $N_f-1$ interior frequency bins contribute $(N_f-1)N_t$ real coefficients, one per cell.
For the two edge bins, DC and Nyquist, half the cells are empty and thus contribute $N_t/2$ coefficients each.
The total is
\begin{equation}
(N_f-1)N_t+(2N_t)/2=N_fN_t=N\,,
\end{equation} 
matching the original time-series degrees of freedom.

The original time series can be recovered exactly via the inverse transform of Sec.~\ref{sec:inverse}.
Equivalently, the WDM transform satisfies Parseval's identity. 
Defining the discrete forward transform matrix $\mathbf W$ from the time to the WDM domain, per Sec.~\ref{sec:logL}, via 
\begin{equation}
    w_{mn}=\sum_{k=0}^{N-1} W_{(mn)k}\,x_k\Rightarrow \mathbf w= \mathbf W \mathbf x\,,
\end{equation}
basis orthonormality implies $\mathbf W^T\mathbf W = \mathbf I$, where $\mathbf I$ is the identity matrix. Therefore
\begin{equation}
\sum_{m=0}^{N_f}\sum_{n=0}^{N_t-1} w_{mn}^2 = \mathbf w^T \mathbf w= \mathbf x^T\mathbf W^T\mathbf W\mathbf x = \mathbf x^T \mathbf x=\sum_{k=0}^{N-1} x_k^2\,,
\end{equation}
so no signal power is lost.

Practical implementations can, however, introduce information loss.
First, if $N$ is a prime number and thus not divisible by $N_f$, the data must be zero-padded or truncated and some boundary samples might be altered.
Second, transforming to the WDM domain from the time-domain directly requires truncation of the filter function.
In the time domain ${\phi}(t)$ decays algebraically as $1/t^{d+1}$ and thus has formally infinite support, while in the frequency domain $\tilde{\phi}(f)$ has compact support.
For this reason, in Sec.~\ref{sec:transform} we first FFT the data to the frequency domain and then transform to the WDM domain.

\subsection{Where does the phase information go if the WDM coefficients are real?}

\begin{figure}
    \centering
    \includegraphics[width=\textwidth]{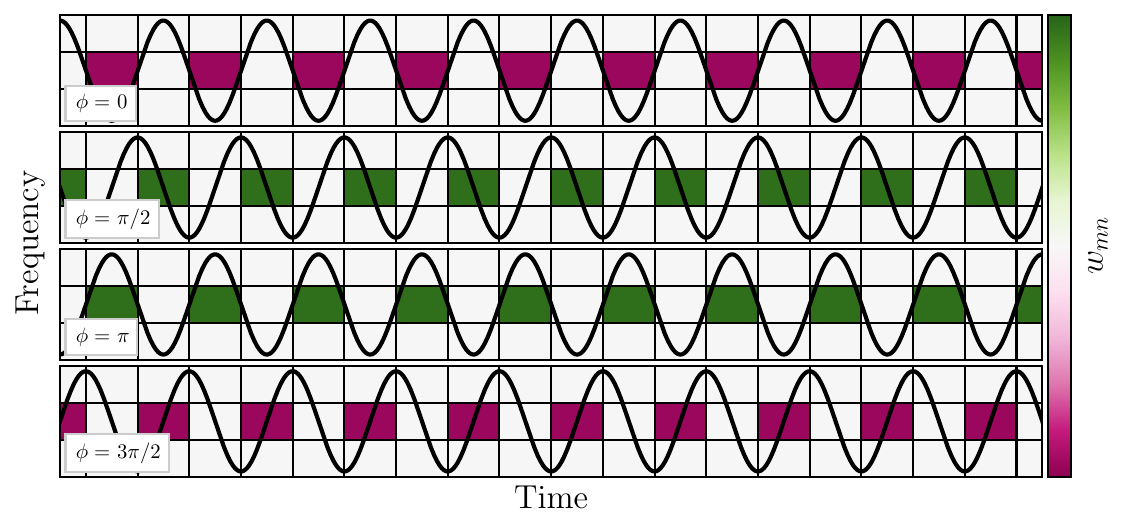}
    \caption{Visualization of how changing the phase value $\phi$ of a simple sinusoid $h(t) = \cos(2\pi f_0 t + \phi)$ affects the corresponding WDM coefficients. 
    Each panel shows the time-domain signal (thick black lines, with arbitrary normalization), the WDM-domain signal (colored boxes), and gridlines demarking the boundary of the WDM cells, for sinusoids with varying phase $\phi$ (top to bottom).
    The relevant frequency index here is $m_0=1$ and therefore per Eq.~\eqref{eq:wdm_phases} odd $n$ have $w_{m_0n} \propto -\cos\phi$ while even $n$ have $w_{m_0n} \propto +\sin\phi$.}
    \label{fig:wdm_phases}
\end{figure}

Consider a simple monochromatic signal given in the time domain by 
\begin{equation}
    h(t) = A \cos (2 \pi f_0 t +\phi)\rightarrow h_k = A \cos (2\pi f_0 t_k +\phi)\,, \quad \mathrm{where~} t_k=k\Delta t, k\in [0,N)\,.
\end{equation}
If the frequency $f_0$ lies on the Fourier grid with $f_0=\ell_0\Delta f$, then a discrete Fourier transform under the unitary convention yields 
\begin{align}
    \tilde{h}[\ell] &= \frac{1}{\sqrt{N}}\sum_{k=0}^{N-1} h_k e^{-2\pi i \ell k /N} = \frac{1}{\sqrt{N}}\sum_{k=0}^{N-1} A \cos (2\pi f_0 t_k +\phi) e^{-2\pi i \ell k /N}= \frac{1}{\sqrt{N}}\sum_{k=0}^{N-1} A \cos (2\pi \ell_0 \Delta f k \Delta t +\phi) e^{-2\pi i \ell k /N}\nonumber\\
    &= \frac{1}{\sqrt{N}}\sum_{k=0}^{N-1} A \cos \left(2\pi \frac{\ell_0 k}{N} +\phi\right) e^{-2\pi i \ell k /N}
    =\frac{A}{2\sqrt{N}} e^{i \phi} \sum_{k=0}^{N-1} e^{2 \pi i\left(\ell_0-\ell\right) k / N}+\frac{A}{2\sqrt{N}} e^{-i \phi} \sum_{k=0}^{N-1} e^{-2 \pi i\left(\ell_0+\ell\right) k / N}\nonumber\\
    &=\frac{A}{2\sqrt{N}} e^{i \phi} N \delta_{\ell, \ell_0}+\frac{A}{2\sqrt{N}} e^{-i \phi} N \delta_{\ell, -\ell_0} = \frac{A \sqrt{N}}{2}\left(e^{i \phi} \delta_{\ell, \ell_0}+ e^{-i \phi} \delta_{\ell, -\ell_0}\right)\,.
\end{align}
Restricting to positive frequencies, $\ell=\ell_0$, the Fourier coefficient for bin $\ell$ is
\begin{equation}
    \tilde{h}[\ell_0] = \frac{A \sqrt{N}}{2}\left(\cos\phi+i\sin\phi\right)\,.
\end{equation}
Therefore the signal amplitude is encoded in the complex coefficient amplitude and the signal phase in the complex coefficient phase. 
A pure cosine signal has a real coefficient and a pure sine signal has an imaginary coefficient.
Two signal degrees of freedom (amplitude and phase) map onto two Fourier degrees of freedom (the real and imaginary part).

In the WDM domain all coefficients are real and the signal phase is instead encoded in adjacent time cells.
The WDM coefficients for the sinusoid signal, assuming interior bands, are obtained by substituting the continuous frequency-domain signal
\begin{equation}
    \tilde{h}(f) = \frac{A }{2}\left[e^{i \phi} \delta\left(f-f_0\right)+e^{-i \phi} \delta\left(f+f_0\right)\right]\,,
\end{equation}
into Eq.~\eqref{eq:wnm_final}:
\begin{equation}
w_{m n}=(-1)^{m n} \frac{A}{\sqrt{2}} \int_{-\Delta F}^{\Delta F} \Re\left[C_{m n}^* \left[e^{i \phi} \delta\left(f+m \Delta F-f_0\right)+e^{-i \phi} \delta\left(f+m \Delta F+f_0\right)\right] \tilde{\phi}(f) e^{2\pi i n f \Delta T}\right] \diff f\,.
\end{equation}
The two delta functions set $f=f_0-m\Delta F$ and $f=-f_0-m\Delta F$ respectively. 
Within the integration limits of $[-\Delta F,\Delta F]$, the two conditions become $(m-1) \Delta F \leq f_0 \leq(m+1) \Delta F$ and $(m-1) \Delta F \leq -f_0 \leq(m+1) \Delta F$ respectively. 
The second condition cannot be satisfied for $m>0$ and $f_0>0$ and thus the integral of the second term vanishes. 
The first condition is satisfied for $m\in \{m_0-1,m_0,m_0+1\}$ where $m_0= f_0/\Delta F$.
Assuming again that $f_0$ falls at the center of a WDM cell, $f_0=m_0\Delta F$, the only contribution is from $m=m_0$ where $\tilde{\phi}(f=f_0-m_0\Delta F=0)=1/\sqrt{\Delta F}$:
\begin{equation} \label{eq:wdm_phases}
w_{m_0 n}=(-1)^{m_0 n} \frac{A}{\sqrt{2\Delta F}} \Re\left[C_{m_0 n}^*e^{i \phi} \right]= (-1)^{m_0 n} \frac{A }{\sqrt{2\Delta F}}
\begin{cases}
    \cos{\phi} & m_0 + n \mathrm{~even}\\
    \sin{\phi} & m_0 + n \mathrm{~odd}
\end{cases}
\,.
\end{equation}
The information that was previously packaged into a single complex Fourier coefficient is now spread over two real coefficients on adjacent time cells of opposite parity.
The signal amplitude is encoded in the sum of the squared coefficients and the signal phase in their ratio.
Likewise, two signal degrees of freedom now map onto two adjacent WDM coefficients of opposite parity at the same band $m_0$.

In Fig.~\ref{fig:wdm_phases}, we provide a visualization of this example monochromatic signal for four phase values: $\phi=0$, $\phi=\pi/2$, $\phi=\pi$, and $\phi=3\pi/2$.
Since the sinusoids have constant frequency in the center of a bin, the nonzero cells are confined to one row, while the phase information enters in each pair of adjacent time cells.
In this example, $\cos\phi$ and $\sin\phi$ are always either zero or one, so only one out of every two time cells has a nonzero coefficient; intermediate phase values would allow for both cells in each pair to be nonzero, with the ratio providing $\tan\phi$.

This calculation also reveals that there is no limit of the WDM series to the frequency series as two real numbers need to be combined into one complex number.
The exact correspondence between the two in the limit $N_f\rightarrow N/2$, $N_t=2$ is derived in Appendix~\ref{app:correspondence}.

\subsection{Is WDM a wavelet or wavelet packet transform?}
\label{sec:wavelet-vs-WDM}

No. Wavelet transforms lie on a non-uniform (typically dyadic) time-frequency grid. 
Wavelet packet transforms are more complicated and allow for flexible (including uniform) tilings, but employ a different construction than the WDM transform.
Wavelet and wavelet packet transforms are associated with the affine group (translations and dilations), while the WDM transform is associated with the Heisenberg group (translations in time and frequency)~\cite{bratteli2002wavelets}.
In Appendix~\ref{app:WDM_Heisenberg} we construct the WDM transform from the Heisenberg group.

In wavelet transforms, the basis is generated through translations and dilations of a ``mother wavelet'' $\psi(t)$, e.g.,
\begin{equation}
\psi_{j n}(t)=2^{j / 2} \psi\left(2^j t-n\right)\,.
\end{equation}
The dyadic dilation produces logarithmically-spaced cells with height $2^j$ and width $2^{-j}$, effectively favoring time resolution over frequency resolution at high frequencies and vice versa. 
The WDM transform, by contrast, tiles the time-frequency plane through translations and frequency modulations
\begin{equation}
g_{m n}(t)=g_m(t-n \Delta T)\,, \qquad g_m(t) \propto \cos (2\pi m \Delta F t) \phi(t) \text { or } \sin (2\pi m \Delta F t) \phi(t)\,,
\end{equation}
see Eq.~\eqref{eq:gmunu_int}.
This results in uniformly-spaced cells with the same height $\Delta F$ and width $\Delta T$, see Fig.~\ref{fig:tf_grid}.

The dyadic-wavelet and WDM grids differ in the quality factor $Q$, which is the ratio of the band center to the bandwidth $\Delta f_\mathrm{bw}$.
A dyadic wavelet scales both by the same factor such that
\begin{equation}
    Q = \frac{f_c}{\Delta f_\mathrm{bw}} = \mathrm{const.}
\end{equation}
However, for the WDM transform we have up to factors of ${\cal{O}}(1)$ due to the rolloff
\begin{equation}
    Q \propto \frac{m\Delta F}{\Delta F} = m\,,
\end{equation}
which grows linearly with band index.
The band-dependent $Q$ shows that the WDM transform is not a dyadic wavelet transform. 

The quality factor does not by itself exclude a wavelet packet transform, which can have uniform bands with a linearly-growing $Q$. 
The distinction is instead in the construction of the basis. 
WDM modulates a single fixed window with sines/cosines (Heisenberg). 
Wavelet packets, by contrast, build bands by repeatedly dilating/halving the frequency axis (affine); halving all bands equally yields a uniform grid resembling WDM's, while other choices yield more customized grids.

The Meyer scaling function appears in both the WDM transform (as the window) and the Meyer wavelet transform (as the mother wavelet).
However, this overlap is nominal as the two transforms employ it differently and the crucial difference is that the WDM transform does not perform dilations.

\begin{acknowledgments}

We thank Michael Katz, Pat Meyers, Tyson Littenberg, and Sophie Bini for discussions and feedback.
We thank Robbie Rosati for comments on the draft and discussions about the WDM transform.
AJ was supported by the NASA LISA Project Office.
KC was supported by NSF Grant PHY-2409001.
JS was supported by the Caltech Walter Burke Institute of Theoretical Physics Graduate Fellowship.
Anthropic's Claude agentic AI was used in the production of this paper for reviewing the manuscript and derivations, producing initial algorithm blocks, and producing software implementations of the algorithms within.
The authors have reviewed everything produced with agentic AI for correctness.
This work has made use of the \textsc{numpy}~\cite{harris2020array}, \textsc{scipy}~\cite{scipy}, and \textsc{matplotlib}~\cite{matplotlib} Python packages.
\end{acknowledgments}

\appendix

\section{Partition of unity}
\label{app:POU}

In this Appendix, we prove the partition-of-unity property of the Meyer scaling function in Eq.~\eqref{eq:Meyer-def}:
\begin{equation}
    {\mathcal{P}}=\sum_{p\in\mathbb{Z}}\tilde{\varphi}(f - p\Delta F)^2 = 1\,,
    \label{eq:POUapp}
\end{equation}
assuming 
\begin{equation}
    \nu_d(x) + \nu_d(1 - x) = 1\,,
\end{equation} 
which follows from Eq.~\eqref{eqn:rolloff}.
Since ${\mathcal{P}}$ is periodic in $\Delta F$, without loss of generality we restrict $f\in [0,\Delta F]$ such that only the $p=0$ and $p=1$ filters have support. 
This area includes half of each filter's constant region plus the rolloff region where they overlap: 
\begin{align}
    \tilde{\varphi}_0(f)\equiv \tilde{\varphi}(f) &=
    \begin{cases}
        1 & \text{if } f < \alpha\\
        \cos \left[\frac{\pi}{2}\nu_d\left(\frac{f-\alpha}{\beta}\right) \right] & \text{if } \alpha < f \leq \alpha + \beta\\
        0 & \text{otherwise}
    \end{cases}\,,\\
    \tilde{\varphi}_1(f) \equiv \tilde{\varphi}(f-\Delta F) &=
    \begin{cases}
        1 & \text{if } f>\Delta F-\alpha\\
        \cos \left[\frac{\pi}{2}\nu_d\left(\frac{\Delta F-f-\alpha}{\beta}\right) \right] & \text{if } \Delta F - (\alpha + \beta)< f \leq \Delta F-\alpha\\
        0 & \text{otherwise}
    \end{cases}\nonumber \\
    &=
    \begin{cases}
        1 & \text{if } f>\alpha+\beta\\
        \cos \left[\frac{\pi}{2}\nu_d\left(1-\frac{f-\alpha}{\beta}\right) \right] & \text{if } \alpha < f \leq \alpha+\beta\\
        0 & \text{otherwise}
    \end{cases}\,,
    \label{eq:Meyer-12}
\end{align}
where in the last line we have used the constraint $2\alpha+\beta=\Delta F$.
For $f<\alpha$ and $f>\alpha+\beta$, Eq.~\eqref{eq:POUapp} is trivially satisfied as ${\mathcal{P}}=\tilde{\varphi}_0^2=1$ and ${\mathcal{P}}=\tilde{\varphi}_1^2=1$ respectively. For $\alpha\leq f \leq \alpha+\beta$ we have
\begin{align}
    {\mathcal{P}}&=\tilde{\varphi}_0^2+\tilde{\varphi}_1^2\nonumber \\
    &=\cos^2 \left[\frac{\pi}{2}\nu_d\left(\frac{f-\alpha}{\beta}\right) \right]+\cos^2 \left[\frac{\pi}{2}\nu_d\left(1-\frac{f-\alpha}{\beta}\right) \right]\nonumber\\
    &=\cos^2 \left[\frac{\pi}{2}\nu_d\left(\frac{f-\alpha}{\beta}\right) \right]+\cos^2 \left[\frac{\pi}{2}\left(1-\nu_d\left(\frac{f-\alpha}{\beta}\right)\right) \right]\nonumber\\
    &=\cos^2 \left[\frac{\pi}{2}\nu_d\left(\frac{f-\alpha}{\beta}\right) \right]+\sin^2 \left[\frac{\pi}{2}\nu_d\left(\frac{f-\alpha}{\beta}\right) \right]=1\,,
\end{align}
where going from the second to third line we have used the property $\nu_d(x) + \nu_d(1 - x) = 1$ and from the third to the fourth line we have applied trigonometric identities. 
Thus Eq.~\eqref{eq:POUapp} is satisfied for all values of $f$.

\section{Orthogonality}
\label{app:orthogonality}

In this Appendix, we show orthogonality of the WDM basis. Substituting Eq.~\eqref{eq:gmunu} into Eq.~\eqref{eqn:orthonormality} we get
\begin{align}
    {\mathcal{O}}&\equiv\int_{-\infty}^{+\infty} \tilde{g}_{mn}(f)\tilde{g}_{pq}^*(f)\, \diff f \nonumber \\
    &= A_m A_p \int_{-\infty}^{+\infty}e^{-2\pi i (n - q) f\Delta T} \left[ C_{mn}^* \tilde{\phi}(f + m\Delta F) + C_{mn}\tilde{\phi}(f - m\Delta F) \right]\left[ C_{pq} \tilde{\phi}(f + p\Delta F) + C_{pq}^*\tilde{\phi}(f - p\Delta F) \right]\diff f \,.
    \label{eq:gmunu_orth1}
    \end{align}
A filter centered at $m\Delta F$ only has support in the bands centered at $(m-1)\Delta F$, $m\Delta F$, and $(m+1)\Delta F$. 
Therefore Eq.~\eqref{eq:gmunu_orth1} vanishes unless $|m-p| \leq 1$, since otherwise the filters have no overlapping support.

Before proceeding, we simplify the above expression. We define
\begin{equation}
I^{+} = \int_{-\infty}^{+\infty} e^{-2\pi i (n - q) f \Delta T} \tilde{\phi}(f + m\Delta F) \tilde{\phi}(f + p\Delta F) \diff f\,,
\end{equation}
make the substitution $x=f + m \Delta F$ and recall that $\Delta F \Delta T=1/2$ to get
\begin{align}
I^{+} &= \int_{-\infty}^{+\infty} e^{-2\pi i (n - q) (x - m \Delta F) \Delta T} \tilde{\phi}(x) \tilde{\phi}(x + (p-m)\Delta F) \diff x\nonumber\\
&= e^{ i (n - q) m \pi} \int_{-\infty}^{+\infty} e^{-2\pi i (n - q) x \Delta T} \tilde{\phi}(x) \tilde{\phi}(x + (p-m)\Delta F) \diff x
\nonumber\\
&= (-1)^{(n - q) m} \int_{-\infty}^{+\infty} e^{-2\pi i (n - q) x \Delta T} \tilde{\phi}(x) \tilde{\phi}(x + (p-m)\Delta F) \diff x\,.
\end{align}
Similar steps lead to
\begin{align}
I^{-} &= \int_{-\infty}^{+\infty} e^{-2\pi i (n - q) f \Delta T} \tilde{\phi}(f - m\Delta F) \tilde{\phi}(f - p\Delta F) \diff f\nonumber\\
&= (-1)^{(n - q) m} \int_{-\infty}^{+\infty} e^{-2\pi i (n - q) x \Delta T} \tilde{\phi}(x) \tilde{\phi}(x - (p-m)\Delta F) \diff x\,.
\end{align}
Further substituting $x\rightarrow-x$ and recalling that $\tilde{\phi}$ is an even function leads to
\begin{align}
I^{-} &= (-1)^{(n - q) m} \int_{-\infty}^{+\infty} e^{2\pi i (n - q) x \Delta T} \tilde{\phi}(x) \tilde{\phi}(x + (p-m)\Delta F) \diff x = \left(I^{+}\right)^*\,.
\end{align}
Similarly, we obtain
\begin{align}
I^{-+} &= \int_{-\infty}^{+\infty} e^{-2\pi i (n - q) f \Delta T} \tilde{\phi}(f - m\Delta F) \tilde{\phi}(f + p\Delta F) \diff f\nonumber\\
&= (-1)^{(n - q) m} \int_{-\infty}^{+\infty} e^{-2\pi i (n - q) x \Delta T} \tilde{\phi}(x) \tilde{\phi}(x + (p+m)\Delta F) \diff x\,,\\
I^{+-} &= \int_{-\infty}^{+\infty} e^{-2\pi i (n - q) f \Delta T} \tilde{\phi}(f + m\Delta F) \tilde{\phi}(f - p\Delta F) \diff f \nonumber\\
&= (-1)^{(n - q) m} \int_{-\infty}^{+\infty} e^{2\pi i (n - q) x \Delta T} \tilde{\phi}(x) \tilde{\phi}(x + (p+m)\Delta F) \diff x = \left(I^{-+}\right)^*\,.
\end{align}
Introducing
\begin{equation}
    F^{\pm} = \int_{-\infty}^{+\infty} e^{-2\pi i (n - q) x \Delta T} \tilde{\phi}(x) \tilde{\phi}(x + (p\pm m)\Delta F) \diff x\,,
    \label{eq:FPM-def}
\end{equation}
Eq.~\eqref{eq:gmunu_orth1} becomes
\begin{align}
    {\mathcal{O}}&=A_m A_p \left[ C_{mn}^*C_{pq} I^+ + C_{mn}^*C_{pq}^*I^{+-} + C_{mn}C_{pq}I^{-+} + C_{mn}C_{pq}^*I^-\right]\nonumber \\
    &=A_m A_p (-1)^{(n - q) m} \left[C_{mn}^*C_{pq} F^- + C_{mn}^*C_{pq}^*\left(F^{+}\right)^* + C_{mn}C_{pq}F^{+} + C_{mn}C_{pq}^*\left(F^{-}\right)^*  \right]\nonumber \\
    &=2 A_m A_p (-1)^{(n - q) m} \left[\Re \left(C_{mn}^*C_{pq} F^- \right)+\Re\left( C_{mn}C_{pq}F^{+}\right) \right]\,.
    \label{eq:gmunu_orth_F2}
\end{align}
For orthogonality we need to verify that this expression vanishes if $|p-m|=1$ or $p=m$, $n\neq q$.

\subsection{Normalization}

For $p=m$, $n= q$ the inner product needs to normalize to unity. Momentarily restricting to $p=m\neq0,N_f$, we find
\begin{align}
    F^+ = \int_{-\infty}^{+\infty} \tilde{\phi}(f) \tilde{\phi}(f + 2 m\Delta F) \diff f=0\,,
\end{align}
as $\tilde{\phi}(f)$ and $\tilde{\phi}(f + 2 m\Delta F)$ have no overlapping support, and 
\begin{align}
    F^- &= \int_{-\infty}^{+\infty} \tilde{\phi}(f)^2 \diff f= \frac{1}{\Delta F}\int_{-\infty}^{+\infty} \tilde{\varphi}(f)^2 \diff f\,.
\end{align}
The integral of the Meyer filter can be split off in the flat and rolloff regions:
\begin{align}
    \int_{-\infty}^{+\infty} \tilde{\varphi}(x)^2 \diff x &= 
    \int^{-\alpha}_{-\alpha-\beta}\cos^2 \left[\frac{\pi}{2}\nu_d\left(\frac{|f|-\alpha}{\beta}\right) \right] \diff f + \int_{-\alpha}^{\alpha} \diff f+\int^{\alpha+\beta}_{\alpha}\cos^2 \left[\frac{\pi}{2}\nu_d\left(\frac{|f|-\alpha}{\beta}\right) \right] \diff f\nonumber \\
    &=2\alpha + 2 \int^{\alpha+\beta}_{\alpha}\cos^2 \left[\frac{\pi}{2}\nu_d\left(\frac{f-\alpha}{\beta}\right) \right] \diff f\,.
\end{align}
Substituting $x=(f-\alpha)/\beta$ yields
\begin{align}
    \int_{-\infty}^{+\infty} \tilde{\varphi}(f)^2 \diff f &=2\alpha + 2 \int^{\alpha+\beta}_{\alpha}\cos^2 \left[\frac{\pi}{2}\nu_d\left(\frac{f-\alpha}{\beta}\right) \right] \diff f 
    = 2\alpha + 2 \beta \int^{1}_{0}\cos^2 \left[\frac{\pi}{2}\nu_d\left(x\right) \right]\diff x\,.
\end{align}
Another substitution $x\rightarrow 1-x$ in the last integral yields
\begin{align}
    \int^{1}_{0}\cos^2 \left[\frac{\pi}{2}\nu_d\left(x\right) \right]\diff x &= \int^{1}_{0}\cos^2 \left[\frac{\pi}{2}\nu_d\left(1-x\right) \right]\diff x 
    = \int^{1}_{0}\cos^2 \left[\frac{\pi}{2}\left(1-\nu_d\left(x\right)\right) \right]\diff x=\int^{1}_{0}\sin^2 \left[\frac{\pi}{2}\nu_d\left(x\right) \right]\diff x = \frac{1}{2}\,,
\end{align}
where we have used the partition-of-unity condition, $\nu_d(x) + \nu_d(1 - x) = 1$, and the fact that $\cos^2x+\sin^2x=1$.
Collecting everything, we find
\begin{equation}
\int_{-\infty}^{+\infty} \tilde{\varphi}(f)^2 \diff f =2\alpha + \beta = \Delta F\,,
\end{equation}
and therefore $F^-=1$.

The inner product then reduces to 
\begin{align}
    {\mathcal{O}}&=2 A_m^2 \left[\Re \left(C_{mn}^*C_{mn} F^- \right)+\Re\left( C_{mn}C_{mn}F^{+}\right) \right]=2  \frac{1}{2} \Re \left(C_{mn}^*C_{mn}\right) = 1
    \label{eq:gmunu_orth_F}
\end{align}
for both $m+n$ even and odd.

For $m=p=0,N_f$, a similar calculation ensues with a few differences. 
Firstly, $F^+$ no longer vanishes, instead $F^+=F^-=1$. 
For $m=0$, this is obvious. 
For $m=N_f$, $F^+$ cannot be obtained from the continuous expression of Eq.~\eqref{eq:FPM-def}.
The result $F^+=F^-=1$ is instead a consequence of aliasing due to finite sampling where the filter centered at $N_f\Delta F$ aliases onto the one at $-N_f\Delta F$, see Appendix~\ref{app:nyquist}.
Second, the amplitude prefactor is smaller by a factor of $\sqrt{1/2}$. 
Overall, the inner product for the DC ($m=0$) and Nyquist ($m=N_f$) bins evaluates to 
\begin{align}
    {\mathcal{O}}&=2 A_{m}^2 \left[\Re \left(C_{mn}^*C_{mn} F^- \right)+\Re\left( C_{mn}C_{mn}F^{+}\right) \right]\nonumber \\
    &=2  \frac{1}{4} \left[\Re \left(C_{mn}^*C_{mn}\right)+\Re\left( C_{mn}C_{mn}\right) \right]\,,
\end{align}
which vanishes for $m+n$ odd and evaluates to $1$ for $m+n$ even.
Thus for the DC and Nyquist bands, half the cells are empty.

\subsection{Orthogonality, $p=m$, $n\neq q$}

Setting $p=m\neq 0, N_f$, $n\neq q$ we again get
\begin{equation}
    F^{+} = \int_{-\infty}^{+\infty} e^{-2\pi i (n - q) x \Delta T} \tilde{\phi}(x) \tilde{\phi}(x + 2 m\Delta F) \diff x=0\,,
\end{equation}
as the filters in the integrand share no support, and 
\begin{equation}
    F^{-} = \int_{-\infty}^{+\infty} e^{-2\pi i (n - q) x \Delta T} \tilde{\phi}(x)^2 \diff x\,.
\end{equation}
This expression amounts to the Fourier transform of $\tilde{\phi}(x)^2$ and thus $F^-$ is real, as $\tilde{\phi}(x)^2$ is real and even.
The inner product reduces to 
\begin{align}
    {\mathcal{O}}&= 2 A_m^2 (-1)^{(n - q) m} \Re \left(C_{mn}^*C_{mq} \right) F^- \,.
    \label{eq:gmunu_orth_peqm}
\end{align}

\begin{itemize}
\item Case 1: $n-q$ is odd. In this case, $m+n$ and $m+q$ have opposite parities and $\Re \left(C_{mn}^*C_{mq}\right)=0$. The inner product thus vanishes for odd $n-q$.
\item Case 2: $n-q$ is even. In this case, defining an integer $n-q=2k$, the integral is
\begin{equation}
    F^{-} = \int_{-\infty}^{+\infty} e^{-2\pi i (2k) x \Delta T} \tilde{\phi}(x)^2 \diff x\,.
\end{equation}
We invoke the partition-of-unity equation 
\begin{equation}
    \sum_{p\in\mathbb{Z}}\tilde{\phi}(f - p\Delta F)^2 = \frac{1}{\Delta F}\,,
\end{equation}
which we Fourier transform to obtain
\begin{align}
    \int e^{-2\pi if t}\sum_{p\in\mathbb{Z}}\tilde{\phi}(f - p\Delta F)^2 \diff f&= \int e^{-2\pi if t}\frac{1}{\Delta F}\diff f \Rightarrow \\
\sum_{p\in\mathbb{Z}}\int e^{-2\pi if t}\tilde{\phi}(f - p\Delta F)^2 \diff f&= \frac{1}{\Delta F}\delta(t) \xrightarrow{u=f-p\Delta F}\\
\sum_{p\in\mathbb{Z}}\int e^{-2\pi i(u+p\Delta F) t}\tilde{\phi}(u)^2 \diff u&= \frac{1}{\Delta F}\delta(t) \rightarrow\\
\sum_{p\in\mathbb{Z}}e^{-2\pi i p\Delta F t}\int e^{-2\pi i ut}\tilde{\phi}(u)^2 \diff u&= \frac{1}{\Delta F}\delta(t) \rightarrow\\
\int e^{-2\pi i ut}\tilde{\phi}(u)^2 \diff u\sum_{p\in\mathbb{Z}}e^{-2\pi i p\Delta F t}&= \frac{1}{\Delta F}\delta(t)\,.
\end{align}
The first term on the left hand side is $F^-$ evaluated at $t=2 k \Delta T=2k/(2\Delta F)$.
The second term for the sum of complex exponentials can be expressed through the Poisson summation formula as
\begin{equation}
    \sum_{p\in\mathbb{Z}}e^{-2 \pi i p\Delta F t} = \frac{1}{\Delta F}\sum_{\ell\in\mathbb{Z}}\delta\left(t-\frac{\ell}{\Delta F} \right)\,, 
\end{equation}
leading to 
\begin{equation}
    \int e^{-2\pi i ut}\tilde{\phi}(u)^2 \diff u\sum_{\ell\in\mathbb{Z}}\delta\left(t-\frac{\ell}{\Delta F}\right)= \delta(t)
\end{equation}
which means that the integral must vanish for all integer multiples of $t=\ell/\Delta F\neq 0$, including $ k/\Delta F$ which corresponds to $F^-$.
The inner product thus vanishes for even $n-q$.
\end{itemize}

For $p=m= 0,N_f$, the calculation is essentially identical, only $F^+$ no longer vanishes but it is equal to $F^-$, similarly to the normality calculation above.
From then, the argument proceeds unaltered and the inner product vanishes by the parity argument for $n-q$ odd and partition-of-unity and Poisson summation resulting in $F^+=F^-=0$ for $n-q$ even.

\subsection{Orthogonality, $|p-m|=1$}

We restrict to $1<p=m+1<N_f$ without loss of generality, as the $m=p+1$ case follows by symmetry. Now
\begin{equation}
    F^{+} = \int_{-\infty}^{+\infty} e^{-2\pi i (n - q) x \Delta T} \tilde{\phi}(x) \tilde{\phi}(x + (2m+1)\Delta F) \diff x=0\,,
\end{equation}
as the filters have no support, and 
\begin{equation}
    F^{-} = \int_{-\infty}^{+\infty} e^{-2\pi i (n - q) x \Delta T} \tilde{\phi}(x) \tilde{\phi}(x + \Delta F) \diff x\,.
\end{equation}
The inner product reduces to
\begin{align}
    {\mathcal{O}}
    &=2 A_m A_{m+1} (-1)^{(n - q) m} \Re \left(C_{mn}^*C_{(m+1)q} F^- \right)\,.
\end{align}
The surviving term only has support in the region where the adjacent windows overlap, $[-\Delta F,0]$.
The change of variables $u=x+\Delta F/2$ centers the functions about the overlap interval and yields
\begin{align}
    F^{-} &=\int_{-\Delta F}^{0} e^{-2\pi i (n - q) x \Delta T} \tilde{\phi}(x) \tilde{\phi}(x + \Delta F) \diff x\nonumber\\
    &= \int_{-\Delta F/2}^{+\Delta F/2} e^{-2\pi i (n - q) \left(u-\frac{\Delta F}{2}\right) \Delta T} \tilde{\phi}\left(u-\frac{\Delta F}{2}\right) \tilde{\phi}\left(u+\frac{\Delta F}{2}\right) \diff u\nonumber\\
    &=e^{2\pi i (n - q) \frac{\Delta F \Delta T}{2}} \int_{-\Delta F/2}^{+\Delta F/2} e^{-2\pi i (n - q) u\Delta T} \tilde{\phi}\left(u-\frac{\Delta F}{2}\right) \tilde{\phi}\left(u+\frac{\Delta F}{2}\right) \diff u\nonumber\\
    &=e^{i (n - q) \frac{\pi}{2}} \int_{-\Delta F/2}^{+\Delta F/2} e^{-2\pi i (n - q) u\Delta T} \tilde{\phi}\left(u-\frac{\Delta F}{2}\right) \tilde{\phi}\left(u+\frac{\Delta F}{2}\right) \diff u\,.
\end{align}
Since $\tilde{\phi}(u)$ is even, so is the product:
\begin{equation}
\tilde{\phi}\left(u-\frac{\Delta F}{2}\right) \tilde{\phi}\left(u+\frac{\Delta F}{2}\right)=\tilde{\phi}\left(-u+\frac{\Delta F}{2}\right) \tilde{\phi}\left(-u-\frac{\Delta F}{2}\right)=\tilde{\phi}\left((-u)-\frac{\Delta F}{2}\right)\tilde{\phi}\left((-u)+\frac{\Delta F}{2}\right)\,.
\end{equation}

\begin{itemize}

\item Case 1: $n-q=2k$ is even. In this case, $C_{mn}$ and $C_{(m+1)q}$ have opposite parity and thus their product is imaginary. The inner product is proportional to the imaginary part of $F^-$,
\begin{align}
    \Im \left[F^{-}\right] &= \sin\left[ k \pi\right] \int_{-\Delta F/2}^{+\Delta F/2} \cos\left[ 2\pi (2k) u\Delta T\right] \tilde{\phi}\left(u-\frac{\Delta F}{2}\right) \tilde{\phi}\left(u+\frac{\Delta F}{2}\right) \diff u\nonumber \\
    &-\cos\left[ k \pi\right] \int_{-\Delta F/2}^{+\Delta F/2} \sin\left[2\pi (2k) u\Delta T\right] \tilde{\phi}\left(u-\frac{\Delta F}{2}\right) \tilde{\phi}\left(u+\frac{\Delta F}{2}\right) \diff u\,.
\end{align}
The first term vanishes due to $\sin\left[ k \pi\right]=0$ for any $k\in\mathbb{Z}$, while the second term vanishes as the integral of an odd function (the even filter product and the odd sine) over a symmetric interval.
Overall, $\Im \left[F^{-}\right]=0$ and the inner product vanishes for even $n-q$.
\item Case 2: $n-q=2k+1$ is odd. In this case, $C_{mn}$ and $C_{(m+1)q}$ have the same parity and thus their product is real. The inner product is proportional to the real part of $F^-$,
\begin{align}
    \Re \left[F^{-}\right] &= \cos\left[ k \pi+\frac{\pi}{2}\right] \int_{-\Delta F/2}^{+\Delta F/2} \cos\left[2\pi (2k+1) u\Delta T\right] \tilde{\phi}\left(u-\frac{\Delta F}{2}\right) \tilde{\phi}\left(u+\frac{\Delta F}{2}\right) \diff u\nonumber \\
    &+\sin\left[ k \pi+\frac{\pi}{2}\right] \int_{-\Delta F/2}^{+\Delta F/2} \sin\left[2\pi (2k+1) u\Delta T\right] \tilde{\phi}\left(u-\frac{\Delta F}{2}\right) \tilde{\phi}\left(u+\frac{\Delta F}{2}\right) \diff u\,.
\end{align}
The first term vanishes due to $\cos\left[ k \pi+\pi/2\right]=0$ for any $k\in\mathbb{Z}$, while the second term vanishes as the integral of an odd function (the even filter product and the odd sine) over a symmetric interval.
Overall, $\Re \left[F^{-}\right]=0$ and the inner product vanishes for odd $n-q$.
\end{itemize}

For the DC ($p=1$, $m=0$) and Nyquist ($p=N_f$, $m=N_f-1$) bins  again $F^+=F^-$, after which the argument proceeds similarly.

\section{Nyquist periodicity}
\label{app:nyquist}

In this Appendix, we show that the finite sampling results in the frequency-domain data having a periodicity of twice the Nyquist frequency $f_\mathrm{Nyq} = N_f\Delta F = 1/(2\Delta t)$.
The time-domain data are sampled at intervals of $\Delta t$
\begin{equation}
    x_s(t) = \sum_{k=0}^{N-1}x_k\delta(t-k\Delta t)\,.
\end{equation}
Substituting the samples, the continuous Fourier transform is
\begin{equation}
    \tilde{x}_{s}(f) = \int_{-\infty}^{\infty}x_s(t)e^{-2\pi if t}\diff t = \int_{-\infty}^{\infty}\sum_{k}x_k\delta(t-k\Delta t)e^{-2\pi if t}\diff t = \sum_{k}x_{k}e^{-2\pi if k\Delta t}\,,
\end{equation}
which is periodic with frequency twice the Nyquist frequency $f_\mathrm{Nyq}$,
\begin{align}
    \tilde{x}_s(f + 2f_\mathrm{Nyq}) &= \sum_{k} x_k e^{-2\pi i(f + 2f_\mathrm{Nyq})k\Delta t} =  \sum_k x_k e^{-2\pi if k \Delta t}e^{-4\pi i f_\mathrm{Nyq} k\Delta t}\nonumber \\
    &=  \sum_k x_k e^{-2\pi if k \Delta t}\left(e^{-2\pi i}\right)^k =  \sum_k x_k e^{-2\pi if k \Delta t} = \tilde{x}_s(f)\,.
    \label{eq:Nyqperiodicity}
\end{align}
This is a direct consequence of finite sampling and of course does not apply for continuous functions.

\section{Fast algorithms for transforms}
\label{app:algorithms}

\begin{algorithm}[H]
\caption{Forward WDM: time series $x$ $\to$ WDM grid $w$.}
\label{alg:fwd}
\begin{algorithmic}[1]
\Require Real time series $x \in \mathbb{R}^{N}$, parameters $(N_f, N_t, \Delta t)$.
\Ensure WDM coefficient grid $w \in \mathbb{R}^{(N_f+1)\times N_t}$.
\State $X \gets \mathrm{RFFT}(x)$ \Comment{length $N/2 + 1$}
\State $\kappa \gets \sqrt{2}/N_f$
\State $w \gets \mathbf{0}_{(N_f+1)\times N_t}$
\For{$m = 0$ to $N_f$}
    \State $u_m[k] \leftarrow \mathbf{0}$
        \Comment{Build spectrum slice $u_m \in \mathbb{C}^{N_t}$}
    \For{$k = 0$ to $N_t - 1$}
        \State $\ell \gets k + (m-1)\,N_t/2$
        \If{$\ell < 0$}
            \State $z \gets {\mathrm{conj}(X[-\ell]})$
            \Comment{only at $m = 0$, $k < N_t/2$: negative index $\to$ conjugate the positive bin}
        \ElsIf{$\ell > N / 2$}
            \State $z \gets \mathrm{conj}({X[N - \ell]})$
            \Comment{only at $m = N_f$, $k > N_t/2$: over-Nyquist $\to$ conjugate the mirror bin}
        \Else
            \State $z \gets X[\ell]$
            \Comment{taken for interior bands $1 \le m \le N_f - 1$}
        \EndIf
        \State $u_m[k] \gets z \cdot \tilde{\phi}[k - N_t/2]$
        \Comment{$\tilde{\phi}$ centered on $k = N_t/2$}
    \EndFor
    \State $y_m \gets \mathrm{IFFT}(u_m)$ \Comment{\textsc{numpy} IFFT convention gives $1/N_t$; accounted for in $\kappa$}
    \For{$n = 0$ to $N_t - 1$}
        \If{$m \in \{0, N_f\}$ and $(m + n) \bmod 2 \ne 0$}
            \State $w[m, n] \gets 0$ \Comment{only even-parity edge pixels nonzero}
            \State \textbf{continue}
        \EndIf
        \State $w[m, n] \gets \kappa\,(-1)^{(m+1) n}\,\Re\!\bigl(C_{mn}^{*}\,y_m[n]\bigr)$
    \EndFor
\EndFor
\State $w[0,:] \gets w[0,:] / \sqrt{2}$ \Comment{rescale edge bands}
\State $w[N_f,:] \gets w[N_f,:] / \sqrt{2}$
\State \Return $w$
\end{algorithmic}
\end{algorithm}

\begin{algorithm}[H]
\caption{Inverse WDM: WDM grid $w$ $\to$ real time series $x$.}
\label{alg:inv}
\begin{algorithmic}[1]
\Require WDM grid $w \in \mathbb{R}^{(N_f+1)\times N_t}$, parameters $(N_f, N_t, \Delta t)$.
\Ensure Real time series $x \in \mathbb{R}^{N}$.
\State $w' \gets w$
\State $w'[0,:] \gets w'[0,:] \cdot \sqrt{2}$ \Comment{undo the forward's edge rescale}
\State $w'[N_f,:] \gets w'[N_f,:] \cdot \sqrt{2}$
\State $\lambda \gets 1/(\sqrt{2}\Delta t)$
\State $X \gets \mathbf{0}_{N/2+1}$ \Comment{complex RFFT vector $X\in\mathbb{C}^{N/2 + 1}$}
\For{$m = 0$ to $N_f$}
    \State Build $g \in \mathbb{C}^{N_t}$:
    \For{$n = 0$ to $N_t - 1$}
        \State $g[n] \gets w'[m, n]\,C_{mn}\,(-1)^{(m+1) n}$
        \Comment{real if $(m+n)$ even, imaginary if odd}
    \EndFor
    \State $W_m \gets \mathrm{FFT}(g)$
    \For{$k = 0$ to $N_t - 1$}
        \State $\ell \gets k + (m-1)\,N_t/2$
        \State $v \gets \lambda\,\tilde{\phi}[k - N_t/2]\,W_m[k]$
        \If{$\ell < 0$ \textbf{or} $\ell > N/2$}
            \State \textbf{continue}
            \Comment{discard at $m \in \{0, N_f\}$; redundant by Hermitian symmetry}
        \EndIf
        \State $X[\ell] \mathrel{{+}{=}} v$
    \EndFor
\EndFor
\State $x \gets \mathrm{IRFFT}(X,\ n = N)$
\State \Return $x$
\end{algorithmic}
\end{algorithm}

\begin{algorithm}[H]
\caption{Stationary forward: PSD $S(f)\to$ WDM covariance $\Lambda$.}
\label{alg:stat}
\begin{algorithmic}[1]
\Require One-sided PSD $S\in\mathbb{R}^{N/2+1}$; parameters $(N_f,N_t,\Delta t)$, $N=N_fN_t$.
\Ensure Diagonal blocks $\Lambda^{(m,m)}$, $m=0,\dots,N_f$, and off-diagonals $\Lambda^{(m,m+1)}$,
        $m=0,\dots,N_f-1$.
\State $\sigma^2[k]\gets S[k]/(2\Delta t)$ \quad for $k=0,\dots,N/2$
\State $\mu\gets 1/N$
\State $C_{mn}\gets i^{(m+n)\bmod 2}$
\State $\tilde\phi\gets$ Meyer window centered at index $0$
       \Comment{band-$m$ window is $\tilde\phi[k-\tfrac{N_t}{2}]$}
\For{each band pair $(m,m')$ with $m'\in\{m,\,m+1\}$}
    \State $\tilde K\gets\mathbf{0}_{N_t}$
    \State fill $\tilde K[k]$ from the $(m,m')$ row of Table~\ref{tab:K_recipes}
           \Comment{zero outside that row's $k$-range}
    \If{$(m,m')\in\{(0,0),\,(N_f,N_f)\}$}
        \For{$k=\tfrac{N_t}{2}+1,\dots,N_t-1$}
            \State $\tilde K[k]\gets\tilde K[N_t-k]$ \Comment{Hermitian extension}
        \EndFor
    \EndIf
    \State $K\gets N_t\,\mathrm{IFFT}(\tilde K)$ \Comment{band lag kernel $K[r]$}
    \For{$n=0,\dots,N_t-1$}
        \For{$n'=0,\dots,N_t-1$}
            \State $r\gets(n-n')\bmod N_t$
            \State $\gamma\gets C^{*}_{mn}\,C_{m'n'}$ 
            \State $s\gets(-1)^{(m+1)(n+n')}$ \Comment{parity sign}
            \State $\Lambda^{(m,m')}_{n,n'}\gets\mu\,s\,\Re[\gamma\,K[r]]$
            \If{$m\in\{0,N_f\}$ \textbf{and} $m+n$ is odd}
                \State $\Lambda^{(m,m')}_{n,n'}\gets 0$ \Comment{inactive edge half-pixel}
            \ElsIf{$m'\in\{0,N_f\}$ \textbf{and} $m'+n'$ is odd}
                \State $\Lambda^{(m,m')}_{n,n'}\gets 0$
            \EndIf
        \EndFor
    \EndFor
\EndFor
\State \Return $\{\Lambda^{(m,m)}\},\ \{\Lambda^{(m,m+1)}\}$
\end{algorithmic}
\end{algorithm}

  \begin{table}[H]
\caption{The six $K$-recipes, math convention ($\tilde\phi$ centered at index $0$). Outside the listed
$k$-range $\tilde K[k]=0$; edge-diagonal rows are Hermitian-extended after filling
(Algorithm~\ref{alg:stat}). The off-diagonal/cross rows use $\tilde\phi[k-N_t]=\tilde\phi[k]$.}
\label{tab:K_recipes}
\centering
\renewcommand{\arraystretch}{1.45}
\begin{tabular}{lll}
\hline
$(m,m')$ type & $\tilde K[k]$ & $k$-range \\
\hline
Interior diag, $1\!\le\! m\!\le\! N_f{-}1$, $m'{=}m$
  & $\tilde\phi[k-\half]^{2}\,\sigma^2[k+(m{-}1)\half]$ & $[0,\,N_t{-}1]$ \\
Interior off-diag, $1\!\le\! m\!\le\! N_f{-}2$, $m'{=}m{+}1$
  & $\tilde\phi[k-\half]\,\tilde\phi[k-N_t]\,\sigma^2[k+(m{-}1)\half]$ & $[\half,\,N_t{-}1]$ \\
Edge diag $(0,0)$
  & $\tilde\phi[k]^{2}\,\sigma^2[k]$ & $[0,\,\half{-}1]$ \\
Edge diag $(N_f,N_f)$
  & $\tilde\phi[k-\half]^{2}\,\sigma^2[k+\tfrac{N}{2}-\half]$ & $[0,\,\half]$ \\
Edge cross $(0,1)$
  & $\sqrt2\,\tilde\phi[k-\half]\,\tilde\phi[k-N_t]\,\sigma^2[k-\half]$ & $[\half,\,N_t{-}1]$ \\
Edge cross $(N_f{-}1,N_f)$
  & $\sqrt2\,\tilde\phi[k-\half]\,\tilde\phi[k-N_t]\,\sigma^2[k+\tfrac{N}{2}-N_t]$ & $[\half,\,N_t{-}1]$ \\
\hline
\end{tabular}
\end{table}

\begin{algorithm}[H]
\caption{Stationary inverse: WDM covariance $\Lambda\to$ PSD $S(f)$ (exact, with off-diagonal blocks).}
\label{alg:psd_backward}
\begin{algorithmic}[1]
\Require Diagonal blocks $\Lambda^{(m,m)}$ and off-diagonals $\Lambda^{(m,m+1)}$ with active-$n$ lists;
         parameters $(N_f,N_t,\Delta t)$, $T=N\Delta t$.
\Ensure One-sided PSD $S\in\mathbb{R}^{N/2+1}$.
\State $\tilde\phi\gets$ Meyer window centered at index $0$
       \Comment{band-$m$ window is $\tilde\phi[k-\half]$}
\State $C_{mn}\gets i^{(m+n)\bmod 2}$
\State $\tilde A_m\gets 1/\sqrt2$ for interior $m$, and $1$ for $m\in\{0,N_f\}$
       \Comment{$\tilde A_m=1/(2A_m)$}
\State $S[\ell]\gets 0$ \quad for all $\ell$
\For{each band pair $(m,m')$ with $m'\in\{m,\,m+1\}$}
    \State $w\gets 1$ if $m'=m$, else $w\gets 2$ \Comment{off-diagonals counted twice}
    \State $\mu\gets N_t$ if $m$ and $m'$ are both interior, else $\mu\gets N_t/2$
    \For{$r=0,\dots,N_t-1$}
        \State pick an active pair $(n,n')$ with $(n-n')\bmod N_t=r$
               \Comment{$t[r]\gets 0$ if none exists}
        \State $t[r]\gets\mu\,C_{mn}\,C^{*}_{m'n'}\,\Lambda^{(m,m')}_{n,n'}$
    \EndFor
    \State $g[r]\gets(-1)^{(m+1)r}\,t[r]$ 
    \State $D\gets\mathrm{FFT}(g)$
    \State $c\gets (2w/T)\,\tilde A_m\,\tilde A_{m'}$ \Comment{prefactor for this band pair}
    \For{$k=0,\dots,N_t-1$}
        \State $\ell\gets k+(m-1)\half$
        \If{$\ell<0$ \textbf{or} $\ell>N/2$}
            \State \textbf{continue}
        \EndIf
        \State $W\gets\tilde\phi[k-\half]\;\tilde\phi[k-(m'{-}m{+}1)\half]$
               \Comment{$=\tilde\phi[k-\half]^2$ (diag); $\tilde\phi[k-\half]\,\tilde\phi[k-N_t]$ (off-diag)}
        \State $S[\ell]\mathrel{{+}{=}}c\,W\,\Re\,D[k]$
    \EndFor
\EndFor
\State \Return $S$
\end{algorithmic}
\end{algorithm}

\section{Correspondence between WDM and the Fourier transform}
\label{app:correspondence}

In this Appendix, we derive the correspondence between the WDM and Fourier transforms in the appropriate limit, $N_t = 2$ and $N_f = N/2$.
Starting with Eq.~\eqref{eq:inverseinterior_final} for the interior bands, the contribution of WDM band $m$ to the frequency domain data is
\begin{equation}
    \tilde{x}_m(f + m\Delta F) = \frac{1}{\sqrt{2}} \tilde{\phi}(f)\sum_{n=0}^{N_t-1}w_{mn}C_{mn}(-1)^{mn}e^{-2\pi inf\Delta T}\,,
\end{equation}
where the discrete values of the frequency series fall at $f = k\Delta f$ where $k = \{-N_t/2,\ldots,N_t/2-1\}$.
Setting $N_t = 2$ and $k = \{-1,0\}$, hence $\Delta f = \Delta F$, we have
\begin{equation}
    \tilde{x}_m(k\Delta F + m\Delta F) = \frac{1}{\sqrt{2}} \tilde{\phi}(k\Delta F)\sum_{n=0}^{1} w_{mn} C_{mn}(-1)^{mn}e^{-2\pi ink\Delta F\Delta T}\,.
\end{equation}
Since $\tilde{\phi}(-\Delta F) = 0$ by the compact support of the filter, only $k = 0$ survives
\begin{equation}
    \tilde{x}_m(m\Delta F) = \frac{1}{\sqrt{2\Delta F}} \left[w_{m0}C_{m0} + (-1)^m w_{m1}C_{m1}\right]\,,
\end{equation}
where we have used $\tilde{\phi}(0) = 1/\sqrt{\Delta F}$.
Finally, noting that $(-1)^m C_{m1}= i C_{m0}$ and that only band $m$ contributes, we can simplify this expression to
\begin{equation}
    \tilde{x}(m\Delta F)=\frac{1}{\sqrt{2\Delta F}} C_{m0}\left(w_{m0} + iw_{m1}\right)\,.
    \label{eq:WDMtofreq}
\end{equation}

The calculation for the edge bands $m=0,N_f$ proceeds similarly and gives
\begin{equation}
    \tilde{x}(0) = \frac{1}{\sqrt{\Delta F}}w_{00}\,, \qquad 
    \tilde{x}(N_f\Delta F) = \frac{1}{\sqrt{\Delta F}}
    \begin{cases}
    w_{N_f 0} & N_f~\mathrm{even}\\
    -w_{N_f 1} & N_f~\mathrm{odd}
    \end{cases}\,,
\end{equation}
both of which are real as expected.

In the limit of $N_t = 2$ and $N_f = N/2$, each complex value of the frequency series is given by a set of two real values in the WDM domain.

\section{The short-time Fourier transform}
\label{app:SFT}

By the Balian-Low theorem, the short-time Fourier transform (SFT) cannot be simultaneously orthonormal and well-localized in both time and frequency.

The common application of searches for continuous waves from isolated neutron stars prioritizes orthonormality~\cite{Jaranowski:1998qm}, which is achieved by making the rectangular SFT windows non-overlapping. 
While this has good localization in time,  the Balian-Low theorem guarantees poor localization in frequency. 
For example, for a rectangular window in the time domain
\begin{equation}
    g_m(t) = \begin{cases} 1 & m\, \Delta T \le t < (m+1)\, \Delta T \\ 0 & \text{otherwise} \end{cases}\,,
\end{equation}
the frequency domain is
\begin{equation}
    \tilde{g}_m(f) = \Delta T \, \frac{\sin\left(2\pi f \Delta T / 2\right)}{2\pi f \Delta T/2}e^{- 2\pi i f\Delta T \left(m + \frac{1}{2}\right)} \,.
\end{equation}
This $\operatorname{sinc}$ function is maximized at $f = 0$ and falls off as $1/f$ and therefore has infinite second frequency moment
\begin{equation}
    \int_{-\infty}^{\infty} f^2 |\tilde{g}_m(f)|^2 \diff f \sim \int_{-\infty}^\infty \diff f= \infty\,.
\end{equation}
As an outcome, the transform of a signal has support beyond the time-frequency bins the signal crosses, smearing to adjacent frequencies as $1/f$.
Isolated neutron star signals are intrinsically compact in frequency.
Even with the frequency spread due to pulsar spin evolution and Doppler modulation, support is confined within a single frequency bin per SFT for typical SFT durations.
For such monochromatic signals, the frequency smearing reduces to simply the window's spectral response, i.e., the $\operatorname{sinc}$ Dirichlet kernel~\cite{Williams:1999nt}.
Sub-bin frequency offsets are recovered in demodulation~\cite{Williams:1999nt}, while the frequency drift across the observation is tracked by demodulating each SFT to the signal's expected frequency.
Compact binary signals have a faster frequency evolution that cannot be contained within a single bin per SFT.
For a linear frequency evolution, the response generalizes to the Fresnel kernel~\cite{Tenorio:2025gci} that combines two contributions: the frequency bins that the signal crosses within an SFT and the $1/f$ rectangular window smearing. 
In both cases, truncation and a corresponding error tolerance are required in the number of frequency bins used to evaluate the transform. 
Alternatively, non-rectangular windows can achieve better frequency localization (steeper than $1/f$ kernel fall-off), but at the expense of overlapping windows in time and non-orthogonality~\cite{Du:2025fes}.

Simultaneous good time and frequency localization can only be achieved by oversampled Gabor frames, $\Delta T \Delta F <1$, thus abandoning orthonormality.
The practical implication is that the covariance matrix, used to evaluate likelihoods and inner products, has two contributions: the noise correlations and the cross-terms between atoms.
For a Gabor frame with atoms $g_{mn}$ centered at time $n\Delta T$ and frequency $m\Delta F$, the data coefficients are
\begin{equation}
d_{mn} = \int d(t)\, g_{mn}^*(t)\, dt \,,
\end{equation}
where
\begin{equation}
\int g_{mn}^*(t) g_{m'n'}(t) \diff t \neq \delta_{mm'}\delta_{nn'}\,.
\end{equation}
Inner products, such as the data/signal term in the likelihood, take the schematic form
\begin{align}
(d \mid h)&\sim\sum_{ij} d^*_i C^{-1}_{ij} h_j\,,\nonumber \\
&\sim \sum_{mn,m'n'} d_{mn}^*\, G_{mn,m'n'}\, h_{m'n'}\,,
\end{align}
where the matrix $G_{mn,m'n'}$
encodes both the noise covariance (present in both orthonormal bases and frames) and the frame correlations, i.e., the overlap between distinct atoms.
The precise form of $G$ depends on whether the frame is tight, with generic frames involving a dual frame that further complicates the expressions. 
In either case $G$ is generally dense, entangling the noise covariance and the frame's structure.
Separating the noise and frame contributions is not possible outside of special cases such as the slowly-varying PSD limit where $\mathbf C^{-1}$ is diagonal in the frequency domain.

\section{Toeplitz derivation}
\label{app:toeplitz}

In this Appendix, we show that Eq.~\eqref{eq:tmmprime} has a Toeplitz structure, i.e., it only depends on $r=n-n'$.
Per Eq.~\eqref{eq:lambda_stationary_discretize} (via a change of variables $f\rightarrow f-m\Delta F$), the WDM domain noise covariance matrix for stationary noise is
\begin{equation}
  \Lambda_{(mn)(m'n')}
  =A_{mm'}(-1)^{rm}\,\Re\!\left[\,C_{m'n'}C_{mn}^{*}
     \int_{(m-1)\Delta F}^{(m+1)\Delta F}\! S(f)\,\tilde\phi(f-m\Delta F)\,\tilde\phi(f-m'\Delta F)\,
       e^{2\pi irf\Delta T}\,\mathrm{d}f\,\right]\,.
  \label{eq:start}
\end{equation}
We introduce the cross-band noise PSD as
\begin{equation}
    \tilde{K}_{mm'}(f) = S(f) \tilde{\phi}(f - m\Delta F) \tilde{\phi}(f - m'\Delta F)\,,
\end{equation}
and its inverse Fourier transform as
\begin{equation}
    K_{mm'}[r] = \int_{(m - 1)\Delta F}^{(m + 1)\Delta F}S(f)\tilde{\phi}(f - m\Delta F) \tilde{\phi}(f - m'\Delta F)e^{2\pi irf\Delta T}\diff f\,,
\end{equation}
which depends only on the lag $r$.
Rewriting Eq.~\eqref{eq:start} with these definitions, we find
\begin{equation}
    \Lambda_{(mn)(m'n')} = A_{mm'}\Re\left(C_{m'n'}C_{mn}^*K_{mm'}[r]\right)\,.
\end{equation}
Equation~\eqref{eq:tmmprime} depends on $C_{mn}C_{m'n'}^*\Lambda_{(mn)(m'n')}$ which is then
\begin{align}
C_{mn}C_{m'n'}^*\Lambda_{(mn)(m'n')} &= A_{mm'} C_{mn}C_{m'n'}^* \Re\left(C_{m'n'}C_{mn}^*K_{mm'}[r]\right)\nonumber \\
& = \frac{A_{mm'}}{2}C_{mn}C_{m'n'}^* \left\{ C_{m'n'}C_{mn}^*K_{mm'}[r] + C^*_{m'n'}C_{mn}K^*_{mm'}[r]\right\}\nonumber \\
&=\frac{A_{mm'}}{2} \left\{ K_{mm'}[r] + \left(C^*_{m'n'}C_{mn}\right)^2K^*_{mm'}[r]\right\}\,,
\end{align}
where we have used $|C_{m'n'}C_{mn}^*|^2 = 1$.
To show that this is Toeplitz, it remains to show that $\left(C^*_{m'n'}C_{mn}\right)^2$ is a function of $r$ only.
We find
\begin{equation}
    (C^*_{m'n'}C_{mn})^2 = (-1)^{m'+n'}(-1)^{m+n}=(-1)^{(m+m') + (n+n')}=(-1)^{(m+m')+(n-n')} = (-1)^{(m+m')+r}\,,
\end{equation}
where from the third to the fourth equality, we use the fact that $n + n'$ and $n - n'$ differ by $2n'$ and have the same parity for all $n, n'$.
Therefore $C_{mn}C_{m'n'}^*\Lambda_{(mn)(m'n')}$ and Eq.~\eqref{eq:tmmprime} depend only on $r$, and therefore have Toeplitz structure.

In other words, for stationary noise the WDM covariance matrix has a parity-modulated Toeplitz structure, which becomes exactly Toeplitz when multiplied by $C_{mn}C_{m'n'}^*$.

\section{WDM and the Heisenberg group}
\label{app:WDM_Heisenberg}
The Gabor atoms of Eq.~\eqref{eq:gabor} can be written in the form of a translation operator $T_{a}$ and a modulation operator $M_{b}$.
Their respective actions are
\begin{align}
    T_ag(t) &= g(t-a)\,,\\
    M_bg(t) &= e^{2\pi i b t} g(t)\,.
\end{align}
Performing $n$ translations and $m$ modulations returns Eq.~\eqref{eq:gabor}:
\begin{equation}
    (M_b)^m(T_a)^ng(t) = (e^{2\pi i b t})^mg(t-na) = e^{2\pi i mbt}g(t-na)\,.
\end{equation}
Individually, the translation and modulation operations are one-parameter groups on $L^2(\mathbb{R})$ with parameters $a$ and $b$.
However, time-frequency shifts of these two operators do not commute and miss closure by a phase as
\begin{equation}
    (M_bT_a)(M_yT_x) = e^{-2 \pi i a y} M_{b + y}T_{a + x}\,.
\end{equation}
Adding an operator $E_\tau$ such that $E_\tau g(t) = \tau g(t)$ with $|\tau|=1$ covers the phase and closes the group.
These three operators, $T_a,M_b,E_\tau$, taken together form the Heisenberg group~\cite{grochenig2001foundations}.

The WDM basis can be written in terms of translations and modulations
\begin{align}
  g_{mn}(t)&=A_m\,T_{n\Delta T}\left[C_{mn}\,M_{+m\Delta F}
  \;+\;C^{*}_{mn}\,M_{-m\Delta F}\right]\,\phi(t)\nonumber\\
  &=A_m\,T_{n\Delta T} \left[C_{mn}e^{2\pi i m \Delta F t} + C_{mn}^*e^{-2\pi i m \Delta F t}\right]\phi(t)\nonumber\\
  &=A_m \left[C_{mn}e^{2\pi i m \Delta F (t - n\Delta T)} + C_{mn}^*e^{-2\pi i m \Delta F(t-n\Delta T)}\right]\phi(t-n\Delta T)\,.
\label{eq:wdm_worked_out}
\end{align}
Breaking this equation into even and odd $m+n$ pieces and identifying the sines and cosines yields
\begin{equation}
    g_{mn}(t)=\begin{cases}
        2A_m\,\cos\left(2\pi i m \Delta F(t-n\Delta T)\right)\phi(t-n\Delta T)&m+n\,\,\mathrm{even}\\
        -2A_m\,\sin\left(2\pi i m \Delta F(t-n\Delta T)\right)\phi(t-n\Delta T)&m+n\,\,\mathrm{odd}\,,
    \end{cases}
\end{equation}
which matches Eq.~\eqref{eq:gmunu_int}.

\bibliography{apssamp}

@unpublished{WDMII,
  author = {Johnson, Aaron and Chatziioannou, Katerina and Rosati, Robert},
  title  = {The WDM time-frequency transform in Gravitational-Wave Data Analysis {I}{I}: Evolutionary Spectrum and Noise Covariance},
  note   = {in preparation},
  year   = {2026}
}

@article{Gabor1946TheoryCommunication,
  author  = {Dennis Gabor},
  title   = {Theory of Communication},
  journal = {Journal of the Institution of Electrical Engineers},
  volume  = {93},
  number  = {26},
  pages   = {429--457},
  year    = {1946}
}

@book{bratteli2002wavelets,
  author    = {Bratteli, Ola and Jorgensen, Palle E. T.},
  title     = {Wavelets Through a Looking Glass: The World of the Spectrum},
  series    = {Applied and Numerical Harmonic Analysis},
  publisher = {Birkh{\"a}user},
  address   = {Boston, MA},
  year      = {2002},
  doi       = {10.1007/978-0-8176-8144-9},
  isbn      = {978-0-8176-4280-8},
}

@book{grochenig2001foundations,
  title     = {Foundations of Time-Frequency Analysis},
  author    = {Gr{\"o}chenig, Karlheinz},
  series    = {Applied and Numerical Harmonic Analysis},
  year      = {2001},
  publisher = {Birkh{\"a}user Boston},
  address   = {Boston, MA},
  edition   = {1},
  doi       = {10.1007/978-1-4612-0003-1},
  isbn      = {978-0-8176-4022-4},
}

@article{Katz:2024oqg,
    author = "Katz, Michael L. and Karnesis, Nikolaos and Korsakova, Natalia and Gair, Jonathan R. and Stergioulas, Nikolaos",
    title = "{Efficient GPU-accelerated multisource global fit pipeline for LISA data analysis}",
    eprint = "2405.04690",
    archivePrefix = "arXiv",
    primaryClass = "gr-qc",
    doi = "10.1103/PhysRevD.111.024060",
    journal = "Phys. Rev. D",
    volume = "111",
    number = "2",
    pages = "024060",
    year = "2025"
}

@article{Pearson:2025wfd,
    author = "Pearson, Noah and Cornish, Neil J.",
    title = "{Handling data gaps for the next generation of gravitational-wave observatories}",
    eprint = "2509.05479",
    archivePrefix = "arXiv",
    primaryClass = "gr-qc",
    doi = "10.1103/htg6-w4yy",
    journal = "Phys. Rev. D",
    volume = "113",
    number = "6",
    pages = "064033",
    year = "2026"
}

@Article{harris2020array,
 title         = {Array programming with {NumPy}},
 author        = {Charles R. Harris and K. Jarrod Millman and St{\'{e}}fan J.
                 van der Walt and Ralf Gommers and Pauli Virtanen and David
                 Cournapeau and Eric Wieser and Julian Taylor and Sebastian
                 Berg and Nathaniel J. Smith and Robert Kern and Matti Picus
                 and Stephan Hoyer and Marten H. van Kerkwijk and Matthew
                 Brett and Allan Haldane and Jaime Fern{\'{a}}ndez del
                 R{\'{i}}o and Mark Wiebe and Pearu Peterson and Pierre
                 G{\'{e}}rard-Marchant and Kevin Sheppard and Tyler Reddy and
                 Warren Weckesser and Hameer Abbasi and Christoph Gohlke and
                 Travis E. Oliphant},
 year          = {2020},
 month         = sep,
 journal       = {Nature},
 volume        = {585},
 number        = {7825},
 pages         = {357--362},
 doi           = {10.1038/s41586-020-2649-2},
 publisher     = {Springer Science and Business Media {LLC}},
 url           = {https://doi.org/10.1038/s41586-020-2649-2}
}

@article{Balian1981UnPrincipe,
  author  = {Roger Balian},
  title   = {Un principe d'incertitude fort en th{'e}orie du signal ou en m{'e}canique quantique},
  journal = {Comptes Rendus de l'Acad{'e}mie des Sciences de Paris, S{'e}rie II},
  volume  = {292},
  pages   = {1357--1362},
  year    = {1981}
}

@incollection{Low1985CompleteSets,
  author    = {Francis E. Low},
  title     = {Complete Sets of Wave Packets},
  booktitle = {A Passion for Physics---Essays in Honor of Geoffrey Chew},
  pages     = {17--22},
  year      = {1985},
  publisher = {World Scientific}
}

@article{Digman:2022jmp,
    author = "Digman, Matthew C. and Cornish, Neil J.",
    title = "{LISA Gravitational Wave Sources in a Time-varying Galactic Stochastic Background}",
    eprint = "2206.14813",
    archivePrefix = "arXiv",
    primaryClass = "astro-ph.IM",
    doi = "10.3847/1538-4357/ac9139",
    journal = "Astrophys. J.",
    volume = "940",
    number = "1",
    pages = "10",
    year = "2022"
}

@article{Digman:2022igm,
    author = "Digman, Matthew C. and Cornish, Neil J.",
    title = "{Parameter estimation for stellar-origin black hole mergers in LISA}",
    eprint = "2212.04600",
    archivePrefix = "arXiv",
    primaryClass = "gr-qc",
    doi = "10.1103/PhysRevD.108.023022",
    journal = "Phys. Rev. D",
    volume = "108",
    number = "2",
    pages = "023022",
    year = "2023"
}

@article{Necula_2012,
doi = {10.1088/1742-6596/363/1/012032},
url = {https://doi.org/10.1088/1742-6596/363/1/012032},
year = {2012},
month = {jun},
publisher = {},
volume = {363},
number = {1},
pages = {012032},
author = {Necula, V and Klimenko, S and Mitselmakher, G},
title = {Transient analysis with fast Wilson-Daubechies time-frequency transform},
journal = {Journal of Physics: Conference Series},
}

@article{doi:10.1137/0522035,
author = {Daubechies, Ingrid and Jaffard, St\'{e}phane and Journ\'{e}, Jean-Lin},
title = {A Simple Wilson Orthonormal Basis with Exponential Decay},
journal = {SIAM Journal on Mathematical Analysis},
volume = {22},
number = {2},
pages = {554-573},
year = {1991},
doi = {10.1137/0522035},
URL = {https://dx.doi.org/10.1137/0522035},
}

@unpublished{Cornish:2025awt,
    author = "Cornish, Neil J.",
    title = "{Non-stationary noise in gravitational wave analyses: The wavelet domain noise covariance matrix}",
    eprint = "2511.10632",
    archivePrefix = "arXiv",
    primaryClass = "gr-qc",
    month = "11",
    journal = "",
    note = "",
    year = "2025"
}

@article{Cornish:2020odn,
    author = "Cornish, Neil J.",
    title = "{Time-Frequency Analysis of Gravitational Wave Data}",
    eprint = "2009.00043",
    archivePrefix = "arXiv",
    primaryClass = "gr-qc",
    doi = "10.1103/PhysRevD.102.124038",
    journal = "Phys. Rev. D",
    volume = "102",
    number = "12",
    pages = "124038",
    month = "8",
    year = "2020"
}

@unpublished{Wilson1987,
  author = {Wilson, Kenneth G.},
  title  = {Generalized {Wannier} Functions},
  note   = {Cornell University preprint},
  year   = {1987}
}

@article{Klimenko:2005xv,
    author = "Klimenko, S. and Mohanty, S. and Rakhmanov, Malik and Mitselmakher, Guenakh",
    title = "{Constraint likelihood analysis for a network of gravitational wave detectors}",
    eprint = "gr-qc/0508068",
    archivePrefix = "arXiv",
    doi = "10.1103/PhysRevD.72.122002",
    journal = "Phys. Rev. D",
    volume = "72",
    pages = "122002",
    year = "2005"
}

@article{Jaranowski:1998qm,
    author = "Jaranowski, Piotr and Krolak, Andrzej and Schutz, Bernard F.",
    title = "{Data analysis of gravitational - wave signals from spinning neutron stars. 1. The Signal and its detection}",
    eprint = "gr-qc/9804014",
    archivePrefix = "arXiv",
    doi = "10.1103/PhysRevD.58.063001",
    journal = "Phys. Rev. D",
    volume = "58",
    pages = "063001",
    year = "1998"
}

@article{Williams:1999nt,
    author = "Williams, Peter R. and Schutz, Bernard F.",
    editor = "Meshkov, S.",
    title = "{An Efficient matched filtering algorithm for the detection of continuous gravitational wave signals}",
    eprint = "gr-qc/9912029",
    archivePrefix = "arXiv",
    doi = "10.1063/1.1291918",
    journal = "AIP Conf. Proc.",
    volume = "523",
    number = "1",
    pages = "473--476",
    year = "2000"
}

@article{Tenorio:2025gci,
    author = "Tenorio, Rodrigo and Gerosa, Davide",
    title = "{Scalable data-analysis framework for long-duration gravitational waves from compact binaries using short Fourier transforms}",
    eprint = "2502.11823",
    archivePrefix = "arXiv",
    primaryClass = "gr-qc",
    doi = "10.1103/PhysRevD.111.104044",
    journal = "Phys. Rev. D",
    volume = "111",
    number = "10",
    pages = "104044",
    year = "2025"
}

@article{Du:2025fes,
    author = "Du, Minghui and Luo, Ziren and Xu, Peng",
    title = "{Enhancing Taiji{\textquoteright}s parameter estimation under nonstationarity: A time-frequency domain framework for Galactic binaries and instrumental noises}",
    eprint = "2506.10599",
    archivePrefix = "arXiv",
    primaryClass = "gr-qc",
    doi = "10.1103/gpmh-1hqx",
    journal = "Phys. Rev. D",
    volume = "112",
    number = "8",
    pages = "083036",
    year = "2025"
}

@unpublished{Isi:2021iql,
    author = "Isi, Maximiliano and Farr, Will M.",
    title = "{Analyzing black-hole ringdowns}",
    eprint = "2107.05609",
    archivePrefix = "arXiv",
    primaryClass = "gr-qc",
    reportNumber = "LIGO-P2100227",
    month = "7",
    year = "2021"
}

@book{Meyer1992,
    author    = "Meyer, Yves",
    title     = "{Wavelets and Operators}",
    series    = "Cambridge Studies in Advanced Mathematics",
    volume    = "37",
    publisher = "Cambridge University Press",
    address   = "Cambridge",
    year      = "1992"
}

@article{Cutler:1994ys,
    author = "Cutler, Curt and Flanagan, Eanna E.",
    title = "{Gravitational waves from merging compact binaries: How accurately can one extract the binary's parameters from the inspiral wave form?}",
    eprint = "gr-qc/9402014",
    archivePrefix = "arXiv",
    doi = "10.1103/PhysRevD.49.2658",
    journal = "Phys. Rev. D",
    volume = "49",
    pages = "2658--2697",
    year = "1994"
}

@article{Whittle1957,
    author  = "Whittle, P.",
    title   = "{Curve and Periodogram Smoothing}",
    journal = "J. Roy. Stat. Soc. B",
    volume  = "19",
    number  = "1",
    pages   = "38--63",
    year    = "1957"
}

@unpublished{Colpi:2024xhw,
    author = "Colpi, Monica and others",
    title = "{LISA Definition Study Report}",
    eprint = "2402.07571",
    archivePrefix = "arXiv",
    primaryClass = "astro-ph.CO",
    month = "2",
    year = "2024"
}

@article{Markel1971,
    author  = "Markel, J. D.",
    title   = "{FFT pruning}",
    journal = "IEEE Trans. Audio Electroacoust.",
    volume  = "19",
    number  = "4",
    pages   = "305--311",
    doi     = "10.1109/TAU.1971.1162205",
    year    = "1971"
}

@article{Wiener1930,
    author  = "Wiener, Norbert",
    title   = "{Generalized harmonic analysis}",
    journal = "Acta Math.",
    volume  = "55",
    pages   = "117--258",
    doi     = "10.1007/BF02546511",
    year    = "1930"
}

@article{Khinchin1934,
    author  = "Khinchin, A.",
    title   = "{Korrelationstheorie der station{\"a}ren stochastischen Prozesse}",
    journal = "Math. Ann.",
    volume  = "109",
    number  = "1",
    pages   = "604--615",
    doi     = "10.1007/BF01449156",
    year    = "1934"
}

@book{Creighton:2011zz,
  author    = {Creighton, Jolien D. E. and Anderson, Warren G.},
  title      = {{Gravitational-Wave Physics and Astronomy: An Introduction to Theory, Experiment and Data Analysis}},
  publisher  = {Wiley},
  year       = {2011},
  isbn       = {978-3-527-40886-3},
}

@misc{lalsuite,
       author         = "{LIGO Scientific Collaboration} and {Virgo Collaboration} and {KAGRA Collaboration}",
       title          = "{LVK} {A}lgorithm {L}ibrary - {LALS}uite",
       howpublished   = "Free software (GPL)",
       doi            = "10.7935/GT1W-FZ16",
       year           = "2018"
 }

@article{numpy,
  author  = {Harris, Charles R. and Millman, K. Jarrod and van der Walt, St{\'e}fan J.
             and Gommers, Ralf and Virtanen, Pauli and Cournapeau, David and Wieser, Eric
             and Taylor, Julian and Berg, Sebastian and Smith, Nathaniel J. and Kern, Robert
             and Picus, Matti and Hoyer, Stephan and van Kerkwijk, Marten H. and Brett, Matthew
             and Haldane, Allan and del R{\'i}o, Jaime Fern{\'a}ndez and Wiebe, Mark
             and Peterson, Pearu and G{\'e}rard-Marchant, Pierre and Sheppard, Kevin
             and Reddy, Tyler and Weckesser, Warren and Abbasi, Hameer and Gohlke, Christoph
             and Oliphant, Travis E.},
  title   = {Array programming with {NumPy}},
  journal = {Nature},
  year    = {2020},
  volume  = {585},
  number  = {7825},
  pages   = {357--362},
  doi     = {10.1038/s41586-020-2649-2},
  url     = {https://doi.org/10.1038/s41586-020-2649-2},
  publisher = {Springer Science and Business Media {LLC}}
}

@article{scipy,
  author  = {Virtanen, Pauli and Gommers, Ralf and Oliphant, Travis E. and Haberland, Matt
             and Reddy, Tyler and Cournapeau, David and Burovski, Evgeni and Peterson, Pearu
             and Weckesser, Warren and Bright, Jonathan and {van der Walt}, St{\'e}fan J.
             and Brett, Matthew and Wilson, Joshua and Millman, K. Jarrod and Mayorov, Nikolay
             and Nelson, Andrew R. J. and Jones, Eric and Kern, Robert and Larson, Eric
             and Carey, C J and Polat, {\.I}lhan and Feng, Yu and Moore, Eric W.
             and {VanderPlas}, Jake and Laxalde, Denis and Perktold, Josef and Cimrman, Robert
             and Henriksen, Ian and Quintero, E. A. and Harris, Charles R. and Archibald, Anne M.
             and Ribeiro, Ant{\^o}nio H. and Pedregosa, Fabian and {van Mulbregt}, Paul
             and {SciPy 1.0 Contributors}},
  title   = {{SciPy} 1.0: Fundamental Algorithms for Scientific Computing in {Python}},
  journal = {Nature Methods},
  year    = {2020},
  volume  = {17},
  pages   = {261--272},
  doi     = {10.1038/s41592-019-0686-2},
  url     = {https://doi.org/10.1038/s41592-019-0686-2},
  publisher = {Springer Science and Business Media {LLC}}
}

@article{matplotlib,
  author  = {Hunter, J. D.},
  title   = {Matplotlib: A 2{D} graphics environment},
  journal = {Computing in Science \& Engineering},
  year    = {2007},
  volume  = {9},
  number  = {3},
  pages   = {90--95},
  doi     = {10.1109/MCSE.2007.55},
  publisher = {IEEE Computer Society}
}

\end{document}